\documentclass[10pt]{article}
    \usepackage{amsmath, amssymb, amsfonts, bm} \usepackage{amsthm} \usepackage{graphicx} \usepackage{natbib}   
    \usepackage{geometry} \usepackage{sectsty} \usepackage[font=small]{caption} \usepackage{mathrsfs} 
    \usepackage{setspace} \usepackage{subcaption} \usepackage{xcolor} \usepackage{multirow}
    \usepackage{pdflscape} 
    \usepackage[
        colorlinks=true,      
        citecolor=blue,       
        linkcolor=black,      
        urlcolor=magenta      
    ]{hyperref}
    \newcommand{\ind}[2]{\mathbf{1}_{#1}\!\left\{#2\right\}}
    \newcommand{\prob}[1]{ \mathbf{Pr}\big( #1\big) }
    \newcommand{\Prob}[1]{ \mathbf{Pr}\left( #1 \right) }
    \newcommand{\expect}[1]{ \mathbb{E}\big[ #1\big] }
    \newcommand{\Expect}[1]{ \mathbb{E}\left[ #1 \right] }
    \newcommand{\condprob}[2]{ \mathbf{Pr}\big( #1 \,\big|\, #2 \big) }
    \newcommand{\condProb}[2]{ \mathbf{Pr}\bigg( #1 \,\bigg|\, #2 \bigg)  }
    \newcommand{\condexpect}[2]{ \mathbb{E}\big[ #1 \,\big|\, #2 \big] }
    \newcommand{\condExpect}[2]{ \mathbb{E}\left[\, #1 \,\middle|\, #2 \right] }
    
    \newcommand{\argmin}[2]{ \underset{#1}{\textnormal{argmin\,}}{#2} }
    \newcommand{\supp}[1]{ \textnormal{supp(}{#1}\textnormal{)} }
    \newcommand{\real }[1]{ \mathbb{R}^{ #1 } }
    \newcommand{\integer }{ \mathbb{Z}_{+} } 
    
    \newcommand{\norm}[1]{ \big\| #1 \big\| }
    \newcommand{\Norm}[1]{ \bigg\| #1 \bigg\| }

    \newcommand{\avg}[1]{ \frac{1}{N}\sum_{i=1}^{N}{#1} }
    \newcommand{\minsingular}[1]{ \sigma_{\textnormal{min}}\left( #1 \right) }

    \newcommand{\examplecont}{  \addtocounter{example}{-1}  \begin{example}[continued] }
    \newcommand{\cald}{ \mathcal{D} }
    
    \newcommand{\caln}{ \mathcal{N}}
    \renewcommand{\iff}{ {\quad\Leftrightarrow\quad} }

    \newcommand{\expectF}[2]{ \mathbb{E}_{#1}\big[ #2\big] }
    \newcommand{\condexpectF}[3]{ \mathbb{E}_{#1}\big[ #2 \,\big|\, #3 \big] }
    \newcommand{\ExpectF}[2]{ \mathbb{E}_{#1}\bigg[ #2 \bigg] }
        
    \newcommand{\classifier}[1]{ \textbf{#1} }
    \newcommand{\smth}[1]{ \widetilde{ \mathscr{#1} } }
    
    \newcommand{\group}{ \pmb{\gamma} }
    \newcommand{\Group}{ \pmb{\Gamma} }

    \newcommand{\Ker}[1]{ \pmb{K}\left( {#1}/{{b}_{N}} \right) }
    \newcommand{\KER}[1]{ \pmb{K}\left( \frac{#1}{{b}_{N}} \right) }
    \newcommand{\discfunc }{ {\pmb{\delta}} }
    \newcommand{\typespace}{ { \mathscr{G} } }
    \newcommand{\plugin}{ {\widetilde{\theta}_{\textnormal{plug-in}} } }
    \newcommand{\oracle}{ {\widetilde{\theta}_{\textnormal{oracle}} } }
    \newcommand{\eucvec}[1]{ \pmb{e}_{#1} }
    \newcommand{\dt}[1]{ \frac{d}{d {\tau}}{#1}\Big|_{\tau = 0} }
    
 \numberwithin{equation}{section}
    
    \theoremstyle{plain}
        \newtheorem{theorem}{Theorem}
        \newtheorem{lemma}{Lemma}
        \newtheorem{proposition}{Proposition}
        
        \newtheorem{definition}{Definition}
        \newtheorem{assumption}{Assumption}
        
    \theoremstyle{definition}
        \newtheorem{example}{Example}
        \newtheorem{remarkk}{Remark}

\begin{document}

\title{
    \textbf{Valid GMM Inference with Non-parametrically Generated Discrete Type}
    \thanks{I am deeply grateful to Andres Aradillas-Lopez for not only insightful feedback but also invaluable guidance and unwavering encouragement. I would also like to thank Patrik Guggenberger, Sungjae Jun, Marc Henry, Keisuke Hirano, Thomas Russell, Amilcar Velez and Lingzhou Xue for the constructive comments that helped sharpen the motivation and theory of this work, as well as its possible extensions. Finally, I am grateful to my classmates, especially Jiaqi Huang, Ryo Kimoto and Juhyun Oh, for their thoughtful comments.}
}
\author{$
    \begin{array}[c]{c}%
    \textnormal{Taegyu Yang}\footnote{tzy5176@psu.edu, taegyuyang94@gmail.com}\\
    \textnormal{Department of Economics}\\
    \textnormal{Pennsylvania State University}%
    \end{array} $}
    \date{\today 
    \vspace{0.25in}}
\maketitle

\begin{abstract}
    The recent progress in machine learning and classification methods has enabled researchers to predict economic discrete attributes that are not observed in standard datasets. These range from missing demographics, such as ethnicity and poverty status, to economic categories, such as innovative firms, metropolitan areas, occupational groups, binary treatment status, and counterfactual status. Although the plug-in procedure is widely used in empirical applications, nonparametric first-stage estimation recovering the labels can introduce regularization bias because of its slow convergence rate. Even if the double/debiased machine learning approach provides a way to correct the regularization bias, the according bias-correction term is not well defined in our problem, because even a small first-stage error can change the type discontinuously. To address this problem, we smooth the discontinuous type indicators using a smooth kernel function and a shrinking tuning parameter $b$. We then construct a well-defined Neyman-orthogonal score and define the appropriate debiased estimator. We show that this debiased estimator satisfies $\sqrt{Nb}$-consistency and asymptotic normality under regularity conditions. The slower $\sqrt{Nb}$ convergence rate is the cost of correcting the regularization bias when the nonparametric first-stage learner enters the moment function through discontinuous functions.
\end{abstract}

\pagebreak\newpage
\section{Introduction}

We study valid inference in generalized method of moments (GMM) models in which discrete latent types are non-parametrically generated regressors. A researcher specifies a moment function $m(\cdot)$ to learn the parameter of interest $\theta$, and its truth $\theta_0$ satisfies the following moment condition:
\[ \expect{ m(\theta, \group_{0,i}, W_{i}) } = 0 \iff \theta = \theta_0, \]
where $W_i$ denotes the data and $\group_{0,i}\in\{0,1\}$ the unobserved discrete type. To recover the latent type, the researcher specifies a classification algorithm, consisting of a loss function identifying a nuisance function $\delta(Z_i)$ representing group-probability or distance to the group-centroid, and a classifier $\mathbf{1}\{ \delta(Z_i) \geq \textnormal{cutoff} \}$.

The recent advances in machine learning (ML) and classification methods have made it possible to predict discrete attributes that are not usually measured in conventional datasets. Applications range from socioeconomic demographics including ethnicity \citep{app_race_bai2024, app_race_howell2024, app_race_sutherland2024}, gender \citep{app_gender_li_liu2025, app_gender_dupas2026}, and poverty status \citep{app_poor_proxy_balboni2022, app_poor_Bau2022, app_poor_aiken2023, app_proxy_unrath2024, app_poor_aiken2026}, to economically defined types, such as innovative firms \citep{app_innovation_lerner2024, app_innovation_leippold2026}, metropolitan areas \citep{app_urban_arribas_bel2021, app_urban_vogel_et_al2023}, product-substitution or occupational groups \citep{app_labor_Dodini2023, app_labor_schuh2026, app_demand_quan_williams2021, app_demand_atalay_et_al2025, app_etc_hansen2023}. Other applications include binary treatment or discrete counterfactual status \citep{app_proxy_kleinberg2018, app_proxy_andini2022,app_proxy_black2023, app_proxy_iselin2023}, as well as complier groups in instrumental variable settings \citep{kennedy_et_al2020}.

A natural and widely used approach in these empirical practices is the naïve \emph{plug-in procedure}: replacing the label $\group_{0,i}$ with the estimated classifier $\mathbf{1}\{ \widehat{\delta}(Z_i) \geq \textnormal{cutoff} \}$, and then minimizing the following GMM criterion:
\[ \plugin = \argmin{\theta}{  \big\| \widehat{\mathbb{E}}[ m(\theta, \mathbf{1}\{\widehat{\delta}(Z_i)\geq \textnormal{cutoff}\}, W_i) ] \big\|^2  }.\]
However, nonparametric first-stage estimators usually results in regularization bias and contaminates downstream inference, unless the first-stage estimator converges to the truth sufficiently fast \citep{dml_ahrens_et_al2026} or unless the machine learner \emph{exactly} recovers all true types with high probability \citep{exact_abbe2017}. Since the seminal paper of \citet{dml_chernozhukov_et_al2018}, double/debiased machine learning (DML) approach has provided a framework to achieve valid inference when the nonparametric or ML-first-stage estimators resulting the regularization bias. A key device is to adapt the specified moment function so that the modified one obeys \emph{Neyman orthogonality} \citep{neyman1959, neyman_newey1994, dml_ahrens_et_al2026}: small perturbations of the first-stage estimate have no first-order effect on the expectation of the moment function.

In our problem, however, the standard DML approach is not directly applicable, because the bias-correction term is not well defined, even if the expectation of the moment function involving the latent type is smooths in the nuisance function $\delta(Z_i)$. Formally, the bias-correction term involves an $\mathcal{L}^2$-Riesz representer that solves a certain functional equation, but this equation has no solution in $\mathcal{L}^2$. The reason is that observations, whose discrete type changes under a small perturbation of the first-stage nuisance, are concentrated near the classification boundary and therefore even a small error can change the type discontinuously. Taking expectations smooths this discontinuity, but the resulting effect depends on the density at the boundary.

In order to construct a well-defined Neyman-orthogonal score, and thereby ensuring the existence of an $\mathcal{L}^2$-Riesz representer, particularly, we approximate the indicator $\ind{}{\delta \geq \textnormal{cutoff}}$ by $k\big( (\delta - \textnormal{cutoff})/b\big)$, where $k(\cdot)$ is a distribution kernel function and $b>0$ is a shrinking tuning parameter. We then define the debiased estimator using the smoothed types and show that it is $\sqrt{Nb}$-consistent and asymptotically normal. We emphasize that smoothing is essential to this construction: without it, the Riesz representer is not defined. Thus, the slower $\sqrt{Nb}$ convergence rate is the cost of correcting the bias when the nonparametric first-stage learner enters the moment function through discontinuous functions.

We contribute to two strands of literature. First, we provide an inference procedure for economic models including latent discrete types. Finite-mixture models \citep{mixture_ichimura_thopson1998, mixture_arcidiacono_jones2003, mixture_chen_khalili2008, mixture_kasahara_shimotsu2009, mixture_henry_kitamura_salanie2014, mixture_compiani_kitamura2016, mixture_jochmans_henry_salanie2017, mixture_manole_khalili2021, mixture_jun_zincenko2026} and grouped-fixed effects models \citep{gfe_bonhomme_manresa2015, gfe_bonhomme_et_al_2022, gfe_lewis_et_al_2023, gfe_mehrabani2023} study related problems. A distinguishing feature of our approach is its separate first stage, which permits nonparametric identification of the groups. Finite-mixture models typically impose parametric assumptions on the distributions of outcomes and treatments conditional on type, which may be restrictive in many applications. In grouped-fixed effects models, the target parameters and group labels jointly minimize a common and known criterion. This joint structure is natural when addressing incidental parameter problems arising from fixed effects in linear panel models. However, if specifying a full model for the structural parameters, latent types, and observed variables is restrictive or even infeasible because the underlying system is too complicated, a separate first-stage approach may be attractive. 

Second, we contribute to the DML literature \citep{dml_chernozhukov_et_al2018,dml_chernozhukov_et_al2022,dml_ahrens_et_al2026} and the growing literature on AI-/ML-generated regressors \citep{generated_battglila_et_al2026, generated_escanciano_perez2026}. We study a nonstandard problem in which the $\mathcal{L}^2$-Riesz representer required for the bias-correction term does not exist. We construct a bias-correction term through smoothing and provide sufficient conditions for valid asymptotic inference. \citet{kennedy_et_al2020} also discuss how estimation error from a first-stage classifier can affect the plug-in estimator. Their result, like ours, shows why plug-in inference may be invalid. Though they establish $\sqrt{N}$-consistency and asymptotic normality of an estimator \emph{under specific rate conditions} on the first-stage error. In contrast, we show how to orthogonalize the effect of first-stage estimation and obtain a normal approximation, and thereby our approach accommodates flexible first-stage estimator. 

We note that the idea of smoothing indicators is not new. For example, \citet{horowitz1992} proposed the smoothed maximum score estimator to address the nonregularity of the maximum score estimator \citep{manski1985}. \citet{dml_whitehouse_et_al2025} and \citet{dml_park2025} consider a related two-stage inference framework in which a first-stage estimator enters the moment function through the $\max\{\delta(Z), 0\}$, which they approximate using a logistic distribution function or a softmax function. Their setting does not face the same problem of existence of an $\mathcal{L}^2$-Riesz representer, because the max function $\max\{\delta(Z),0\}$ is continuous in $\delta(Z)$. On the other hand, we consider the case under which $\delta(Z)$ changes the moment function $m(\theta, \textbf{1}\{\delta(Z)\geq \textnormal{cutoff}\}, W_i)$ in discontinuous way.

This paper is organized as follows. In Section \ref{sec02}, we use a simple model to show why plug-in inference is invalid, why an $\mathcal{L}^2$-Riesz representer does not exist without modification, and how smoothing the indicator addresses this problem. In Section \ref{sec03}, we introduce the econometric framework, formalize the problem, and state the basic assumptions. In Section \ref{sec04}, we impose restrictions on the smoothing kernel and the rate at which the tuning parameter $b$ shrinks, derive the Neyman-orthogonal score, and define the debiased estimator. We then establish its $\sqrt{Nb}$-consistency and asymptotic normality. Finally, in Section \ref{sec05}, we report Monte Carlo simulation results and show the debiased estimator achieves the nominal coverage probability.


\section{A simple example}\label{sec02}

In this section, we provide a simple example to briefly illustrate what this paper does: how the plug-in procedure may lead to invalid inference; why the standard way to construct the Neyman-orthogonal score may not work; and how to modify the problem to make the DML approach applicable.

\subsection{Setup}

    Let a binary latent type and an outcome of interest be denoted by $\group_{0,i} \in \{0,1\}$ and $Y_i$, respectively, with $\pi_0 \equiv \expect{\group_{0,i}} \in (0,1)$. We assume that the pairs $(Y_i, \group_{0,i})$ are independent across $i=1,...,N$.

    A researcher is interested in the effect of the $\group_{0,i}$ on the outcome $Y_i$. For the sake of simplicity, we consider the simplest linear model without an intercept and any covariates, as follow:
    \[ Y_i = \theta_0\group_{0,i} + \epsilon_i\quad\textnormal{where}\quad \condexpect{\epsilon_i}{\group_{0,i}} = 0\,,\,\condexpect{\epsilon_i^2}{\group_{0,i}} = 1\quad\textnormal{almost surely.} \]

    Because the type is not observed in the data, we have to predict or estimate the latent type. To do so, we assume that the latent type can be expressed with a discrete classifier $\ind{}{\delta_0(Z_i) \geq 0}$ with high probability, where $\delta_0(\cdot)$ indicates a normalized discriminant function. We assume that the signal $Z$ is a 1-dimensional continuous random variable, whose Lebesgue density is denoted by $f_{Z}(\cdot)$ with $f_Z(0) > 0$.

    We assume that the true value $\delta_0(Z)$ is a solution to:
    \begin{equation}\label{sec02 - eq01 - first-stage}
        \expect{ \zeta(Z)(T - \delta_0(Z)) } = 0 \quad\textnormal{for any }\zeta(\cdot)\in\cald,         
    \end{equation}
    where the random variable $T$ is an observed proxy variable and $\cald$ denotes a nuisance space. To simplify our discussion, we further assume $\delta_0(Z) = Z$ in this example. A nonparametric estimator for $\delta_0(\cdot)$, denoted by $\widehat{\delta}(\cdot)$, is a solution to the sample analogue of the problem \eqref{sec02 - eq01 - first-stage}. We simply assume the estimated type $\widehat{\group}_i \equiv \mathbf{1}\{\widehat{\delta}(Z_i) \geq 0\}$ satisfies the following condition: for each $i = 1,...,N$,
    \begin{equation}\label{sec02 - eq01}
        \widehat{\group}_i - \group_{0,i} = \begin{cases}
            1   & \textnormal{ with probability }\rho^{+}_{N} \\ 
            -1  & \textnormal{ with probability }\rho^{-}_{N} \\ 
            0   & \textnormal{ with probability }1 - \rho^{+}_{N} - \rho^{-}_{N}, \end{cases}
    \end{equation}
    with $\rho^{+}_{N} \downarrow 0$ and $\rho^{-}_{N} \downarrow 0$. In this example, we rule out the miraculous case that $\rho^{+}_{N} = \rho^{-}_{N}$ holds for every $N\in{\integer}$, implying that the direction of misclassification error is symmetrically distributed.

    For a moment contribution $m_i(\theta,\delta) \equiv Y_i - \theta\ind{}{ \delta(Z_i) > 0 }$, we have a moment restriction:
    \[ \expect{ m_{i}(\theta, \delta_0) } = 0 \iff \theta = \theta_0.\]
    For a given type profile $\Group \equiv [\group_{1}, ..., \group_{N}]$, an estimator $\Group \mapsto \widetilde{\theta}(\Group)$ is defined as follow:
    \[ \widetilde{\theta}(\Group) = \left( \sum_{i=1}^{N}\group_{i} \right)^{-1}\sum_{i=1}^{N}Y_i. \]
    An oracle estimator and the plug-in estimator are defined by $\oracle \equiv \widetilde{\theta}( {\Group}_0 )$ and $\plugin \equiv \widetilde{\theta}(\widehat{\Group})$, respectively.

    \subsection{The plug-in procedure may not be valid}

    Our first interest is the asymptotic behavior of a statistic $\sqrt{N}\big( \plugin - \theta_0 \big)$, in particular, under what conditions this statistic converges in distribution to a normal distribution with consistently estimable mean and variance. By adding and subtracting $\oracle$, the statistic is expressed as:
    \[ \sqrt{N}\big( \plugin - \theta_0 \big) = \sqrt{N}\big( \oracle - \theta_0 \big) - \sqrt{N}R_N, \]
    where $R_N \equiv \oracle - \plugin$.

    Because $\sqrt{N}\big( \oracle - \theta_0 \big) \overset{d}{\longrightarrow} \caln(0, \pi_0^{-2} )$ by the Lindeberg-Feller central limit theorem, asymptotic normality of the target statistic relies on the convergence speed of the second remainder term $\sqrt{N} R_N$. We obtain:
    \[  \sqrt{N}R_N
            =   \sqrt{N} \times \frac{\expect{ Y } + o_{P}(1)}{ \big\{\expect{ \group_{0,i} } \big\}^2 + o_{P}(1)}\times \frac{1}{N}\sum_{i=1}^{N}\big( \widehat{\group}_i - \group_{0,i} \big)
            =   \frac{\theta_0}{\pi_0}\sqrt{N}\big( \rho^{+}_{N} - \rho^{-}_{N} \big) + o_{P}(1). \]
    Therefore, the remainder term depends on $\sqrt{N}(\rho^{+}_{N} - \rho^{-}_{N})$ and there can be three scenarios. First, if $\sqrt{N}\big( \rho^{+}_{N} - \rho^{-}_{N} \big) \to 0$, then $\sqrt{N}\big( \plugin - \theta_0 \big) \overset{d}{\longrightarrow} \caln(0, \pi_0^{-2})$, meaning the plug-in procedure is oracle-valid. On the other hand, if $\sqrt{N}\big( \rho^{+}_{N} - \rho^{-}_{N} \big) \to \pm\infty$, then $\sqrt{N}\big( \plugin - \theta_0 \big)$ diverges, meaning the plug-in inference is invalid. The last case is when $\sqrt{N}R_N$ converges to a nonzero constant $c$. In this case, though $\sqrt{N}(\plugin - \theta_0) \overset{d}{\longrightarrow} \caln(-c, \pi_0^{-2})$ holds, the shifter $c$ may not be consistently estimable, because $c$ is a parameter relevant to the misclassification.

    Therefore, the misclassification probabilities can induce the first-stage bias. Moreover, if the misclassification rate converges more slowly than the parametric rate, the plug-in procedure may not provide valid inference.

    \subsection{Standard debiasing procedure may not work}

    In order to eliminate the impact of the first-stage error on the second-stage estimator, we can construct a bias-correction term invoking the standard DML approach. We first define a score $\psi_i(\theta, \delta, \beta)$ by
    \[ \psi_i(\theta, \delta, \beta) \equiv Y_i - \theta \ind{}{\delta(Z_i)\geq 0} + \beta(\theta,Z_i)(T_i - \delta(Z_i)).\]
    We note that $\expect{\psi_i(\theta,\delta,\beta)}$ is differentiable in $\delta$, even if $\psi_i(\theta,\delta,\beta)$ is not differentiable at $\delta(\cdot)=0$.
    
    We consider a local path $\delta_\tau(Z) \equiv \delta_0(Z) + \tau\cdot d(Z)$ for $\tau\in[0,1]$. The path is assumed to be linear in $\tau$ to simplify our discussion. We do not restrict the function $d(Z)$ except that it lies in $\cald$. Our goal is to find some $\mathcal{L}^2(P)$-function $\beta_0(Z)$ satisfying the following condition:
    \begin{equation}\label{sec02 - eq02 - neyman}
        \dt{ \expect{ \psi_i(\theta, \delta_\tau, \beta_\tau) } } = 0,        
    \end{equation}
    which is called \emph{Neyman-orthogonality}. The score $\psi_i(\cdot)$ obeying the condition \eqref{sec02 - eq02 - neyman} is called the \emph{Neyman-orthogonal score}.

    Many standard problems without such discrete transformations usually solve a functional equation in the form of $\expect{ \beta_0(\theta, Z) d(Z) } = \expect{ B(Z)d(Z) }$ with the known function $B(z)\in\mathcal{L}^2(P)$, and the $\mathcal{L}^2(P)$-function $\beta_0(\cdot)$ can be constructed by the Riesz representation theorem. 

    In contrast, $\beta_0(\cdot)\in\mathcal{L}^2(P)$, satisfying the Neyman-orthogonality in our problem, must be a solution to the following functional equation:
    \begin{equation}\label{sec02 - eq03 - FE01}
        \expect{ \beta_0(\theta, Z)d(Z) } =  -\theta f_{Z}(0)d(0)\quad\textnormal{for $\forall\theta\in\Theta$ and $\forall d(\cdot)\in\cald$.}
    \end{equation}
    Its derivation is provided in the Appendix \ref{appendix a-1}.
    
    The intuition is as follows: While types for observations far from the classification boundary are not affected by any local perturbation from the true value $\delta_0(\cdot)$, misclassified observations under the local perturbation are concentrated around the perturbed classification boundary. So the aggregate impact of the first-stage error for estimating $\delta$ is expressed as a probability density of $\delta(Z)$ at the boundary.  

    However, there is no $\mathcal{L}^2(P)$-function solving the functional equation \eqref{sec02 - eq03 - FE01}. Its rigorous proof is provided in the Appendix \ref{appendix a-1}, particularly see Proposition \ref{appendix a-1 - proposition}. Therefore, we need additional modifications to get a well-defined Neyman-orthogonal score in our problem.
    
    \subsection{Smoothing the discrete type}

    Instead of directly using the discrete type, we approximate it by a smoothed indicator. We choose a small positive number $b>0$ and a CDF (cumulative distribution function) kernel $k:\mathbb{R}\to [0,1]$ satisfying $k(t/b) \to \mathbf{1}\{t \geq 0\}$ as $b\to 0$ for every $t\ne 0$. We impose standard conditions on monotonicity, smoothness, and integrability. We additionally assume that the distribution of the true discriminant function has no atom at zero. Specific restrictions are discussed in the next section.
    
    Then we define a smoothed score as:
    \[ \psi_{bi}(\theta, \delta, \beta_b) \equiv Y_i - \theta k\left(\frac{\delta(Z_i)}{b}\right) + \beta_b(\theta,Z_i)(T_i - \delta(Z_i)).\]
    In this case, $\beta_{0,b}(\cdot)\in\mathcal{L}^2(P)$, which ensures that the score obeys Neyman-orthogonality, is a solution to the following functional equation: 
    \begin{equation}\label{sec02 - eq04 - FE02}
        \expect{ \beta_{0,b}(\theta,Z)d(Z) } = -\Expect{ \frac{\theta}{b} k'\left( \frac{\delta_0(Z)}{b} \right) d(Z)}\quad\textnormal{ for }\forall d(\cdot)\in\cald.        
    \end{equation}
    where $k'(\cdot)$ denote the derivative of $k(\cdot)$. Therefore, $\beta_{0,b}(\theta,Z) \equiv -\theta b^{-1}k'(\delta_0(Z)/b)$ is $\mathcal{L}^2(P)$-function and it gives the Neyman-orthogonal score. Detailed derivation is explained in Appendix \ref{appendix a-2}.
    
    We note that the solution to \eqref{sec02 - eq04 - FE02} is valid only for $b>0$. When $b$ depends on $N$, it therefore cannot shrink to zero arbitrarily fast. It will turn out that $\expect{\psi_{bi}(\theta_0,\delta_0,\beta_{0,b})^2} = O(b^{-1})$, meaning the debiased estimator $\widehat{\theta}_N$, defined below, is scaled by $\sqrt{Nb}$ rather than $\sqrt{N}$.

    We provide a preview of our main results. Using the smoothed Neyman-orthogonal score $\psi_{bi}(\cdot)$ and cross-fitting, we construct the debiased estimator $\widehat{\theta}_N$; see Definition \ref{sec04 - definition01} for its formal definition. Under suitable regularity conditions, the estimator is consistent for a pseudo-true value $\theta_b$ solving $\expect{\psi_{bi}(\theta, \delta_0, \beta_{0,b})} = 0$ for a given $b>0$. We impose restrictions on the rate at which $b$ shrinks to zero. These restrictions ensure that the smoothing bias vanishes sufficiently fast, so that $\theta_b-\theta_0=o\big(1/\sqrt{Nb}\big)$, and that the second-order remainder in the asymptotic linear representation is negligible. This approach does not require the Riesz representer for the original problem to be well-defined. We establish the following limit distribution and provide a consistent estimator of an asymptotic variance $\Omega_0$ (formally defined later):
    \[ \sqrt{Nb}(\widehat{\theta}_{N} - \theta_0) = \sqrt{Nb}(\widehat{\theta}_{N} - \theta_b) + \underset{ = o(1)}{ \underline{ \sqrt{Nb}( \theta_{b} - \theta_0) } } \overset{d}{\to} \mathcal{N}(0, \Omega_0), \]
    

\section{Econometric framework}\label{sec03}

\subsection*{Basic Notation}

We begin by introducing the basic notation used throughout the paper. For a generic $d$-dimensional vector $\pmb{v}\in \real{d}$, let $\norm{ \pmb{v} }$ denote the $d$-dimensional Euclidean norm. Let $\eucvec{1}, ..., \eucvec{d} \in \{0,1\}^{d}$ denote the column vectors of the $d$-by-$d$ identity matrix. For a generic matrix $\pmb{D} \in \real{d_1 \times d_2}$, $D_{ij}$ and $\pmb{D}_{j}$ denote its $(i,j)$-th entry and its $j$-th column, respectively. The smallest singular value of a matrix $\pmb{D}$ is denoted by $\minsingular{\pmb{D}}$. 

For a positive integer $N\in\integer$, an index set $[N]$ denotes $\{1,...,N\}$. Generic probability density and distribution functions are denoted by $f(\cdot)$ and $F(\cdot)$, and $F_0(\cdot)$ denotes the true distribution function generating the underlying random variables. $\expectF{F}{\cdot}$ and $\expect{\cdot}$ (or $\expectF{F_0}{\cdot}$) denote expectations where the underlying probability measure is given by $F$ and $F_0$, respectively. Let $\norm{\cdot}_{Q,p}$ denote the $\mathcal{L}^p(Q)$-norm with $p\in[1,+\infty]$ and an underlying measure $Q$. If the underlying measure is the true joint distribution, then we simply write $\norm{\cdot}_{p}$. Particularly, $\norm{f(\cdot)}_{\infty}$ denotes $\sup_{x}\norm{f(x)}$.
 
\subsection{Latent discrete type and data generating process}

Let a positive integer $N\in \integer$ denote a sample size. Let $G_0\in{\integer}$ with $G_0<\infty$ denote the true value of the number of types, which is assumed to be known. We do not consider cases in which $G_0$ is unknown or depends on the sample size. The type space is denoted by $\typespace  = \left\{ \group\in\{0,1\}^{G_0} : \sum_{g=1}^{G_0}\gamma_{g} = 1 \right\}$. A generic element of $\typespace$ is denoted by $\group$. The true value of each latent type for $i\in[N]$ and the type profile are denoted by $\group_{0,i}$ and $\Group_{0}$, respectively. For example, if $G_0=2$ and $\group_{0,i} = [0,1]^\top$ for $i\in[N]$, it means that the observation $i$ belongs to the second-labeled type. Throughout this paper, we assume this label is normalized. We also assume that every observation $i\in [N]$ is assigned to only one of the types with probability one. 

\newcommand{\vect}{ \pmb{T} }
A researcher observes realizations of random vectors $(W_{i}, Z_{i}, \vect_{i}) \in \supp{W} \times \supp{Z} \times \supp{\vect} \subset {\real{d_{W}}} \times {\real{d_{Z}}} \times \mathbb{R}^{d_{T}}$ for $i\in[N]$. The vector $W_{i}$ includes the outcome, variable of interest and controls, and the characteristics $Z_{i}$ are signals for classification. $\vect_{i}$ is a vector of proxy variables in order to predict the latent discrete type. Its dimension $d_{T}$ is clarified in the subsection describing the first-stage procedure. We allow the vectors $Z_{i}$ and $W_{i}$ to share components.

The underlying data-generating process and sampling scheme are described in the following assumption:

\begin{assumption}\textnormal{(Data and type)}\label{sec03 - assumption01}
\begin{itemize}
    \item[(a)] For each $N\in\mathbb{N}$, $\{(W_i, Z_i, \vect_i, U_i,\group_{0,i})\}_{i\in[N]}$  take values in a measurable space with law determined by the true probability measure. The observed data $\{(W_i,Z_i, \vect_i)\}_{i\in[N]}$ are independently drawn from the true joint distribution $F_0$.
    \item[(b)] Let $Z_{1,i}, Z_{2,i}$ be subvectors of the $Z_{i}$. Conditional on $\{(Z_{1,i},U_i)\}_{i\in[N]}$, the latent types $\{\group_{0,i}\}_{i\in[N]}$ are independently generated across $i$. Conditional on $\{(Z_{1,i},U_i,\group_{0,i})\}_{i\in[N]}$, the complementary signals $\{Z_{2,i}\}_{i\in[N]}$ are independently generated across $i$.
    \item[(c)] $\underset{g\in [G_0]}{\textnormal{inf}}\condprob{ \group_{0,i} = \eucvec{g} }{Z_i} > 0$ almost surely.
\end{itemize}
\end{assumption}\vspace{+0.5cm}

Throughout this article, the underlying data-generating process induces triangular-array data indexed by $N$, although we suppress this dependence for notational simplicity. Accordingly, all conditional and unconditional expectations and distributions, including $F_0$ and its alternatives $F$, should be understood as depending on $N$.

This framework allows different components of the characteristics $Z_i$ to be realized at different stages relative to the realization of type. For instance, if type corresponds to college graduation status, then $Z_{1,i}$ may consist of predetermined characteristics such as gender, race, family background, and high-school activities; her unobserved attributes $U_i$ may include inborn ability, genetic attributes and other unobserved endowments; and the complementary characteristics $Z_{2,i}$ may include post-type variables such as her occupation and income. 

\subsection{Moment condition}

A target parameter is denoted by $\theta \in \Theta \subseteq \mathbb{R}^{d_{\theta}}$ and its dimension $d_{\theta}$ is assumed to be low relative to the sample size. The researcher's economic model and substantive theory provide a valid moment contribution $m : \Theta \times  \typespace \times \supp{W}  \to \mathbb{R}^{d_{M}}$ with $d_{M} \geq d_{\theta}$, and the true value $\theta_{0} \in \Theta$ is defined as a solution to the following moment equation: 
\begin{equation}\label{sec03 - def01 - moment equation}
    \expect{ m( \theta, \group_{0,i}, W_{i}) } = 0.
\end{equation}
We do not assume that the true values $\theta_0$ and $\group_{0,i}$ jointly solve \eqref{sec03 - def01 - moment equation}. All underlying random variables follow a common joint distribution, and misspecifying $\group_i$ may fail to preserve its dependence on the observed data $W_i$. Jointly determining these quantities may also require characterizing the general equilibrium of the economy or underlying system, which can be too restrictive or even infeasible in many applications. Instead, we assume that the researcher specifies a nonparametric first-stage that determines $\group_{0,i}$ separately from the final stage.

In economic models with spatial or network interference, the latent types of other observations may also enter the moment contribution for observation $i$. Accommodating such dependence typically requires additional restrictions on how observations interact, introducing further complexity. We leave this extension to our future work.

We impose the following conditions to the moment function:

\begin{assumption}\textnormal{(Regularities on moment function)}\label{sec03 - assumption02}

    \begin{itemize}
        \item[(a)] $\Theta$ is a compact subset of $\mathbb{R}^{d_{\theta}}$, $\theta_{0}$ lies in the interior of $\Theta$. For any $\epsilon>0$, 
        \begin{equation}
            \inf_{\theta\in \Theta\backslash B_{\epsilon}(\theta_0)} \norm{ \expect{ m(\theta, \group_{0,i}, W_i)} }^2 > \norm{ \expect{ m(\theta_0, \group_{0,i}, W_i)} }^2 = 0.
        \end{equation}
        \item[(b)] $m(\theta, \group_{i}, W_i)$ is affine in $\group_{i}$, meaning there are known functions $M_{1}(\theta, W_i), ..., M_{G_0}(\theta, W_i)$ such that
        \[ m(\theta, \group_{i}, W_i) = \sum_{g=1}^{G_0}M_{g}(\theta, W_i)\eucvec{g}'\group_{i}. \]
        \item[(c)] There exists a neighborhood $\Theta_0$ of $\theta_0$  such that each $M_{g}(\theta, W_{i})$ is twice continuously differentiable in $\theta$ on $\Theta_0$. Furthermore, there are positive constants $\overline{M}^{(1)}, \overline{M}^{(2)}$ and $\underline{\sigma}$ such that:
        \begin{align}
            \sup_{\theta\in\Theta_0}\sup_{g\in[G_0]}\norm{ \nabla_\theta M_{g}(\theta, \cdot) }_{2} < \overline{M}^{(1)} < \infty\\
            \sup_{\theta\in\Theta_0}\sup_{g\in[G_0]}\norm{ \nabla_\theta^2 M_{g}(\theta, \cdot) }_{2} < \overline{M}^{(2)} < \infty\\
            \inf_{\theta\in\Theta_0}\minsingular{ \expect{ \nabla_{\theta} m(\theta, \group_{0,i}, W_i) } } > \underline{\sigma}.
        \end{align}
        \newcommand{\calm}{ {\mathscr{M}} }
        \item[(d)] Each $M_{g}(\theta, W_{i})$ lies in the function class $\calm \equiv \{ M(\theta, \cdot) : \theta\in \Theta \}$ whose uniform covering entropy obeys:
        \begin{equation}
            \sup_{Q}\log{ N( \epsilon \norm{\overline{M}}_{Q,2}, \calm, \norm{\cdot}_{Q,2} ) } \leq A_0\log(A_1/\epsilon)\quad\textnormal{for all }0 < \epsilon \leq 1,
        \end{equation}
        where $\overline{M}$ is a measurable envelope for $\calm$ with $\norm{\overline{M}}_2 < \infty$. 
    \end{itemize}    
\end{assumption}\vspace{+0.5cm}

Assumption \ref{sec03 - assumption02} imposes standard regularity conditions for the GMM problem. Assumption \ref{sec03 - assumption02}(a) requires that $\theta_0$ is the unique minimizer of the GMM criterion \citep{Newey1994}. Assumption \ref{sec03 - assumption02}(b) is a natural restriction because only the type of observation $i$ enters its moment contribution $m(\theta,\group_i,W_i)$. For example, if $\typespace = \{\eucvec{1}, \eucvec{2}\}$ and only $i$'s type changes her moment contribution, the researcher can set $M_{1}(\theta, W_{i}) \equiv m(\theta, \eucvec{1}, W_{i})$, $M_{2}(\theta, W_{i}) \equiv m(\theta, \eucvec{2}, W_{i})$, and then $m(\theta, \group_{i}, W_{i})$ satisfies the following identity:
\[ m(\theta, \group_{i}, W_{i}) = m(\theta, \eucvec{1}, W_i)\ind{}{\group_{i} = \eucvec{1}} + m(\theta, \eucvec{2}, W_i)\ind{}{\group_{i} = \eucvec{2}} = \sum_{g=1}^{2}M_{g}(\theta,W_i)\eucvec{g}^\top \group_{i}. \]
Assumption \ref{sec03 - assumption02}(c) can be relaxed by assuming $\expect{M_{g}(\theta,W_i)}$ is directionally differentiable in $\theta$. It also requires its Jacobian to have full column rank, providing an identification condition. Assumption \ref{sec03 - assumption02}(d) is standard condition for a GMM problem with nonlinear moment contributions, ensuring the uniform weak law of large number and the uniform central limit theorem (See, e.g., \citet{pakespollard1989, vw2023}). If every $M_{g}(\theta,W)$ is linear in $\theta$, then Assumption \ref{sec03 - assumption02}(d) is automatically satisfied.

\subsection{First-stage procedure}

In order to estimate the missing type profile, the researcher specifies a first-stage learner, consisting of the following components: \begin{itemize}
    \item[i.]{Discriminant function}

    For each ordered pair $g,h\in [G_0]$, let $\delta_{g,h}: \supp{Z} \to \mathbb{R}$ denote a pairwise discriminant function, normalized so that $\delta_{g,h}(z)=-\delta_{h,g}(z)$ for all $g,h\in [G_0]$ and $z\in \supp{Z}$.  For each $g\in[G_0]$, we define:
    \[  \discfunc_{g}(z) \equiv \big\{ \delta_{g,h}(z) : h\ne g \big\} \in \mathbb{R}^{G_0-1}\quad\textnormal{and}\quad
        \discfunc(z) \equiv \big[ \discfunc_{1}(z) \quad ... \quad \discfunc_{G_0}(z)  \big] \in \mathbb R^{(G_0-1)\times G_0}. \]
    A space of discriminant functions is denoted by:
    \[ \cald \equiv \left\{ \discfunc: \supp{Z} \to \mathbb R^{(G_0-1)\times G_0} \textnormal{ such that } \delta_{g,h}(z)=-\delta_{h,g}(z) \textnormal{ for all } g,h\in [G_0] \textnormal{ and } z\in\supp{Z} \right\}. \]
    
    \item[ii.] {Proxy variable and score function}

    The proxy variables satisfy the following conditions: $d_{T} = G_0 \times (G_0 - 1)$, $\vect_{i,g} = \{T_{i,gh} : g < h\}$ where $T_{i,hg} = -T_{i,gh}$.
    For any joint distribution $F$, the pseudo true value of the discriminant function $\discfunc[F](\cdot) \in \cald$ is a solution to :
    \begin{equation}\label{sec03 - def02 - first-stage}
        \expectF{F}{ \zeta(Z_i)( \vect_{i,g} - \discfunc_{g}(Z)) } = 0 \quad\textnormal{for any measurable }\zeta(Z)\in\cald.   
    \end{equation}
    Then, we write the true value of the discriminant function $\discfunc_{0}(\cdot) \equiv \discfunc[F_0](\cdot)$, and $\delta_{gh}[F](Z_i)$ identifies $\condexpectF{F}{T_{i,gh}}{Z_i}$ for each $g,h\in[G_0]$. Throughout this paper, let $\cald_N \subset \cald$ denote a shrinking neighborhood of $\discfunc_0$ at a specified rate. Its nonparametric estimator is denoted by $\widehat{\discfunc}(Z)$, and we assume that the nuisance estimator $\widehat{\discfunc}$ belongs to $\cald_N$ with probability approaching one. Appendix \ref{appendix a-3} provides two examples illustrating how the pairwise proxy variables $\vect$ and the corresponding discriminant function $\discfunc$ can be constructed.
    
    \item[iii.] {Classifier}
    
    A classifier $\classifier{A} : \mathbb{R}^{(G_0-1)\times G_0} \mapsto \typespace$ is defined by $\classifier{A}(\discfunc) \equiv \big\{ \ind{}{ \delta_{g,h} \geq 0 \textnormal{ for }h \ne g} \big\}_{g\in[G_0]}$. To simplify the notation, we write $\classifier{A}_{i}(\discfunc) \equiv  \classifier{A}\big( \discfunc(Z_{i}) \big)$ for $\forall i\in[N]$.
\end{itemize}

The specified first-stage procedure requires proxy variables. We exclude settings in which the first-stage involves an unsupervised clustering problem. The score function associated with such a problem is nonlinear and nonsmooth in $\discfunc(Z)$, resulting in additional technical challenges. We leave this case for future research.

We next present applications in which observed labels or outcomes are used to construct economically meaningful groups. There are possible three cases with latent labels recovered from the observable proxy variables. 

\begin{example}\label{sec03 - example01-1}
First, the label may be observed only for a small subset of the population/sample. \citet{app_proxy_black2023} analyze how a college admissions reform affects enrollment, graduation, and earnings. The true label is $\group_{0i} = \mathbf{1}\{$student $i$ belongs to the top ten percent of their own high-school class$\}$, and the proxy is $T_i = \mathbf{1}\{$student $i$ is observed to be eligible for automatic admission under the top-ten-percent rule$\}$. This proxy is observed only for students who applied to a Texas public higher education institution.$\|$
\end{example}

\begin{example}\label{sec03 - example01-2}
Second, the true label may represent a latent or counterfactual object inferred from observed outcomes. 
\citet{app_proxy_kleinberg2018} assess whether risk-based release decisions can reduce both detention and failures to appear. In their application, the true label is $\group_{0i} = \mathbf{1}\{$defendant $i$ would fail to appear if released$\}$, and the proxy is $T_i = \mathbf{1}\{$defendant $i$ failed to appear after being released$\}$, which is observed only for defendants who were actually released. \citet{app_proxy_iselin2023} estimate how many households eligible for a refundable tax credit for low-income workers fail to claim it. They define the true label by $\group_{0i} = \mathbf{1}\{$non-filing recipient $i$ would appear as a head or spouse if a tax return were filed$\}$, and the proxy is $T_i = \mathbf{1}\{$individual $i$ appears as a head or spouse on an observed tax-year return$\}$.$\|$
\end{example}

\begin{example}\label{sec03 - example01-3}
Lastly, the true label may depend on an unknown cutoff that must be estimated. \citet{app_poor_proxy_balboni2022} examine the effect of whether households are above or below a poverty-trap threshold. The true type is defined by $\group_{0i} = \mathbf{1}\{$household $i$'s initial post-transfer log productive assets exceed $c_0\}$, and the proxy $T_i$ is household $i$'s log productive assets observed four years after the transfer. The latent cutoff $c_0$ is estimated using $T_i$.$\|$
\end{example}\vspace{+0.5cm}

The specified first-stage procedure is assumed to satisfy the following conditions:

\begin{assumption}\textnormal{(Regularities of first-stage procedure)}\label{sec03 - assumption03}

    \begin{itemize}
        \item[(a)] $\cald$ is a convex and complete normed space. The true value $\discfunc_0$ is $F_0$-measurable, solves \eqref{sec03 - def02 - first-stage}, and lies in the interior of $\cald$. For every $\discfunc(\cdot)\in\cald_N$, $\norm{\discfunc}_2 < \infty$ and $\supp{\discfunc(Z)}$ is compact. 
        \item[(b)] $\condexpect{ \norm{\pmb{T}_{i}}^2 }{Z_i = z} \leq C_{T} < \infty$ for almost every $z\in\supp{Z}$. 
        \item[(c)] For a sequence $\lambda_{N}\downarrow 0$, $\underset{i\in[N]}{\textnormal{max}}\,\prob{ \group_{0,i} \ne \classifier{A}_i\big( \discfunc_{0} \big) } = O\big(  \lambda_{N} \big)$.
        \item[(d)] For every $\discfunc\in \cald_N$, there is a sequence $\rho_{N}\downarrow 0$ converging to zero such that $\| \discfunc - \discfunc_{0} \|_{\infty} = o_{P}\big( \rho_{N} \big)$.
        \item[(e)] There is a positive constant $c_0$ and a neighborhood of zero such that for every $t$ in the neighborhood:
        \begin{equation}\label{sec03 - assumption03 - eq03}
            \sup_{g,h\in[G_0]}\Prob{ \big|\delta_{0,gh}(Z)\big| \leq t } \leq c_0 t.
        \end{equation}
    \end{itemize}
\end{assumption}\vspace{+0.5cm}

Assumption \ref{sec03 - assumption03}(a) is a mild standard condition on the nuisance space and nuisance function in the DML literature, and the specified first-stage procedure can nonparametrically identify the true value $\discfunc_{0}$. Assumption \ref{sec03 - assumption03}(c) requires the classification rule to be correctly specified in the sense that the truth $\discfunc_0(\cdot)$ recovers the true type with probability approaching one. If $\lambda_N=0$, the assumption requires $\classifier{A}_{i}(\delta_0)=\group_{0,i}$ almost surely. This condition is often implicit but essential in applications that use classification methods to recover discrete latent types.

The probability mass near the classification boundary $\{ z\in \supp{Z} : \discfunc(z) =0 \}$ measures the extent to which observations are difficult to classify. Assumption \ref{sec03 - assumption03}(e) rules out an extreme concentration of probability mass at particular points around the boundary. This is a standard margin condition in clustering or classification theory \citep{cluster_levrard2013, cluster_levrard2018, kennedy_et_al2020}, and any random variable $\discfunc_{0}(Z)$ admitting a well-defined Lebesgue density around the boundary automatically satisfies the condition.


\section{Debiased estimator and its large sample property}\label{sec04}

In this section, we first study the asymptotic behavior of the plug-in procedure. Next, we define the debiased estimator, by introducing the smoothing kernel function and constructing the Neyman-orthogonal score accordingly. Then we analyze the asymptotic properties of the debiased estimator, and determine what restrictions on the smoothing tuning parameter are needed in order to achieve valid inference.

\subsection{Plug-in estimator}

Estimators of each type and the type profile are denoted by $\widehat{\group}_{i} \equiv \classifier{A}\big( \widehat{\discfunc}(Z_i) \big)$ and $\widehat{\Group}_N \equiv \big[ \widehat{\group}_1, ..., \widehat{\group}_N \big]$, respectively. For a generic type profile $\Group\in \mathscr{G}^{N}$, an estimator $\widetilde{\theta} : \typespace^{N} \to \Theta_0$ denotes a minimizer of $\norm{N^{-1}\sum_{i=1}^{N}m(\theta, \group_i, W_i)}^2$, and the plug-in estimator $\plugin$ is defined by $\widetilde{\theta}(\widehat{\Group}_N)$. The oracle estimator $\oracle$ is defined by $\widetilde{\theta}( \Group_0 )$. Under standard regularity conditions, the oracle estimator achieves the desirable $\sqrt{N}$-consistency. 

If $\prob{ \widehat{\Group}_{N} \ne \Group_{0} } = o(1)$, the estimated type profile is said to achieve \emph{exact recovery} \citep{exact_abbe2017}. In this case, the plug-in estimator is asymptotically equivalent to its oracle counterpart. By contrast, Proposition \ref{sec04 - proposition01} establishes an observation-level error bound associated with \emph{almost-exact recovery}.

\begin{proposition}\label{sec04 - proposition01}
    Under Assumptions \ref{sec03 - assumption01} and \ref{sec03 - assumption03}, for sufficiently large $N>0$,
    \[
        \underset{i\in[N]}{\textnormal{max}}\prob{\widehat{\group}_i\ne\group_{0,i}}
            \leq O(\lambda_N)+\binom{G_0}{2}c_0\rho_N+\prob{\norm{\widehat{\discfunc}-\discfunc_0}_\infty>\rho_N}.
    \]
\end{proposition}

\textbf{Proof of Proposition \ref{sec04 - proposition01}} : See Appendix \ref{sec04 - proposition01 - proof}.\\

Proposition \ref{sec04 - proposition01} does not imply that the first-stage procedure exactly recovers the entire type profile. Instead, it shows that $\widehat{\group}_{i} = \classifier{A}( \widehat{\discfunc}(Z_i) )$ recovers the true type $\group_{0,i}$ at the observation level, provided that the first-stage procedure is asymptotically correctly specified ($\lambda_N\to 0$) and the first-stage estimator is consistent ($\rho_N\to 0$).

For illustration purposes, we suppose that, for given $N$, the classification outcomes are independent across $i$, $\lambda_N = 0$, $\prob{\norm{\widehat{\discfunc}-\discfunc_0}_\infty>\rho_N} = 0$, and therefore $\prob{\widehat{\group}_{i} \ne \group_{0,i}} \leq \rho_N$. Then, Proposition \ref{sec04 - proposition01} implies that:
\[ \prob{\widehat{\Group}_N = \Group_{0}}
        =   \prod_{i=1}^{N} \left\{1-\prob{\widehat{\group}_{i}\neq\group_{0,i}}\right\} 
        \geq (1 -\rho_N)^N \geq 1 - N\cdot \rho_N. \]
Thus, the condition $\rho_N\to0$ alone does not generally guarantee exact recovery of the entire type profile. One sufficient condition is $N\rho_N\to0$, which is typically too restrictive for flexible first-stage estimators. Alternatively, strong group separation can ensure exact recovery when the probability mass near the classification boundary is sufficiently small (see, e.g., \citet{exact_audibert_tsybakov2007, exact_fei_chen2007, exact_awasthi2015, exact_giraud_verzelen2018, exact_le_gouic_paris2018, exact_li_et_al2020, exact_ndaoud2022, exact_de_rosa_khajavirad2022}). However, we do not impose such a strong group-separation condition in this paper.

Under the almost-exact recovery regime, the remainder term of the plug-in estimator does not vanish because first-stage error usually affects the plug-in procedure through the resulting misclassification rate. Proposition \ref{sec04 - proposition02} characterizes the effect of this first-stage error on the plug-in estimator, which is similar to the result of \citet{kennedy_et_al2020}.

\begin{proposition}\label{sec04 - proposition02}
    Under Assumptions \ref{sec03 - assumption01}, \ref{sec03 - assumption02} and \ref{sec03 - assumption03},
    \[ \norm{ \plugin - \theta_0 } =
            \underset{ \substack{\textnormal{misspecified}\\\textnormal{first-stage}} }{ \underline{ O_{P}\big( \lambda_{N}^{1/2} \big) } } 
        +   \underset{ \substack{\textnormal{first-stage}\\ \textnormal{estimation error}} }{ \underline{ O_{P}\big( \rho_{N}^{1/2} \big) } }
        +   \underset{ \substack{\textnormal{second stage}\\ \textnormal{estimation error}} }{ \underline{ O_{P}\big( N^{-1/2} \big) } } .
    \]
\end{proposition}

\textbf{Proof of Proposition \ref{sec04 - proposition02} : } See Appendix \ref{sec04 - proposition02 - proof}.\\

\subsection{Kernel smoothing}

Proposition \ref{sec04 - proposition02} shows that a slowly vanishing first-stage estimation error can contaminate downstream inference. However, as discussed in Section \ref{sec02}, the Riesz representer, required to construct a bias-correction term, may not be well defined in this setting. To avoid this challenge, we approximate $\classifier{A}_{i}(\discfunc)$ using kernel smoothing.

Let $k:\mathbb{R}\mapsto [0,1]$ denote a univariate CDF kernel function, and $k'(t)$ denote its first-order derivative. We define a product kernel $\pmb{K}:\mathbb{R}^{G_0 - 1}\to [0,1]$ such that:
\[ \KER{\discfunc_{g}(Z_i)} = \prod_{h \ne g}k\left( \frac{\delta_{g,h}(Z_i)}{b_N} \right). \]
Its first-order derivative $\nabla \pmb{K}( \discfunc(Z_i) ) \in \mathbb{R}^{G_0 - 1}$ is defined by:
\[ \nabla \pmb{K}(\discfunc_g(Z_i)) \equiv 
    \left\{ k'\left( \delta_{g,g'}(Z_i) \right) \prod_{h \ne g,g'} k\left( \delta_{g,h}(Z_i) \right) \right\}_{g'\ne g }. \]
We define a smoothed classifier $\smth{A}\big( \discfunc(Z_i) \big)\in\mathbb{R}^{G_0}$ by:
\begin{equation}
    \smth{A}\big( \discfunc(Z_i) \big) = \eucvec{G_0} + \sum_{g=1}^{G_0 - 1}\KER{ \discfunc_{g}(Z_{i}) }\big( \eucvec{g} - \eucvec{G_0} \big)
\end{equation}
To simplify our notation, $\smth{A}_i(\discfunc)$ denotes $\smth{A}( \discfunc(Z_i) )$. We impose the following restrictions on the univariate kernel $k(\cdot)$:

\begin{assumption}{(Restrictions on Kernel Function)}\label{sec04 - assumption04}

\begin{itemize}
    \item[(a)] $k(t) = 1$ for $t \geq S$ and $k(t) = 0$ for $t \leq -S$ for a specified constant $S>0$, and $k(t) \in [0,1]$ is continuous and strictly increasing in $t \in [-S, S]$.
    \item[(b)] $k(t)$ is twice continuously differentiable in $t \in (-S,S)$.
    \item[(c)] For $k'(t) \equiv \frac{d}{dt}k(t)$, $k'(t)=k'(-t)$ for every $t\in\mathbb{R}$, and $\underset{|t|\leq S}{\textnormal{sup}}|k'(t)| < \infty$, \item[(d)] For $k''(t)\equiv \frac{d^2}{dt^2}k(t)$, $\underset{|t|\leq S}{\textnormal{sup}}|k''(t)| < \infty$.
\end{itemize}
\end{assumption}\vspace{+0.5cm}

Assumption \ref{sec04 - assumption04} imposes standard smoothness and compact-support conditions on the smoothing kernel commonly used in kernel-based nonparametric estimation \citep{kernel_fan1996,kernel_pagan1999,kernel_li2007}. A similar approach to approximating nonsmooth functions is employed by \citet{horowitz1992}, \citet{dml_park2025} and \citet{dml_whitehouse_et_al2025}.

Because the smoothing function is chosen by the researcher, its differentiability can be imposed directly without restricting the underlying data-generating process. Assumption \ref{sec04 - assumption04} further implies that $k'$ is a symmetric density kernel supported on $[-S,S]$, where $S>0$ is chosen by the researcher, and that $k$ can be interpreted as the corresponding cumulative distribution function. Although the choice of $S$ typically does not affect the asymptotic theory, it may influence finite-sample performance. Compact support is particularly useful to simplify the derivation of stochastic bounds. The boundedness of $k'$ and $k''$ controls the curvature of the smoothed classifier and the induced smoothed score. A possible kernel function satisfying Assumption \ref{sec04 - assumption04} is illustrated in the next example.

\begin{example}\label{sec04 - example04}
    For $S>0$, let $q(t) \equiv {(t + S)}/{2S}$ and define:
    \begin{equation}\label{sec04 - example04 - eq01 - kernel}
    k(t) = \begin{cases}
        0 & \textnormal{ if }t < -S\\
        6 q(t)^5 - 15 q(t)^4 + 10 q(t)^3 & \textnormal{ if }t\in[-S, S] \\
        1 & \textnormal{ if }t > S.
    \end{cases}        
    \end{equation}
    Then, $k'(t) = \frac{15}{S} q(t)^2(1 - q(t))^2\ind{}{|t|\leq S}$, $k''(t) = \frac{15}{S^2}q(t)(1-q(t))(1 - 2q(t))\ind{}{|t|\leq S}$.$\|$
\end{example}

\subsection{Neyman-orthogonal score}

We describe how to construct the orthogonal score with $m( \theta, \smth{A}_i(\discfunc), W_i )$. For an index $\tau\in [0,1]$, we consider a mixture distribution $F_{\tau} = F_0 + \tau(F - F_0)$, where $F$ is an alternative distribution. We define $\discfunc_\tau(Z) \equiv \discfunc[F_{\tau}](Z)$, which is a solution to equation \eqref{sec03 - def02 - first-stage} when $F = F_{\tau}$. The alternative distribution $F$ is unrestricted, except that it ensures the existence of the solution $\discfunc_\tau(Z)$ and all moments we analyze \citep{dml_chernozhukov_et_al2022, semi_ichimura_newey2022}.
\newcommand{ \vecmu}{ \pmb{\mu} }

We aim to find a bias-correction term $\phi_N(\cdot)$, so that a function $\psi_{N}(\theta, \discfunc_\tau, \vecmu_\tau, W_i) = m(\theta, \smth{A}_{i}(\discfunc_\tau), W_i) + \phi_{N}(\theta, \discfunc_\tau, \vecmu_\tau, W_i)$ satisfies the following Neyman-orthogonality conditions: \citep{dml_chernozhukov_et_al2018, dml_chernozhukov_et_al2022, dml_ahrens_et_al2026}:
\begin{equation}\label{sec04 - neyman}
    \begin{cases}
        \expectF{F_{\tau}}{\phi_{N}(\theta, \discfunc_\tau, \vecmu_\tau, W_i)} = 0 &\textnormal{for }\forall \tau\in[0,\bar{\tau}), \bar{\tau}\in[0,1]\textnormal{ and for }\forall \theta\in\Theta\\
        \dt{\expectF{F_0}{\psi_{N}(\theta, \discfunc_\tau, \vecmu_\tau, W_i)} } = 0  &\textnormal{for }\forall \theta\in\Theta,
    \end{cases}
\end{equation}
where $\vecmu_\tau(\cdot)$ is an additional nuisance function which can be introduced to appropriately define $\phi_{N}$.

Key ingredients to define the orthogonal score are as follows:
\begin{equation}\label{sec04 - def051 - mu}
    \mu_{0,g}(\theta, Z_i) \equiv \condexpectF{F_0}{M_{g}(\theta, W_i) - M_{G_0}(\theta,W_i)}{Z_i},    
\end{equation}
\begin{equation}\label{sec04 - def052 - representer}
    B_{N,g}[\vecmu, \discfunc](\theta,Z_i) \equiv \mu_{g}(\theta,Z_i)\frac{1}{b_N}\nabla \KER{ \discfunc_{g} }^\top,\quad \beta_{N,g}(\theta, Z_i) \equiv B_{N,g}[\vecmu_0, \discfunc_0](\theta, Z_i),
\end{equation}
\begin{equation}\label{sec04 - def053 - phi}
    \phi_N(\theta, \discfunc, \pmb{\mu}, W_i) \equiv \sum_{g=1}^{G_0-1}B_{N,g}[\vecmu, \discfunc](\theta, Z_i)(\vect_{i,g} - \discfunc_{g}(Z_i)).
\end{equation}
Then, the following proposition shows that a new score $\psi_N(\theta, \discfunc, \pmb{\mu}, W_i) \equiv m( \theta, \smth{A}_i(\discfunc), W_i ) + \phi_N(\theta, \discfunc, \pmb{\mu}, W_i)$ obeys Neyman-orthogonality.

\begin{proposition}\label{sec04 - proposition03}
    Under Assumptions \ref{sec03 - assumption01}, \ref{sec03 - assumption02}, \ref{sec03 - assumption03} and \ref{sec04 - assumption04}, the score $\psi_N(\theta, \discfunc, \pmb{\mu}, W_i)$ obeys \eqref{sec04 - neyman}.
\end{proposition}

\textbf{Proof of Proposition \ref{sec04 - proposition03}} : See Appendix \ref{sec04 - proposition03proof}.\\

Smoothing replaces the original population moment condition $\expect{m(\theta, \classifier{A}_i(\discfunc_0), W_i)}=0$ with the smoothed moment condition $\expect{ m(\theta, \smth{A}_i(\discfunc_0), W_i) } = 0$ and the solutions to these two equations need not coincide, in general. As in many nonparametric methods, the smoothing procedure requires additional regularity conditions \citep{kernel_nadaraya1964,kernel_fan1996,kernel_pagan1999,kernel_li2007,horowitz1992}, particularly smoothness assumptions on the relevant density and conditional expectations. We need the following high-level conditions:

\begin{assumption}\label{sec04 - assumption05}
\textnormal{(Smoothness condition around the classification boundary)}

\begin{itemize}
    \item[(a)] For every $g\in[G_0-1]$, $\discfunc_{0,g}(Z_i)$ has a Lebesgue density $f_{\discfunc_{0,g}}(\pmb t)$ satisfying:
    \[ \sup_{g\in[G_0-1]}\sup_{\pmb t\in\mathbb R^{G_0-1}}f_{\discfunc_{0,g}}(\pmb t)\leq C_f<\infty. \]

    \item[(b)] For each $g\in[G_0-1]$, there exist a constant $C_M<\infty$ and an open neighborhood $\mathcal U_g$ of the boundary set $\underset{h\ne g}{\cup}\big\{ \pmb t\in\mathbb{R}^{G_{0} - 1} : t_h=0,\ t_{\ell} \geq 0, \ell\ne h \big\}$ such that, for $f_{\discfunc_{0,g}}$-almost every $\pmb t\in\mathcal U_g$:
    \[ \condexpect{\norm{M_g(\theta_0,W_i)-M_{G_0}(\theta_0,W_i)}^2}{\discfunc_{0,g}(Z_i)=\pmb t}\leq C_M \]
    
    \item[(c)]For every $\theta\in\Theta_0$, the functions
    \[ r_g^{(0)}(\theta,\pmb t)\equiv\condexpect{M_g(\theta,W_i)-M_{G_0}(\theta,W_i)}{\discfunc_{0,g}(Z)=\pmb t}f_{\discfunc_{0,g}}(\pmb t), \]
    \[ r_g^{(1)}(\theta,\pmb t)\equiv\condexpect{\nabla_\theta M_g(\theta,W_i)-\nabla_\theta M_{G_0}(\theta,W_i)}{\discfunc_{0,g}(Z)=\pmb t}f_{\discfunc_{0,g}}(\pmb t) \]
    are continuously differentiable over $\mathcal U_g$. Moreover,
    \[ \sup_{g\in[G_0-1]}\sup_{\theta\in\Theta_0}\sup_{\pmb t\in\mathcal U_g}\left\{\norm{\nabla_{\pmb t}r_g^{(0)}(\theta,\pmb t)}+\norm{r_g^{(1)}(\theta,\pmb t)}+\norm{\nabla_{\pmb t}r_g^{(1)}(\theta,\pmb t)}\right\}<\infty. \]
\end{itemize}
\end{assumption}\vspace{+0.5cm}

Assumption \ref{sec04 - assumption05} imposes standard mild regularity conditions commonly used in nonparametric analysis. Assumption \ref{sec04 - assumption05}(a) requires that $\discfunc_{0,g}(Z)$ admits a bounded density with respect to Lebesgue measure. Assumption \ref{sec04 - assumption05}(b) is particularly relevant to the additional nuisance function $\vecmu_0(\cdot)$. Assumption \ref{sec04 - assumption05}(c) requires first-order differentiability around the classification boundary, together with boundedness of the corresponding first-order derivative. Importantly, it does not require smoothness over the entire support of $\discfunc_{0,g}$, because our approach focuses on the local property around the classification boundary.

The following proposition characterizes the order of the smoothing bias.

\begin{proposition}\label{sec04 - proposition04}
    Under Assumptions \ref{sec03 - assumption01}, \ref{sec03 - assumption02}, \ref{sec03 - assumption03}, \ref{sec04 - assumption04} and \ref{sec04 - assumption05}, then
    \begin{itemize}
        \item[(i)] $\expect{m(\theta_0, \smth{A}_i(\discfunc_0), W_i)} = \expect{m(\theta_0, \classifier{A}_i(\discfunc_0), W_i)} + O(b_{N}^{2} )$.
        \item[(ii)] $\expect{ \nabla_\theta \psi_N(\theta_0, \discfunc_{0}, \vecmu_{0}, W_i)} = \expect{ \nabla_\theta m(\theta_0, \classifier{A}_i(\discfunc_0), W_i)} + O(b_{N}^{2} )$.
        \item[(iii)] $\minsingular{ \expect{ \nabla_\theta \psi_N(\theta_0, \discfunc_{0}, \vecmu_{0}, W_i)} }\,\, \geq \,\, \underline{\sigma} - O(b_{N}^{2})$.
    \end{itemize}
\end{proposition}

\textbf{Proof of Proposition \ref{sec04 - proposition04}} : See Appendix \ref{sec04 - proposition04 - proof}.\\

Under the smoothness assumption, the resulting smoothing bias, which vanishes, is of order $O(b_{N}^2)$. This ensures that the pseudo parameter solving the $\expect{\psi_{N}(\theta,\discfunc_0, \vecmu_0, W_i)} = 0$ converges to the truth $\theta_0$. The Jacobian of $\psi_{N}$ is also nonsingular for sufficiently large $N>0$. This Jacobian centers the asymptotic variance of the debiased estimator, formally defined in the next subsection.

\subsection{Debiased estimator and its large sample property}

We now define the debiased estimator for this problem using the standard cross-fitting procedure \citep{dml_chernozhukov_et_al2018, dml_chernozhukov_et_al2022, dml_ahrens_et_al2026}. We adopt the DML2-type estimator because it usually outperforms the DML1-type estimator \citep{dml_velez2026}. 

\begin{definition}\textnormal{(Debiased Estimator)}\label{sec04 - definition01}

\begin{itemize}
    \item[(1)] Sample $[N]$ is split into $L$-folds, such that $I_{1}\cup...\cup I_{L} = [N]$ and $I_{\ell}\cap I_{\ell'} = \emptyset$. We require $N/|I_{\ell}| \to L$ for every $\ell\in[L]$. Let $I_{-\ell} \equiv [N] \backslash I_{\ell}$. For each $\ell\in[L]$, a primitive plug-in estimator $\widetilde{\theta}_\ell$ and nuisance estimates $\widehat{\discfunc}_\ell(\cdot)$ and $\widehat{\vecmu}_\ell(\theta,\cdot)$ are estimated by using observations in $I_{-\ell}$.
    \item[(2)] For each $\ell\in[L]$ and for observations $i\in I_{\ell}$, we obtain the orthogonal score $\psi_{N,\ell}(\theta,W_i)$,
    \begin{equation}
        \psi_{N,\ell}(\theta,W_i) \equiv m\big( \theta, \smth{A}_{i}(\widehat{\discfunc}_{\ell}), W_i \big) + \phi_N(\widetilde{\theta}_\ell, \widehat{\discfunc}_\ell, \widehat{\vecmu}_\ell(\widetilde{\theta}_{\ell}, \cdot), W_i).
    \end{equation} 
    \item[(3)] The debiased estimator $\widehat{\theta}_N$ is a minimizer of the following criterion:
    \begin{equation}
        \widehat{\theta}_N \in \argmin{\theta\in \Theta}{ \Norm{\frac{1}{L}\sum_{\ell=1}^{L}\frac{1}{|I_{\ell}|}\sum_{i \in I_{\ell}}\psi_{N,\ell}(\theta, W_i)}^2}.
    \end{equation}
\end{itemize}
\end{definition}

We additionally impose the following conditions to achieve the desired asymptotic result.

\begin{assumption}\label{sec04 - assumption06}\textnormal{(Restrictions on $b_{N}$, $\lambda_N$ and $\rho_N$)}

    \begin{itemize}
        \item[(a)] There exist deterministic sequences $\rho_N\downarrow0$ and ${\nu_{N}}\downarrow0$ such that
        \[ \max_{\ell\in[L]}\norm{ \widehat{\discfunc}_{\ell} - \discfunc_0}_{\infty} = o_P(\rho_N)\quad\textnormal{and}\quad\max_{\ell\in[L]}\norm{\widehat{\vecmu}_{\ell}-\vecmu_0}_{\infty}=o_P({\nu_{N}}). \]
        \item[(b)] $Nb_{N}\to\infty$ and $Nb_{N}^5 \to 0$.
        \item[(c)] $\rho_N = O(b_N)$ and $N\rho_N^3 = O(1)$.
        \item[(d)] $\sqrt{Nb_{N}}\rho_N\nu_{N} = O(1)$.
        \item[(e)] $Nb_{N}\lambda_{N} = o(1)$.
\end{itemize}
\end{assumption}\vspace{+0.5cm}

Assumption \ref{sec04 - assumption06}(a) is a standard algorithmic stability condition in the DML literature, requiring the first-stage nuisance estimator to be consistent in each fold. Possible first-stage learners satisfying the condition (a) are briefly discussed in Example \ref{sec04 - example01 - KRR} and \ref{sec04 - example02 - NN}. Assumption \ref{sec04 - assumption06}(b) imposes restrictions on $b_N$ that determine the scaling of the debiased estimator. In particular, the condition $Nb_N^5 \to 0$ ensures that the pseudo-true parameter solving $\expect{\psi_{N}(\theta, \discfunc_0, \vecmu_0, W_i)} = 0$ converges to $\theta_0$ at a rate of $o(1/\sqrt{Nb_N})$. Assumption \ref{sec04 - assumption06}(c) and (d) ensure that the second-order remainder in the asymptotic linear representation is asymptotically negligible. These conditions are discussed further in Remark \ref{sec04 - remark02}. Assumption \ref{sec04 - assumption06}(e) requires the classification algorithm to be asymptotically correctly specified at a suitable rate.

The remaining quantities enter the asymptotic variance matrix of the debiased estimator. Let $\Upsilon_{0}$ denote the Jacobian of the population moment evaluated at $\theta_0$, defined as follows:
\begin{equation}\label{sec04 - upsilon}
    \Upsilon_{0} \equiv \expect{ \nabla_{\theta}m(\theta_0, \classifier{A}_{i}(\discfunc_0), W_i) }.
\end{equation}
The scaled second-moment matrix $\expect{ b_{N}\psi_N(\theta_0, \discfunc_0, \vecmu_0(\theta_0,\cdot), W_i)\psi_N(\theta_0, \discfunc_0, \vecmu_0(\theta_0,\cdot), W_i)^\top }$ also enters the asymptotic variance. Up to an $O(\sqrt{b_N})$ remainder, it is given by:
\[ \sum_{g, h=1}^{G_0-1}\frac{1}{b_N}\Expect{\mu_{0,g}(\theta_0,Z_i)\nabla\KER{\discfunc_{0,g}}^\top\condexpect{(\vect_{i,g}-\discfunc_{0,g})(\vect_{i,h}-\discfunc_{0,h})^\top}{Z_i}\nabla\KER{\discfunc_{0,h}}\mu_{0,h}(\theta_0,Z_i)^\top}. \]
Although the current assumptions imply that this matrix is $O(1)$, they do not guarantee convergence to a positive definite limit. We impose the following assumption. 
\begin{assumption}\textnormal{(Variance Regularity)}\label{sec04 - assumption07}\\
    There exists a positive definite matrix $\Sigma_{0}\in \mathbb{R}^{d_{M}\times d_{M}}$ such that:
    \[ \expect{ b_{N}\psi_N(\theta_0, \discfunc_0, \vecmu_0(\theta_0,\cdot), W_i)\psi_N(\theta_0, \discfunc_0, \vecmu_0(\theta_0,\cdot), W_i)^\top }  \to \Sigma_{0}. \]
\end{assumption}\vspace{+0.5cm}

Now we are ready to provide the main asymptotic result of the debiased estimator:

\begin{theorem}\label{sec04 - theorem01}
    Suppose Assumptions \ref{sec03 - assumption01}, \ref{sec03 - assumption02}, \ref{sec03 - assumption03}, \ref{sec04 - assumption04}, \ref{sec04 - assumption05},  \ref{sec04 - assumption06}, and  \ref{sec04 - assumption07} hold. Then,
    \begin{equation}
        \sqrt{Nb_N}(\widehat{\theta}_N - \theta_0)
            \overset{d}{\to}    \caln\left(0\,,\,\, \Omega_0 \right),  
    \end{equation}
    where $\Omega_0    \equiv  \big\{\Upsilon_{0}^{\top}\Upsilon_{0}\big\}^{-1}\Upsilon_{0}^\top\Sigma_{0}\Upsilon_{0}\big\{\Upsilon_{0}^{\top}\Upsilon_{0}\big\}^{-1}.$
\end{theorem}

\textbf{Proof of Theorem \ref{sec04 - theorem01}} : See Appendix \ref{sec04 - theorem01 - proof}.\\

\begin{remarkk}\label{sec04 - remark01}\textnormal{(About conditions on Bandwidth)}

Assumption \ref{sec04 - assumption06}(b) characterizes the admissible order of $b_N$ and restricts the convergence rate of $\norm{\widehat{\discfunc}-\discfunc_0}_{\infty}$. To illustrate, we consider polynomial rates. For some $R,B>0$, let $\rho_N=N^{-R}$ and $b_N=N^{-B}$. Assumption \ref{sec04 - assumption06}(b) requires $Nb_N=N^{1-B}\to\infty$ and $b_N^2\sqrt{Nb_N}=N^{(1-5B)/2}\to0$ as $N\to\infty$, which together imply $1/5<B<1$. Moreover, the condition $\rho_N=O(b_N)$ requires $\rho_N/b_N=N^{-R+B}$ to remain bounded, implying $B\leq R$. The condition $N\rho_{N}^3 = O(1)$ implies $R \geq 1/3$. Thus, $B$ and $R$ must satisfy
\[ \frac{1}{5}<B<1 \quad\textnormal{and}\quad \max\{B, 1/3\} \leq R. \]
Equivalently, for these polynomial sequences,
\[ N^{-1}<b_N<N^{-1/5} \quad\textnormal{and}\quad \rho_N\leq \min\{b_N, N^{-1/3}\}.\,\, \| \]
\end{remarkk}\vspace{+0.5cm}

\begin{remarkk}\label{sec04 - remark02}\textnormal{(About the condition that $N\rho_N^3 = O(1)$)}

In usual DML estimators without kernel smoothing, a sufficient rate condition is $\sqrt{N}\rho_N^2 = O(1)$, or equivalently $\rho_N = O(N^{-1/4})$ \citep{dml_chernozhukov_et_al2018, dml_chernozhukov_et_al2022, dml_ahrens_et_al2026}, when the first-stage estimation error is $o_P(\rho_N)$. This condition controls the second-order remainder in the asymptotic linear representation. In our setting, the corresponding rate factor is of order $O\big( \sqrt{Nb_{N}} \times (\rho_N^2/b_N) \big)$. Since Assumption \ref{sec04 - assumption06}(b) imposes $\rho_N/b_N = O(1)$, we have:
\[ O\left(  \sqrt{Nb_{N}}\frac{\rho_{N}^2}{b_{N}} \right) = O\left( \sqrt{ N\rho_{N}^3 \frac{\rho_{N}}{b_{N}} } \right). \]
Thus, $N\rho_N^3 = O(1)$ is sufficient to ensure this rate.$\|$
\end{remarkk}\vspace{+0.5cm}

The rate condition $N\rho_N^3 = O(1)$ usually requires stronger sufficient smoothness conditions on the first-stage nuisance $\discfunc_0(\cdot)$ than the condition $N\rho_N^4 = O(1)$ in the usual DML setup without smoothing. We illustrate this difference using two first-stage learners: kernel ridge regression (KRR) and neural network regression (NN).

\begin{example}\textnormal{(Kernel ridge regression)}\label{sec04 - example01 - KRR}

We first consider KRR as the first-stage learner. Let $B_{2,2}^{s}$ denote the Sobolev space of smoothness $s$ with integrability index $2$. Choose a KRR kernel, distinct from the smoothing kernel, whose reproducing kernel Hilbert space is norm-equivalent to $B_{2,2}^{r}$ with $r>s$, and set the regularization parameter to $\lambda_{\mathrm{KRR},N}\asymp N^{-2r/(2s+d_Z)}$. Suppose that $\delta_{0,gh}(\cdot)\in B_{2,2}^{s}$ for every $g<h$. Under the regularity conditions of \citet{fischer_steinwart2020}, their Corollary 4.1 gives:
\[ \norm{ \widehat{\discfunc}_{\ell} - \discfunc_0  }_{\infty}=O_P\left(N^{-(s-t)/(2s+d_Z)}\right)\quad\textnormal{for }d_Z/2<t<s. \]
\noindent If $s > \frac{3}{2}d_Z$, it yields $\norm{ \widehat{\discfunc}_{\ell} - \discfunc_0  }_{\infty}=o_P(N^{-1/4})$ by choosing $d_Z/2<t<(2s-d_Z)/4$.

\noindent If $s > \frac{5}{2}d_Z$, it yields $\norm{ \widehat{\discfunc}_{\ell} - \discfunc_0  }_{\infty}=o_P(N^{-1/3})$ by choosing $d_Z/2<t<(s-d_Z)/3$. $\|$
\end{example}\vspace{+0.5cm}

\begin{example}\textnormal{(Neural network regression)}\label{sec04 - example02 - NN}

We next consider the NN estimator of \citet{imaizumi2020} as the first-stage learner. Suppose that each $\delta_{0,gh}$ belongs to an $s$-Hölder ball on $[0,1]^{d_Z}$ with $s>0$. Under the regularity conditions and tuning choices in that paper, for some fixed $a\geq 0$, Theorem 3 of \citet{imaizumi2020} implies:
\[ \norm{ \widehat{\discfunc}_{\ell} - \discfunc_0  }_{\infty}=O_P\left(N^{-s/(2s+d_Z)}(\log N)^{\max\{2,a\}/2}\right), \]
\noindent If $s > \frac{1}{2}d_Z$, it yields $\norm{ \widehat{\discfunc}_{\ell} - \discfunc_0  }_{\infty}=o_P(N^{-1/4})$.

\noindent If $s > d_Z$, it yields $\norm{ \widehat{\discfunc}_{\ell} - \discfunc_0  }_{\infty}=o_P(N^{-1/3})$.$\|$
\end{example}\vspace{+0.5cm}

For valid inference, the asymptotic variance must be consistently estimated. To this end, define:
\begin{equation}
    \widehat{\Upsilon} = \frac{1}{L}\sum_{\ell=1}^{L}\frac{1}{|I_{\ell}|}\sum_{i\in I_{\ell}}\nabla_{\theta}\psi_{N,\ell}( \widehat{\theta}_{N}, W_i),
\end{equation}
\begin{equation}
    \widehat{\Sigma} = \frac{1}{L}\sum_{\ell=1}^{L}\frac{1}{|I_{\ell}|}\sum_{i\in I_{\ell}}b_{N}\psi_{N,\ell}( \widehat{\theta}_{N}, W_i)\psi_{N,\ell}( \widehat{\theta}_{N},  W_i)^\top.
\end{equation}
The following proposition establishes that $\widehat{\Upsilon}$ and $\widehat{\Sigma}$ are consistent estimators of $\Upsilon_{0}$ and $\Sigma_0$, respectively.

\begin{proposition}\label{sec04 - proposition05}
    Under Assumptions \ref{sec03 - assumption01}, \ref{sec03 - assumption02}, \ref{sec03 - assumption03}, \ref{sec04 - assumption04}, \ref{sec04 - assumption05}, \ref{sec04 - assumption06}, and \ref{sec04 - assumption07},  
    $\widehat{\Upsilon}\overset{P}{\to} \Upsilon_{0}$ and $\widehat{\Sigma}\overset{P}{\to} \Sigma_{0}$ hold, and therefore:
    \[ \big( \widehat{\Upsilon}^\top\widehat{\Upsilon} \big)^{-1} \widehat{\Upsilon}^\top \widehat{\Sigma} \widehat{\Upsilon} \big( \widehat{\Upsilon}^\top\widehat{\Upsilon} \big)^{-1} \overset{P}{\to} \Omega_0. \]
\end{proposition}

\textbf{Proof of Proposition \ref{sec04 - proposition05}} : See Appendix \ref{sec04 - proposition05 - proof}.\\


\begin{remarkk}\label{sec04 - remark04}\textnormal{(Choice of Bandwidth via Cross-Validation)}

We provide a cross-validation rule for choosing a bandwidth. We emphasize that \emph{this is a practical tuning rule for bandwidth selection} and that \emph{we have not established uniformly valid post-selection inference}. Although this would be a useful extension of the paper, it is beyond our scope.

We set $b_N=\widetilde{b}_0N^{-B}$, where $B>0$ is fixed and satisfies Assumption \ref{sec04 - assumption06}, and consider candidates $\widetilde{b}_0\in\overline{\mathscr{B}}$. We divide the full sample $[N]$ into $L$ folds, $I_1,\ldots,I_L$. For each fold $I_\ell$ and each candidate $\widetilde{b}_0\in\overline{\mathscr{B}}$, we use the observations in $I_{-\ell}$ to obtain the debiased estimator in Definition \ref{sec04 - definition01}, denoted by $\widetilde{\theta}_{\textnormal{dml},\ell}(\widetilde{b}_0)$. We then construct the following objective functions:
\[ \mathcal{R}_{\ell}(\widetilde{b}_{0}) = \frac{1}{|I_\ell|}\sum_{i\in I_{\ell}} \norm{m\bigl(\widetilde{\theta}_{\textnormal{dml},\ell}(\widetilde{b}_{0}), \classifier{A}_i(\widehat{\discfunc}_\ell),W_i\bigr)}^2 \quad\textnormal{and}\quad \overline{\mathcal{R}}(\widetilde{b}_{0}) \equiv \frac{1}{L}\sum_{\ell=1}^{L}\mathcal{R}_\ell(\widetilde{b}_{0}), \]
where $\widehat{\discfunc}_{\ell}$ is estimated using the observations in $I_{-\ell}$.

One choice is $\widetilde{b}_{\min} =\operatorname*{argmin}_{\widetilde{b}_{0}\in\overline{\mathscr{B}}} \overline{\mathcal{R}}(\widetilde{b}_{0})$. In our simulation experiments, however, $\overline{\mathcal{R}}(\widetilde{b}_0)$ is sometimes decreasing in $\widetilde{b}_0$, so $\widetilde{b}_{\min}$ frequently selects a large value from the candidate set. Because a large $\widetilde{b}_0$ may introduce smoothing bias, we instead use a one-standard-error rule. Specifically, we choose
\[ \widetilde{b}_{\textnormal{CV}} \equiv \min\left\{ \widetilde{b}_{0}\in\overline{\mathscr{B}}: \overline{\mathcal{R}}(\widetilde{b}_{0}) \leq \overline{\mathcal{R}}(\widetilde{b}_{\min}) +\textnormal{SE}_{\mathcal{R}}(\widetilde{b}_{\min}) \right\}, \]
where $\textnormal{SE}_{\mathcal{R}}(\widetilde{b}_{0}) \equiv L^{-1/2}\textnormal{SD}\bigl( \{\mathcal{R}_\ell(\widetilde{b}_{0}):\ell\in[L]\} \bigr)$. Finally, we obtain the main debiased estimator $\widehat{\theta}_N$ according to Definition \ref{sec04 - definition01}, using $b_N=\widetilde{b}_{\textnormal{CV}}N^{-B}$.$\|$
\end{remarkk}

\section{Simulation Experiment}\label{sec05}

In this section, we conduct a simple numerical experiment to evaluate the performance of the plug-in and debiased estimators, focusing on bias, mean squared error, and coverage probability for different choices of $b_N$. We also show that the debiased estimator with $b_N$ chosen by the tuning rule in Remark \ref{sec04 - remark04} achieves valid inference.

\subsection{Setup}

We generate data that are independent across $i\in[N]$, consisting of signals $Z_{i} = (Z_{i1}, ..., Z_{i4})^{\top}\in \mathbb{R}^{4}$, proxies $(T_{i,12}, T_{i,13}, T_{i,23})^\top \in \mathbb{R}^3$, covariate $X_{i}\in \mathbb{R}$ and outcome $Y_{i}\in\mathbb{R}$. The number of groups is three and $\group_{0i}$ takes one of the values $\eucvec{1}, \eucvec{2}$ and $\eucvec{3}$. The detailed data-generating process is described in Appendix \ref{appendix - a-4}. The true value of $(\theta_{0,1},\theta_{0,2},\theta_{0,3})^\top$ is specified to be $(1, 2, 3)^\top$, and the true model is:
\[  Y_i = \sum_{g=1}^{3}\theta_{0,g}X_i \eucvec{g}^\top \group_{0i}+ U_i\quad\textnormal{where}\quad\condexpect{U_i}{X_i, \group_{0i}} = 0.\]
Then the valid moment contribution is $m(\theta, \classifier{A}_{i}(\discfunc), W_i) = \sum_{g=1}^{3}M_{g}(\theta, W_i)\eucvec{g}^\top \classifier{A}_i(\discfunc)$, where:
\[  M_{1}(\theta,W_i) = X_i(Y_i - \theta_1 X_i)\eucvec{1},\quad
    M_{2}(\theta,W_i) = X_i(Y_i - \theta_2 X_i)\eucvec{2}\quad\textnormal{and}\quad
    M_{3}(\theta,W_i) = X_i(Y_i - \theta_3 X_i)\eucvec{3}.\]
The additional nuisance $\mu_{0,1}(\widetilde{\theta},Z_i)$ and $\mu_{0,2}(\widetilde{\theta},Z_i)$ are then defined by:
\[ \mu_{0,1}(\widetilde{\theta}, Z_i) = \begin{bmatrix} 
    +\condexpect{X_iY_i}{Z_i} - \widetilde{\theta}_{1}\condexpect{X_i^2}{Z_i} \\ 
    0 \\
    -\condexpect{X_iY_i}{Z_i} + \widetilde{\theta}_{3}\condexpect{X_i^2}{Z_i} \end{bmatrix}\quad\textnormal{and}\quad
    \mu_{0,2}(\widetilde{\theta}, Z_i) = \begin{bmatrix} 
    0 \\ 
    +\condexpect{X_iY_i}{Z_i} - \widetilde{\theta}_{2}\condexpect{X_i^2}{Z_i} \\
    -\condexpect{X_iY_i}{Z_i} + \widetilde{\theta}_{3}\condexpect{X_i^2}{Z_i} \end{bmatrix}. \]
We generate a nonlinear function $\eta_0(Z_i) = Z_{i1} + Z_{i2} + \sin(Z_{i1} + Z_{i2})$ and a random variable $S_i = \eta_0(Z_i) + \varepsilon_{S,i}$ with $\varepsilon_{S,i} \overset{\textnormal{iid}}{\sim}\caln(0,1)$, and the true type is generated by:
\[ \group_{0,i} =  \eucvec{1}\ind{}{\eta_0(Z_i) \in (-\infty, 0.5]} + \eucvec{2}\ind{}{\eta_0(Z_i) \in [0.5, 1.75)}+\eucvec{3}\ind{}{\eta_0(Z_i) \in [1.75, \infty)}. \]
The proxy variables $\vect$ and the discriminant functions $\discfunc$ are defined accordingly (See Appendix \ref{appendix - a-4}).

The univariate kernel is specified to be \eqref{sec04 - example04 - eq01 - kernel} with $S = 1$. The debiased estimators are obtained using $10$-fold cross-fitting. 

\subsection{Experiment 1: Impact of the first-stage convergence rate}

We first examine how the convergence rate of $\rho_N$ relative to $b_N$ affects the bias and the coverage probability. We set the sample size to $N \in \{500, 1,000, 2,000, 5,000\}$, and use $10,000$ of simulation replications. We set the bandwidth to $b_N=\widetilde{b}_0N^{-0.25}$, with $\widetilde{b}_0\in\{0.5,0.6,\ldots,1.9, 2.0\}$, and consider first-stage rates $R\in\{0.10, 0.35\}$. Theorem \ref{sec04 - theorem01} predicts correct coverage for $R=0.35$, but not for $R=0.10$. We use the first-stage learner $\widehat{\eta}(Z_i)=\eta_0(Z_i)+N^{-R}\xi_i$, where $N^{-R}\xi_i$ is a random drift (see Appendix \ref{appendix - a-4}), while fixing $\condexpect{XY}{Z}$ and $\condexpect{X^2}{Z}$ at their oracle values.

Table \ref{sec05 - table01} compares the plug-in biases, the DML biases with a bandwidth selected by cross-validation, and the DML biases with the fixed bandwidth, and the root mean squared error (RMSE) of each estimate. The unscaled RMSE averages squared estimation errors across the three parameters and $10,000$ simulation replications:
\[ \textnormal{RMSE} = \sqrt{ \frac{1}{10{,}000} \sum_{s=1}^{10{,}000} \frac{1}{3} \sum_{g=1}^{3} \bigl(\widehat{\theta}_{g}^{(s)}-\theta_{0,g}\bigr)^2 }. \]
For both values of $R$, the plug-in estimates has larger bias in absolute value than the debiased estimates. When $R=0.10$, the debiased biases are generally larger than those for $R=0.35$ at smaller bandwidths, and some remain visible even at $N=5,000$. As $\widetilde{b}_0$ increases, the DML biases tend to grow in absolute value, particularly when $R=0.10$. This pattern is consistent with greater smoothing bias at larger bandwidths. The plug-in bias declines steadily with $N$; the DML biases are already small and generally tend to decline with $N$.  

For the scaled RMSE, the squared errors inside this expression are multiplied by $N$ for the plug-in estimates and by $Nb_N$ for the debiased estimates. If first-stage error does not affect the limiting distribution, the scaled RMSE should remain stable as $N$ increases. The plug-in estimate does not exhibit this pattern and its scaled RMSE increases, though its unscaled RMSE decreases with $N$. Both RMSEs are smaller at larger values of $R$, but the scaled RMSE still rises with $N$ for every value of $R$ considered. In contrast, the scaled RMSEs of the debiased estimates are relatively stable when $\widetilde{b}_0$ is small, including in the cases with $R>0.21$. The debiased estimate that selects $\widetilde{b}_0$ by cross-validation shows a similarly stable pattern. With a larger fixed bandwidth, particularly $\widetilde{b}_0=2.0$, the scaled RMSE increases with $N$ for some values of $R$.

Table \ref{sec05 - table02} reports the empirical coverage probabilities of nominal 95\% confidence intervals for $R=0.35$ and $R=0.10$. The plug-in estimates has undercoverage in both cases. Its bias, when multiplied by $\sqrt{N}$ in the $t$-statistic, shifts the center of the statistic away from zero. When $R=0.10$, the debiased estimates also tend to undercover, although their coverage probabilities are much closer to the nominal level than that of the plug-in estimates. When $R=0.35$, DML coverage is generally close to the nominal level, except when $\widetilde{b}_0$ is large, which is consistent with smoothing bias induced by a larger $\widetilde{b}_0$.

\subsection{Experiment 2: Finite-sample performance with the specified first-stage learners}

We additionally investigate the finite-sample performance of the plug-in and debiased estimates with the specified first-stage learners. We set the sample size to $N = 500$, and use $10,000$ of simulation replications. We set the bandwidth to $b_N=\widetilde{b}_0 N^{-0.25}$, where $\widetilde{b}_0\in\{0.5, 0.6,\ldots, 2.0\}$.

Gaussian kernel ridge regression (KRR) estimates the nuisance functions $\discfunc$ and $\condexpect{X^2}{Z}$ by learning how each varies with $Z$. We use the Gaussian kernel $\exp\big(-\norm{z-z'}^2/2\big)$ with $\lambda_N^{\textnormal{penalty}} \asymp N^{-1}$, $ c_\textnormal{ridge} = 0.2$, and a ridge parameter of $c_\textnormal{ridge}N\lambda_N^{\textnormal{penalty}} = c_\textnormal{ridge}$.\footnote{Since $d_{Z} = 4$, we choose $s = 12$ and $r = 14$. Therefore, $-2r/(2s + d_Z) = -28/(24 + 4) = -1$.} We estimate $\condexpect{XY}{Z}$ using neural networks (NN) with hidden layer structure $[16,8]$ to accommodate the relatively complicated structure of $\condexpect{XY}{Z} = \sum_g \theta_{0,g} \condexpect{X^2\eucvec{g}^{\top}\group_0}{Z}$. Though we primarily exploit KRR to minimize the computation time, other flexible methods can also be used.

Table~\ref{sec05 - table03} reports the empirical frequency with which cross-validation selects each value of $\widetilde{b}_0$. The rule most often selects $0.5$ or $0.6$, while large candidate values are rarely selected.

Table~\ref{sec05 - table04} reports the bias and the empirical coverage probabilities of the $t$-statistic. Although the plug-in bias is modest in absolute terms, it is generally larger than the DML bias. The plug-in estimate is different from zero, whereas the DML estimates are generally much closer to zero, suggesting that the intended debiasing works in this design. As seen in Tables \ref{sec05 - table01} and \ref{sec05 - table03}, scaling the plug-in bias by $\sqrt{N}$ shifts the distribution of its $t$-statistic away from zero, leading to undercoverage. In contrast, the DML coverage probabilities are generally close to their nominal levels. When $\widetilde{b}_0$ is large, their finite-sample performance is slightly worse, which is consistent with the presence of smoothing bias.

\section{Concluding Remarks}

In this paper, we provide valid inference for GMM problems in which discrete latent types are estimated using machine learning or classification methods. Because the first-stage nuisance functions enter the moment function through discontinuous indicators, the $\mathcal{L}^2$-Riesz representer required for bias correction is not well defined. By approximating the indicators with a smooth kernel function, we construct the Neyman-orthogonal score and define the debiased estimator. We establish its $\sqrt{Nb}$-consistency and asymptotic normality under regularity conditions on the first-stage learner, the moment function, the kernel, and the shrinking tuning parameter $b$.

Our approach applies to many modern empirical problems involving unobserved discrete types, including demographic characteristics such as gender and ethnicity, as well as economic categories such as innovative firms, occupational groups, and counterfactual status. We do not consider settings in which the first-stage learner uses unsupervised learning such as clustering, or settings in which the moment function depends on group averages of other observations under spatial or peer interactions. Nevertheless, we anticipate that our approach could be extended to these settings with suitable modifications, which we leave for future research.

\begin{landscape}
\begin{table}[p]
    \centering
    \begin{tabular}{c|c|ccc|cc|ccc|cc}
    \hline\hline
    \multirow{3}{*}{Estimator}
    & \multirow{3}{*}{$N$} & \multicolumn{5}{c|}{$R=0.10$} & \multicolumn{5}{c}{$R=0.35$} \\
    \cline{3-12}
    & & \multicolumn{3}{c|}{Bias} & \multicolumn{2}{c|}{RMSE}& \multicolumn{3}{c|}{Bias} & \multicolumn{2}{c}{RMSE}\\
    & & $(\theta_1)$ & $(\theta_2)$ & $(\theta_3)$ & (Unscaled) & (Scaled)
      & $(\theta_1)$ & $(\theta_2)$ & $(\theta_3)$ & (Unscaled) & (Scaled)\\
    \hline
    \multirow{4}{*}{Plug-in}
      & 500   & 0.217 & 0.095 & -0.153 & 0.191 & 4.280
              & 0.044 & 0.017 & -0.033 & 0.074 & 1.653 \\
      & 1,000 & 0.201 & 0.087 & -0.144 & 0.172 & 5.435
              & 0.034 & 0.013 & -0.027 & 0.053 & 1.678 \\
      & 2,000 & 0.187 & 0.080 & -0.134 & 0.157 & 7.037
              & 0.027 & 0.009 & -0.021 & 0.039 & 1.731 \\
      & 5,000 & 0.172 & 0.073 & -0.123 & 0.143 & 10.080
              & 0.020 & 0.007 & -0.016 & 0.026 & 1.825 \\
    \hline
    \multirow{4}{*}{DML (CV)}
      & 500   & -0.011 & -0.017 & -0.005 & 0.211 & 1.568
              & 0.005 & 0.005 & -0.007 & 0.188 & 1.423 \\
      & 1,000 & -0.010 & -0.018 & -0.007 & 0.159 & 1.511
              & 0.007 & 0.004 & -0.006 & 0.145 & 1.397 \\
      & 2,000 & -0.005 & -0.019 & -0.005 & 0.119 & 1.463
              & 0.003 & 0.001 & -0.005 & 0.111 & 1.380 \\
      & 5,000 & -0.004 & -0.018 & -0.008 & 0.083 & 1.439
              & 0.004 & 0.002 & -0.003 & 0.080 & 1.384 \\
    \hline
    \multirow{4}{*}{DML ($\widetilde{b}_0=0.5$)}
      & 500   & -0.011 & -0.017 & -0.005 & 0.215 & 1.563
              & 0.003 & 0.004 & -0.006 & 0.194 & 1.411 \\
      & 1,000 & -0.010 & -0.018 & -0.007 & 0.160 & 1.511
              & 0.006 & 0.004 & -0.005 & 0.148 & 1.391 \\
      & 2,000 & -0.005 & -0.019 & -0.005 & 0.120 & 1.463
              & 0.003 & 0.001 & -0.005 & 0.112 & 1.375 \\
      & 5,000 & -0.004 & -0.018 & -0.008 & 0.083 & 1.439
              & 0.004 & 0.002 & -0.003 & 0.080 & 1.381 \\
    \hline
    \multirow{4}{*}{DML ($\widetilde{b}_0=1.0$)}
      & 500   & -0.007 & -0.016 & -0.007 & 0.159 & 1.636
              & 0.019 & 0.009 & -0.017 & 0.145 & 1.487 \\
      & 1,000 & -0.006 & -0.018 & -0.009 & 0.118 & 1.573
              & 0.019 & 0.007 & -0.015 & 0.109 & 1.449 \\
      & 2,000 & -0.003 & -0.018 & -0.008 & 0.088 & 1.528
              & 0.015 & 0.005 & -0.012 & 0.084 & 1.447 \\
      & 5,000 & -0.002 & -0.017 & -0.009 & 0.061 & 1.497
              & 0.013 & 0.005 & -0.011 & 0.058 & 1.425 \\
    \hline
    \multirow{4}{*}{DML ($\widetilde{b}_0=2.0$)}
      & 500   & 0.007 & -0.004 & -0.020 & 0.121 & 1.755
              & 0.058 & 0.024 & -0.048 & 0.118 & 1.723 \\
      & 1,000 & 0.006 & -0.010 & -0.018 & 0.089 & 1.687
              & 0.051 & 0.020 & -0.040 & 0.090 & 1.703 \\
      & 2,000 & 0.005 & -0.013 & -0.015 & 0.067 & 1.639
              & 0.042 & 0.016 & -0.034 & 0.070 & 1.711 \\
      & 5,000 & 0.004 & -0.014 & -0.013 & 0.047 & 1.608
              & 0.034 & 0.012 & -0.028 & 0.050 & 1.737 \\
    \hline\hline
    \end{tabular}
    \caption{Monte Carlo biases and root mean squared errors of the plug-in and DML estimates for $R=0.10$ and $R=0.35$. Unscaled RMSEs are computed across the three parameters and simulation replications. For scaled RMSEs, plug-in estimation errors are scaled by $\sqrt{N}$, while DML estimation errors are scaled by $\sqrt{N b_N}$.}
    \label{sec05 - table01}
\end{table}
\end{landscape}

\begin{landscape}

\begin{table}[htbp]
\centering
\begin{tabular}{c|c|c|c|cccc|c|cccc}
\hline\hline
\multirow{3}{*}{Coverage}
& \multirow{3}{*}{$N$}
& \multirow{3}{*}{$\theta$}
& \multicolumn{5}{c|}{$R=0.10$}
& \multicolumn{5}{c}{$R=0.35$} \\
\cline{4-13}
& & 
& \multirow{2}{*}{Plug-in} & \multicolumn{4}{c|}{DML}
& \multirow{2}{*}{Plug-in} & \multicolumn{4}{c}{DML}\\
&&& 
& (CV) & ($\widetilde{b}_0=0.5$) & ($\widetilde{b}_0=1.0$) & ($\widetilde{b}_0=2.0$) &
& (CV) & ($\widetilde{b}_0=0.5$) & ($\widetilde{b}_0=1.0$) & ($\widetilde{b}_0=2.0$) \\
\hline
\multirow{12}{*}{90\%}
& \multirow{3}{*}{500}
& $\theta_1$
& 0.228 & 0.874 & 0.873 & 0.881 & 0.877
& 0.831 & 0.905 & 0.906 & 0.892 & 0.851 \\
& & $\theta_2$
& 0.583 & 0.903 & 0.903 & 0.900 & 0.899
& 0.866 & 0.904 & 0.907 & 0.901 & 0.895 \\
& & $\theta_3$
& 0.241 & 0.890 & 0.889 & 0.892 & 0.885
& 0.822 & 0.906 & 0.908 & 0.892 & 0.847 \\
\cline{2-13}
& \multirow{3}{*}{1,000}
& $\theta_1$
& 0.115 & 0.879 & 0.879 & 0.885 & 0.879
& 0.820 & 0.904 & 0.904 & 0.896 & 0.845 \\
& & $\theta_2$
& 0.478 & 0.899 & 0.899 & 0.898 & 0.897
& 0.857 & 0.900 & 0.904 & 0.899 & 0.895 \\
& & $\theta_3$
& 0.109 & 0.890 & 0.889 & 0.894 & 0.881
& 0.808 & 0.903 & 0.904 & 0.895 & 0.835 \\
\cline{2-13}
& \multirow{3}{*}{2,000}
& $\theta_1$
& 0.051 & 0.893 & 0.893 & 0.890 & 0.884
& 0.799 & 0.897 & 0.897 & 0.889 & 0.830 \\
& & $\theta_2$
& 0.392 & 0.899 & 0.899 & 0.894 & 0.887
& 0.843 & 0.896 & 0.898 & 0.897 & 0.891 \\
& & $\theta_3$
& 0.044 & 0.888 & 0.888 & 0.889 & 0.875
& 0.780 & 0.903 & 0.904 & 0.893 & 0.824 \\
\cline{2-13}
& \multirow{3}{*}{5,000}
& $\theta_1$
& 0.021 & 0.891 & 0.890 & 0.891 & 0.882
& 0.763 & 0.899 & 0.900 & 0.896 & 0.810 \\
& & $\theta_2$
& 0.276 & 0.895 & 0.895 & 0.890 & 0.878
& 0.833 & 0.896 & 0.897 & 0.901 & 0.888 \\
& & $\theta_3$
& 0.011 & 0.886 & 0.886 & 0.885 & 0.861
& 0.733 & 0.899 & 0.899 & 0.892 & 0.790 \\
\hline
\multirow{12}{*}{95\%}
& \multirow{3}{*}{500}
& $\theta_1$
& 0.311 & 0.928 & 0.927 & 0.937 & 0.932
& 0.900 & 0.960 & 0.962 & 0.951 & 0.915 \\
& & $\theta_2$
& 0.670 & 0.957 & 0.958 & 0.955 & 0.950
& 0.927 & 0.956 & 0.959 & 0.954 & 0.945 \\
& & $\theta_3$
& 0.334 & 0.945 & 0.943 & 0.947 & 0.940
& 0.899 & 0.960 & 0.963 & 0.949 & 0.914 \\
\cline{2-13}
& \multirow{3}{*}{1,000}
& $\theta_1$
& 0.162 & 0.933 & 0.933 & 0.940 & 0.935
& 0.895 & 0.960 & 0.960 & 0.953 & 0.914 \\
& & $\theta_2$
& 0.553 & 0.953 & 0.953 & 0.949 & 0.947
& 0.919 & 0.953 & 0.954 & 0.953 & 0.948 \\
& & $\theta_3$
& 0.160 & 0.942 & 0.942 & 0.945 & 0.938
& 0.884 & 0.957 & 0.958 & 0.951 & 0.905 \\
\cline{2-13}
& \multirow{3}{*}{2,000}
& $\theta_1$
& 0.074 & 0.939 & 0.939 & 0.943 & 0.940
& 0.875 & 0.955 & 0.955 & 0.947 & 0.901 \\
& & $\theta_2$
& 0.457 & 0.954 & 0.954 & 0.949 & 0.942
& 0.909 & 0.951 & 0.952 & 0.951 & 0.946 \\
& & $\theta_3$
& 0.066 & 0.943 & 0.943 & 0.942 & 0.934
& 0.862 & 0.957 & 0.957 & 0.949 & 0.895 \\
\cline{2-13}
& \multirow{3}{*}{5,000}
& $\theta_1$
& 0.027 & 0.942 & 0.942 & 0.945 & 0.937
& 0.845 & 0.952 & 0.952 & 0.950 & 0.888 \\
& & $\theta_2$
& 0.324 & 0.949 & 0.949 & 0.943 & 0.934
& 0.901 & 0.948 & 0.949 & 0.946 & 0.943 \\
& & $\theta_3$
& 0.016 & 0.939 & 0.939 & 0.938 & 0.919
& 0.823 & 0.949 & 0.950 & 0.942 & 0.869 \\
\hline
\multirow{12}{*}{99\%}
& \multirow{3}{*}{500}
& $\theta_1$
& 0.498 & 0.976 & 0.975 & 0.982 & 0.981
& 0.971 & 0.995 & 0.995 & 0.994 & 0.980 \\
& & $\theta_2$
& 0.803 & 0.995 & 0.995 & 0.993 & 0.991
& 0.981 & 0.995 & 0.995 & 0.994 & 0.991 \\
& & $\theta_3$
& 0.540 & 0.985 & 0.985 & 0.989 & 0.987
& 0.970 & 0.995 & 0.996 & 0.993 & 0.979 \\
\cline{2-13}
& \multirow{3}{*}{1,000}
& $\theta_1$
& 0.284 & 0.980 & 0.980 & 0.986 & 0.986
& 0.971 & 0.995 & 0.996 & 0.993 & 0.978 \\
& & $\theta_2$
& 0.687 & 0.993 & 0.993 & 0.990 & 0.988
& 0.979 & 0.993 & 0.994 & 0.992 & 0.990 \\
& & $\theta_3$
& 0.299 & 0.985 & 0.985 & 0.989 & 0.987
& 0.966 & 0.995 & 0.995 & 0.992 & 0.977 \\
\cline{2-13}
& \multirow{3}{*}{2,000}
& $\theta_1$
& 0.137 & 0.985 & 0.985 & 0.986 & 0.987
& 0.959 & 0.993 & 0.993 & 0.991 & 0.974 \\
& & $\theta_2$
& 0.575 & 0.991 & 0.991 & 0.990 & 0.989
& 0.974 & 0.994 & 0.994 & 0.991 & 0.990 \\
& & $\theta_3$
& 0.127 & 0.986 & 0.986 & 0.988 & 0.984
& 0.957 & 0.993 & 0.993 & 0.990 & 0.970 \\
\cline{2-13}
& \multirow{3}{*}{5,000}
& $\theta_1$
& 0.046 & 0.987 & 0.987 & 0.988 & 0.986
& 0.943 & 0.992 & 0.992 & 0.989 & 0.971 \\
& & $\theta_2$
& 0.414 & 0.990 & 0.990 & 0.987 & 0.985
& 0.969 & 0.991 & 0.991 & 0.990 & 0.988 \\
& & $\theta_3$
& 0.032 & 0.985 & 0.985 & 0.983 & 0.979
& 0.936 & 0.992 & 0.992 & 0.989 & 0.958 \\
\hline\hline
\end{tabular}
    \caption{Empirical coverage probabilities of $t$-statistics for the plug-in and DML estimates.}
    \label{sec05 - table02}
    \end{table}
\end{landscape}

\begin{table}[htbp]
    \centering
    \begin{tabular}{c|cccccccccccc}
    \hline\hline
    $\widetilde{b}_0$ & $0.5$ & $0.6$ & $0.7$ & $0.8$ & $0.9$ & $1.0$ & $1.1$ & $1.2$ & $1.3$ & $1.4$ & $1.5$ & $\geq 1.6$ \\
    \hline
    Frequency & 0.198 & 0.076 & 0.086 & 0.090 & 0.091 & 0.082 & 0.087 & 0.084 & 0.072 & 0.061 & 0.043 & 0.030 \\
    \hline\hline
    \end{tabular}
    \caption{Empirical frequencies of $\widetilde{b}_0$ selected by cross-validation.}
    \label{sec05 - table03}
\end{table}
\vspace{1em}
\begin{table}[htbp]
    \centering
    \begin{tabular}{c|c|c|cccc}
    \hline\hline
    &  \multirow{2}{*}{$\theta$}& \multirow{2}{*}{Plug-in} & \multicolumn{4}{c}{DML}\\
    & & & (CV) & ($\widetilde{b}_0=0.5$) & ($\widetilde{b}_0=0.8$) & ($\widetilde{b}_0=1.0$)\\
    \hline
    \multirow{3}{*}{Bias}
      & $\theta_1$  & 0.080 & 0.032 & 0.023 & 0.028 & 0.032 \\
      & $\theta_2$  & 0.040 & -0.008 & -0.013 & -0.014 & -0.013 \\
      & $\theta_3$  & -0.057 & -0.020 & -0.014 & -0.015 & -0.018 \\
    \hline
    \multirow{3}{*}{90\% Coverage}
      & $\theta_1$ & 0.709 & 0.893 & 0.896 & 0.892 & 0.891 \\
      & $\theta_2$ & 0.772 & 0.906 & 0.909 & 0.904 & 0.904 \\
      & $\theta_3$ & 0.714 & 0.899 & 0.901 & 0.899 & 0.899 \\
    \hline
    \multirow{3}{*}{95\% Coverage}
      & $\theta_1$ & 0.804 & 0.949 & 0.953 & 0.949 & 0.949 \\
      & $\theta_2$ & 0.851 & 0.961 & 0.962 & 0.960 & 0.959 \\
      & $\theta_3$ & 0.806 & 0.956 & 0.962 & 0.956 & 0.953 \\
    \hline
    \multirow{3}{*}{99\% Coverage}
      & $\theta_1$  & 0.920 & 0.991 & 0.993 & 0.993 & 0.991 \\
      & $\theta_2$  & 0.944 & 0.994 & 0.996 & 0.994 & 0.994 \\
      & $\theta_3$  & 0.923 & 0.994 & 0.996 & 0.994 & 0.994 \\
    \hline
    \multicolumn{2}{c|}{Unscaled RMSE}
        & 0.097 & 0.350 & 0.469 & 0.370 & 0.331 \\
    \hline\hline
    \end{tabular}
    \caption{Finite sample performance ($N=500$) of the plug-in and the debiased estimates. The nuisances are estimated by KRR and NN.}
    \label{sec05 - table04}
\end{table}



\bibliographystyle{apalike}  
\bibliography{references}    

\pagebreak\newpage
\begin{appendix}

    \section{Supplementary materials}\label{appendix a}
    
    \subsection{Derivation of the functional equation \eqref{sec02 - eq03 - FE01}}\label{appendix a-1}
    For $\tau\in [0,1]$, we consider a mixture distribution $F_{\tau} = F_0 + \tau(F - F_0)$, where $F$ is an admissible alternative distribution. We define $\delta_\tau(Z) \equiv \delta[F_{\tau}](Z)$ such that:
    \begin{equation}
        \expectF{F_\tau}{ \zeta(Z)(T - \delta_\tau(Z)) } = 0 \quad\textnormal{for any admissible measurable test function }\zeta(\cdot)\in\cald.
    \end{equation}
    As in Section \ref{sec02}, we assume $\delta_0(Z)=Z$. To simplify the derivation, we consider paths satisfying $\delta_\tau(Z) = \delta_0(Z) + \tau\cdot d(Z)$, where $d(\cdot)\in\cald$. This exact linearity is an additional restriction on the paths considered here. Hadamard differentiability of $\delta[F](\cdot)$ at $F_0$ instead gives a first-order expansion along admissible distributional paths, although it does not, by itself, imply exact linearity.

    For the boundary calculation below, we consider continuously differentiable directions $d(\cdot)$ with bounded derivatives. We assume that $Z$ has a density $f_Z$ that is continuous at zero, with $f_Z(0)>0$. We implicitly assume that there is no strong group-seperation. We also assume that the admissible directions include the smooth, compactly supported functions used in the nonexistence proof below.

    The original moment contribution is $m(\theta, \classifier{A}_i(\delta_\tau), W_i) = Y_i - \theta\classifier{A}_i(\delta_\tau)$, where $\classifier{A}_i(\delta_\tau) \equiv \mathbf{1}\{\delta_\tau(Z_i) \geq 0\}$. We aim to find a bias-correction term $\phi(\theta, \delta_\tau, \beta_\tau, W_i)$ with an additional nuisance function $\beta(\theta,\cdot)$, so that $\psi(\cdot) = m(\cdot) + \phi(\cdot)$ satisfies:
    \begin{equation}
        \begin{cases}
            \expectF{F_{\tau}}{\phi(\theta, \delta_\tau, \beta_\tau, W_i)} = 0
                &\textnormal{for all }\tau\in[0,\bar{\tau})\textnormal{ and }\theta\in\Theta,\\
            \dt{\expectF{F_0}{\psi(\theta, \delta_\tau, \beta_\tau, W_i)}} = 0
                &\textnormal{for all }\theta\in\Theta,
        \end{cases}
    \end{equation}
    where $\bar{\tau}\in(0,1]$. Differentiating the first condition with respect to both the distribution and the nuisance functions gives:
    \begin{equation*}
        0 = \dt{\expectF{F_\tau}{\phi(\theta,\delta_0,\beta_0,W_i)}}
            + \dt{\expectF{F_0}{\phi(\theta,\delta_\tau,\beta_\tau,W_i)}}.
    \end{equation*}
    Combining this identity with the second condition yields:
    \begin{equation}
        \dt{\expectF{F_0}{m(\theta, \classifier{A}_{i}(\delta_\tau), W_i)}}
            = \dt{\expectF{F_{\tau}}{\phi(\theta, \delta_0, \beta_0, W_i)}}.
    \end{equation}

    We set $\phi(\theta,\delta,\beta,W_i) \equiv \beta(\theta,Z_i)(T_i - \delta(Z_i))$, where $\beta(\theta,\cdot)$ is square-integrable and is an admissible test function in the first-stage moment condition. Then $\expectF{F_{\tau}}{\phi(\theta, \delta_\tau, \beta_\tau, W_i)} = 0$, and we have:
    \begin{align*}
        \expectF{F_\tau}{\beta_0(\theta,Z_i)(T_i-\delta_0(Z_i))}
            &= \expectF{F_\tau}{\beta_0(\theta,Z_i)(\delta_\tau(Z_i)-\delta_0(Z_i))}\\
            &= \tau\expectF{F_\tau}{\beta_0(\theta,Z_i)d(Z_i)},
    \end{align*}
    and therefore:
    \begin{equation*}
        \dt{\expectF{F_{\tau}}{\phi(\theta, \delta_0, \beta_0, W_i)}}
            = \expectF{F_0}{\beta_0(\theta,Z)d(Z)}.
    \end{equation*}

    We next calculate $\dt{\expectF{F_0}{m(\theta, \classifier{A}_{i}(\delta_\tau), W_i)}}$. For sufficiently small $\tau$, $z + \tau d(z)$ is strictly increasing in $\tau$. Let $Z_\tau^*$ denote its unique zero, which is a classification boundary for a fixed path. Then $\classifier{A}_i(\delta_\tau)=\ind{}{Z_i\geq Z_\tau^*}$ and $Z_0^*=0$. Consequently,
    \begin{equation*}
        \dt{\expectF{F_0}{m(\theta,\classifier{A}_i(\delta_\tau),W_i)}}
            = -\theta\dt{\big(1-F_Z(Z_\tau^*)\big)}  = \theta f_Z(0)\dt{Z_\tau^*}.
    \end{equation*}
    Differentiating $Z_\tau^*+\tau d(Z_\tau^*)=0$ at $\tau=0$ gives $d(0) + \dt{Z_\tau^*}=0$, and it follows $\dt{Z_\tau^*}=-d(0)$. Therefore,
    \begin{equation*}
        \dt{\expectF{F_0}{m(\theta,\classifier{A}_i(\delta_\tau),W_i)}}
            = -\theta f_Z(0)d(0).
    \end{equation*}

    The same functional equation follows directly from Neyman orthogonality. Since $\condexpectF{F_0}{T_i-\delta_0(Z_i)}{Z_i}=0$, the derivative associated with $\beta_\tau$ vanishes, and:
    \begin{align*}
        \dt{\expectF{F_0}{\psi(\theta,\delta_\tau,\beta_\tau,W_i)}}
            &= \dt{\expectF{F_0}{m(\theta,\classifier{A}_i(\delta_\tau),W_i)}}
                -\expectF{F_0}{\beta_0(\theta,Z)d(Z)}\\
            &= -\theta f_Z(0)d(0)-\expectF{F_0}{\beta_0(\theta,Z)d(Z)}.
    \end{align*}
    Thus, $\beta_0(\theta,\cdot)$ must solve:
    \begin{equation}\label{appendix a-1 - FE01}
        \expectF{F_0}{\beta_0(\theta,Z)d(Z)}=-\theta f_Z(0)d(0),
    \end{equation}
    for every $\theta\in\Theta$ and every admissible direction $d(\cdot)\in\cald$.

    The next proposition shows that this functional equation has no square-integrable solution for nonzero $\theta$. At $\theta=0$, the choice $\beta_0(0,Z)=0$ gives a trivial bias correction. The implication of proposition is that a rich class of such directions can prevent the existence of a square-integrable solution to the functional equation, even when the admissible directions include the bounded and continuously differentiable functions, which are common in nonparametric estimations.

    \begin{proposition}\label{appendix a-1 - proposition}
        Suppose that $Z$ has a Lebesgue density $f_Z$ that is continuous at zero, with $f_Z(0)>0$, and that every bounded, continuously differentiable function on $\mathbb{R}$ is an admissible direction in $\cald$. Then, for each $\theta\in\Theta\backslash\{0\}$, there is no
        $\mathcal{L}^2(P)$-function $\beta_0(\theta,Z)$ solving \eqref{appendix a-1 - FE01} for every admissible direction $d(\cdot)\in\cald$.
    \end{proposition}

    \textbf{Proof of Proposition \ref{appendix a-1 - proposition}:}
    Fix any $\theta\in\Theta\backslash\{0\}$. To obtain a contradiction, suppose that there exists a square-integrable solution $\beta_0(\theta,\cdot)$. 
    Let $\overline{B}\equiv\norm{\beta_0(\theta,\cdot)}_{2}<\infty$. The Cauchy--Schwarz inequality gives:
    \begin{equation}\label{appendix a-1 - proposition - eq01}
        |\theta \cdot f_Z(0) d(0)| = \left|\expectF{F_0}{\beta_0(\theta,Z)d(Z)}\right| \leq \overline{B}\,\norm{d}_{2}.
    \end{equation}

    Take $d_n(Z)=\exp(-nZ^2)$. Since each $d_n$ is bounded and continuously differentiable on $\mathbb{R}$, it is an admissible direction
    in $\cald$. By construction, $d_n(0)=1$ for every $n\in\integer$ and $0\leq d_n(Z)^2\leq1$ for every $n\in\integer$. Since $Z$ has a Lebesgue density, $\prob{Z=0}=0$ so that $d_n(Z)^2=\exp(-2nZ^2) \to 0$ almost surely. Then, the dominated convergence theorem implies:
    \[ \norm{d_n}_{2}^2 = \expectF{F_0}{\exp(-2nZ^2)} \to 0. \]
    Substituting $d_n$ into
    \eqref{appendix a-1 - proposition - eq01} gives:
    \[ 0 < |\theta|f_Z(0) \leq\overline{B}\,\norm{d_n}_{2} \to0, \]
    which yields a contradiction. Thus, no such $\beta_0(\theta,\cdot)\in\mathcal{L}^2(P)$ exists.\qed

    \subsection{Derivation of the Neyman-orthogonal score with a smoothed indicator}\label{appendix a-2}

    The boundary evaluation in \eqref{appendix a-1 - FE01} can be approximated by a local average. For a direction $d$ continuous at zero,
    \[ \theta\int_{-b}^{b}f_Z(u)d(u)\frac{du}{2b}
            \to \theta f_Z(0)d(0)
            \quad\textnormal{as }b\to 0. \]
    For each fixed $b>0$, this local average can be represented by a square-integrable function.
    
    Motivated by this observation, we approximate $\ind{}{\delta(Z)\geq0}$ by $k(\delta(Z)/b)$, where $k$ is a smooth CDF kernel. We fix $b>0$ and assume that assumes that $k$ is twice differentiable, $k'$ is bounded, and that the relevant expectations are finite. The smoothed moment contribution and the score are:
    \[  m_{b,i}(\theta,\delta)=Y_i-\theta k\left(\frac{\delta(Z_i)}{b}\right)
        \quad\textnormal{and}\quad
        \psi_{b,i}(\theta,\delta,\beta)
            =m_{b,i}(\theta,\delta)+\beta(\theta,Z_i)(T_i-\delta(Z_i)). \]
    The derivative of the expected smoothed moment is:
    \[ \dt{\expectF{F_0}{m_{b,i}(\theta,\delta_\tau)}}
            =-\ExpectF{F_0}{\frac{\theta}{b}k'\left(\frac{\delta_0(Z)}{b}\right)d(Z)}. \]
    Therefore, Neyman-orthogonality requires:
    \[ \expectF{F_0}{\beta_{0,b}(\theta,Z)d(Z)}
            =-\ExpectF{F_0}{\frac{\theta}{b}k'\left(\frac{\delta_0(Z)}{b}\right)d(Z)}
        \quad\textnormal{for every admissible }d(\cdot)\in\cald. \]
    A solution is:
    \[ \beta_{0,b}(\theta,Z)  =-\frac{\theta}{b}k'\left(\frac{\delta_0(Z)}{b}\right). \]
    For each fixed $b>0$, boundedness of $k'$ gives
    $\norm{\beta_{0,b}(\theta,\cdot)}_{2}\leq |\theta|b^{-1}\sup_{u\in\mathbb{R}}|k'(u)|<\infty$.
    Substituting the corresponding correction into the score yields:
    \[ \psi_{b,i}(\theta,\delta)
            =Y_i-\theta k\left(\frac{\delta(Z_i)}{b}\right) -\frac{\theta}{b}k'\left(\frac{\delta(Z_i)}{b}\right)(T_i-\delta(Z_i)). \]
    This score satisfies Neyman orthogonality for each fixed $b>0$. Specifically,
    \begin{align*}
        \dt{\expectF{F_0}{\psi_{b,i}(\theta,\delta_\tau)}}
            =   &-\ExpectF{F_0}{\frac{\theta}{b}k'\left(\frac{\delta_0(Z)}{b}\right)d(Z)}&\\
                &+ \ExpectF{F_0}{\frac{\theta}{b}k'\left(\frac{\delta_0(Z)}{b}\right)d(Z)} -\ExpectF{F_0}{\frac{\theta}{b^2}k''\left(\frac{\delta_0(Z)}{b}\right)d(Z)(T-\delta_0(Z))}&\\
            =   &0,&
    \end{align*}
    where the second term is zero because $\condexpectF{F_0}{T-\delta_0(Z)}{Z}=0$.

    \subsection{Construction of Proxy Variable $\vect$ and Discriminant Function $\discfunc$}\label{appendix a-3}
    In this section, we provide two examples of how to construct the pairwise proxy variables $\vect$ and the corresponding discriminant function $\discfunc$. We mainly focus on two cases that practitioners frequently encounter.
    
    First, we suppose that a single proxy $S_i$ is observed and that its conditional expectation $\condexpect{S}{Z}$ generates $G_0$ groups. In particular, $\group_{0,i} = \eucvec{g}$ if $c_{g-1}<\condexpect{S}{Z}\leq c_{g}$, where the known cutoffs satisfy 
    \[- \infty = c_0 < c_1 < ... < c_{G_0-1} < c_{G_0} = \infty.\]
    We then obtain $\vect$ and $\discfunc$ by transforming $S_i$ and $\condexpect{S}{Z}$ as follows:
    \[ \delta_{0,gh}(Z) \equiv (g - h)\condexpect{S}{Z} +\sum_{j = g}^{h-1}c_{j} \quad\textnormal{and}\quad T_{i,gh} \equiv (g- h)S_{i} + \sum_{j=g}^{h-1} c_{j}, \]
    with $\delta_{0,hg}(Z) = -\delta_{0,gh}$ and $T_{i,hg} = -T_{i,gh}$ for $g < h$.

    Second, we consider a case in which two proxies $S_{1i}, S_{2i}\in \mathbb{R}$ are observed. Suppose the researcher knows that
    \[ \group_{0,i} = \begin{cases}
    \eucvec{1}&\textnormal{ if }\condexpect{S_{1i}}{Z_{i}} \leq 0, \condexpect{S_{2i}}{Z_{i}} \leq 0\\
    \eucvec{2}&\textnormal{ if }\condexpect{S_{1i}}{Z_{i}} > 0, \condexpect{S_{2i}}{Z_{i}} \leq 0\\
    \eucvec{3}&\textnormal{ if }\condexpect{S_{1i}}{Z_{i}} \leq 0, \condexpect{S_{2i}}{Z_{i}} > 0\\
    \eucvec{4}&\textnormal{ if }\condexpect{S_{1i}}{Z_{i}} > 0, \condexpect{S_{2i}}{Z_{i}} > 0.
    \end{cases} \]
    For $g=1,2,3,4$, define the following two functions:
    \[ u_1(g):=(g-1)-2\left\lfloor\frac{g-1}{2}\right\rfloor, \qquad u_2(g):=\left\lfloor\frac{g-1}{2}\right\rfloor, \]
    where $\lfloor{n}\rfloor$ denotes the largest integer not exceeding $n$. Using these functions, we construct the proxy variables and discriminant functions as
    \begin{align*}
        \delta_{0,gh}(Z_i)
            =   &\{u_1(g) - u_1(h)\} \cdot \condexpect{S_{1i}}{Z_i} + \{u_2(g) - u_2(h)\} \cdot\condexpect{S_{2i}}{Z_{i}},&\\
        T_{i,gh}
            =   &\{u_1(g) - u_1(h)\}\cdot S_{1i} + \{u_2(g) - u_2(h)\}\cdot S_{2i},&
    \end{align*}
    respectively. Lastly, let $\delta_{0,hg}(Z) = -\delta_{0,gh}$ and $T_{i,hg} = -T_{i,gh}$ for $g < h$.

\subsection{Simulation Setup}\label{appendix - a-4}
Two signals $Z_{i1}$ and $Z_{i2}$ are generated as follows: for $Z_{i1}^{a}\overset{\textnormal{iid}}{\sim}\caln(-1, 1)$ and $Z_{i1}^{b}\overset{\textnormal{iid}}{\sim}\caln( 1.5, 0.75^2 )$:
\[ Z_{i1} = 0.5 Z_{i1}^{a} + 0.5 Z_{i1}^{b}, \]
and for $Z_{i2}^{a}\overset{\textnormal{iid}}{\sim}\caln(-1.5, 1)$, $Z_{i2}^{b}\overset{\textnormal{iid}}{\sim}\caln(0.5, 0.75^2)$ and $Z_{i2}^{c}\overset{\textnormal{iid}}{\sim}\caln(1, 0.75^2)$:
\[ Z_{i2} = 0.2 Z_{i2}^{a} + 0.3 Z_{i2}^{b} + 0.5 Z_{i2}^{c}.  \]
The remaining signals are generated by $(Z_{i3}, Z_{i4})^\top \overset{\textnormal{iid}}{\sim}\caln\left(0, \begin{bmatrix}1 & 0.5\\ 0.5 & 1\end{bmatrix} \right)$. For $\varepsilon_{X,i} \overset{\textnormal{iid}}{\sim}\caln(0, 0.5^2)$, the covariate $X$ is generated by:
\[ X_{i} = Z_{i2} + \cos(Z_{i1}) + 0.1\cdot\big(Z_{i3} + Z_{i4}\big) + \varepsilon_{X,i}.\]
For a nonlinear function $\eta_0(Z_i) = Z_{i1} + Z_{i2} + \sin(Z_{i1} + Z_{i2})$ and a random variable $S_i = \eta_0(Z_i) + \varepsilon_{S,i}$ with $\varepsilon_{S,i} \overset{\textnormal{iid}}{\sim}\caln(0,1)$, the proxy variables and the discriminant functions are defined accordingly by:
\[ \begin{bmatrix} T_{i,12}\\ T_{i,13}\\ T_{i,23} \end{bmatrix} = \begin{bmatrix}0.5 - S_i\\ 0.5 - S_i  + 1.75 - S_i\\ 1.75 - S_i, \end{bmatrix} 
\quad\textnormal{and}\quad
\begin{bmatrix} \delta_{0,12}(Z) \\ \delta_{0,13}(Z) \\ \delta_{0,23}(Z)  \end{bmatrix} =
\begin{bmatrix} 0.5 - \eta_0(Z) \\ 0.5 - \eta_0(Z) + 1.75 - \eta_0(Z) \\1.75 - \eta_0(Z) \end{bmatrix}. \]
For $\varepsilon_{\xi,i} \overset{ \textnormal{iid} }{\sim} \caln\big( 0, \mathbb{I}_{13}\big)$, the random drift term $\xi_{i}$ is generated as follows:
\[ \widetilde{\xi}_{i} \equiv [ Z_{i}^\top, \sin(Z_{i})^\top, \cos(Z_i)^\top, Z_{i1}Z_{i2} ] \cdot \varepsilon_{\xi,i} \in \mathbb{R}\quad\textnormal{and}\quad \xi_{i} \equiv \big(\widetilde{\xi}_{i} - \expect{\widetilde{\xi}_{i}}\big)/{ \textnormal{SD}(\widetilde{\xi}_{i}) }. \]

\section{Proofs of Results}
\subsection{Proof of Proposition \ref{sec04 - proposition01}}\label{sec04 - proposition01 - proof}
For each $i\in[N]$, Assumption \ref{sec03 - assumption03}(c) gives:
\begin{align*}
    \prob{\widehat{\group}_i\ne\group_{0,i}}
        \leq    &\prob{\group_{0,i}\ne\classifier{A}_i(\discfunc_0)}+\prob{\classifier{A}_i(\discfunc_0)\ne\classifier{A}_i(\widehat{\discfunc})}&\\
        \leq    &O(\lambda_N)+\prob{\classifier{A}_i(\discfunc_0)\ne\classifier{A}_i(\widehat{\discfunc})},&
\end{align*}

On the event $\norm{\widehat{\discfunc}-\discfunc_0}_\infty\leq\rho_N$, a change in the classifier requires a change in at least one pairwise comparison, meaning:
\[ \ind{}{ \widehat{\group}_i\ne\classifier{A}_i(\discfunc_0)}
    \leq    \sum_{g=1}^{G_0-1} \sum_{h=g+1}^{G_0} \ind{}{|\delta_{0,gh}(Z_i)| \leq \rho_N}, \]
and therefore:
\[ \ind{}{ \widehat{\group}_i\ne\classifier{A}_i(\discfunc_0)}
    \leq    \sum_{g=1}^{G_0-1} \sum_{h=g+1}^{G_0} \ind{}{|\delta_{0,gh}(Z_i)| \leq \rho_N} + \ind{}{\norm{\widehat{\discfunc}-\discfunc_0}_\infty>\rho_N}. \]
Taking expectations and applying the margin condition in Assumption \ref{sec03 - assumption03}(e), for sufficiently large $N>0$, we obtain:
\begin{align*}
    \prob{\classifier{A}_i(\discfunc_0)\ne\classifier{A}_i(\widehat{\discfunc})}
        \leq    &\sum_{g=1}^{G_0-1}\sum_{h=g+1}^{G_0}\prob{|\delta_{0,gh}(Z_i)|\leq\rho_N}+\prob{\norm{\widehat{\discfunc}-\discfunc_0}_\infty>\rho_N}&\\
        \leq    &\binom{G_0}{2}c_0\rho_N+\prob{\norm{\widehat{\discfunc}-\discfunc_0}_\infty>\rho_N}.&
\end{align*}
Combining these bounds uniformly over $i\in[N]$ gives:
\[
    \underset{i\in[N]}{\textnormal{max}}\prob{\widehat{\group}_i\ne\group_{0,i}}\leq O(\lambda_N)+\binom{G_0}{2}c_0\rho_N+\prob{\norm{\widehat{\discfunc}-\discfunc_0}_\infty>\rho_N}.
\]
\qed

\subsection{Proof of Proposition \ref{sec04 - proposition02}}\label{sec04 - proposition02 - proof}

For sufficiently large $N>0$, the event $\norm{\widehat{\discfunc}-\discfunc_0}_\infty\leq\rho_N$ has probability close to one by Assumption \ref{sec03 - assumption03}(d). On this event, it follows:
\[ \ind{}{ \widehat{\group}_i \ne \classifier{A}_i(\discfunc_0)} \leq \sum_{g=1}^{G_0-1}\sum_{h=g+1}^{G_0}\ind{}{|\delta_{0,gh}(Z_i)|\leq\rho_N}.
\]
Since both classifiers take values in $\typespace$, Assumption \ref{sec03 - assumption02}(b) gives:
\begin{align*}
    \sup_{\theta\in\Theta}\Norm{\frac{1}{N}\sum_{i=1}^{N}\left\{m(\theta,\widehat{\group}_i,W_i)-m(\theta,\classifier{A}_i(\discfunc_0),W_i)\right\}}
        \leq    &\frac{2}{N}\sum_{i=1}^{N}\overline{M}(W_i)\sum_{g=1}^{G_0-1}\sum_{h=g+1}^{G_0}\ind{}{|\delta_{0,gh}(Z_i)|\leq\rho_N}.&
\end{align*}
By the Cauchy-Schwarz inequality and Assumption \ref{sec03 - assumption03}(e):
\begin{align*}
    \Expect{\frac{2}{N}\sum_{i=1}^{N}\overline{M}(W_i)\sum_{g=1}^{G_0-1}\sum_{h=g+1}^{G_0}\ind{}{|\delta_{0,gh}(Z_i)|\leq\rho_N}}
        \leq    &2\norm{\overline{M}}_2\sum_{g=1}^{G_0-1}\sum_{h=g+1}^{G_0}\prob{|\delta_{0,gh}(Z_i)|\leq\rho_N}^{1/2}&\\
        \leq    &G_0(G_0-1)\norm{\overline{M}}_2\sqrt{c_0\rho_N}.&
\end{align*}
The Markov inequality implies:
\begin{equation}\label{sec04 - proposition02 - proof - eq01}
    \sup_{\theta\in\Theta}\Norm{\frac{1}{N}\sum_{i=1}^{N}\left\{m(\theta,\widehat{\group}_i,W_i)-m(\theta,\classifier{A}_i(\discfunc_0),W_i)\right\}}=O_P(\rho_N^{1/2}).
\end{equation}
For the population difference, Assumption \ref{sec03 - assumption03}(c) and the Cauchy-Schwarz inequality give:
\begin{equation}\label{sec04 - proposition02 - proof - eq02}
    \begin{aligned}
        \sup_{\theta\in\Theta}\Norm{\expect{m(\theta,\classifier{A}_i(\discfunc_0),W_i)-m(\theta,\group_{0,i},W_i)}}
            \leq    &2\expect{\overline{M}(W_i)\ind{}{\classifier{A}_i(\discfunc_0)\ne\group_{0,i}}}&\\
            \leq    &2\norm{\overline{M}}_2\prob{\classifier{A}_i(\discfunc_0)\ne\group_{0,i}}^{1/2}&\\
            =       &O(\lambda_N^{1/2}).&
    \end{aligned}
\end{equation}
Assumption \ref{sec03 - assumption02}(d) imply the uniform bound:
\begin{equation}\label{sec04 - proposition02 - proof - eq03}
    \sup_{\theta\in\Theta} \Norm{\frac{1}{N}\sum_{i=1}^{N} m(\theta,\classifier{A}_i(\discfunc_0),W_i)-\expect{m(\theta,\classifier{A}_i(\discfunc_0),W_i)}} = O_P(N^{-1/2}).    
\end{equation}
Combining \eqref{sec04 - proposition02 - proof - eq01}, \eqref{sec04 - proposition02 - proof - eq02}, and \eqref{sec04 - proposition02 - proof - eq03} yields:
\begin{equation}\label{sec04 - proposition02 - proof - eq04}
    \sup_{\theta\in\Theta}\Norm{\frac{1}{N}\sum_{i=1}^{N}m(\theta,\widehat{\group}_i,W_i)-\expect{m(\theta,\group_{0,i},W_i)}}=O_P(\lambda_N^{1/2}+\rho_N^{1/2}+N^{-1/2}).
\end{equation}

By definition of $\plugin$, Assumption \ref{sec03 - assumption02}(a), and \eqref{sec04 - proposition02 - proof - eq04},
\begin{align*}
    \Norm{\expect{m(\plugin,\group_{0,i},W_i)}}
        \leq    &\Norm{\expect{m(\plugin,\group_{0,i},W_i)}-\frac{1}{N}\sum_{i=1}^{N}m(\plugin,\widehat{\group}_i,W_i)}+\Norm{\frac{1}{N}\sum_{i=1}^{N}m(\theta_0,\widehat{\group}_i,W_i)}&\\
        \leq    &2\sup_{\theta\in\Theta}\Norm{\frac{1}{N}\sum_{i=1}^{N}m(\theta,\widehat{\group}_i,W_i)-\expect{m(\theta,\group_{0,i},W_i)}}&\\
        =       &O_P(\lambda_N^{1/2}+\rho_N^{1/2}+N^{-1/2}).&
\end{align*}
Since the right-hand side is $o_P(1)$, the identification condition in Assumption \ref{sec03 - assumption02}(a) implies $\plugin\overset{P}{\to}\theta_0$.

For $\theta$ sufficiently close to $\theta_0$, Taylor's theorem and the second-order derivative bound in Assumption \ref{sec03 - assumption02}(c) give:
\[  \Norm{\expect{m(\theta,\group_{0,i},W_i)} - \expect{\nabla_\theta m(\theta_0,\group_{0,i},W_i)}(\theta-\theta_0)} 
        \leq    \frac{G_0\overline{M}^{(2)}}{2}\norm{\theta-\theta_0}^2. \]
The singular-value condition of Assumption \ref{sec03 - assumption02}(c) therefore implies:
\[  \Norm{\expect{m( \theta,\group_{0,i},W_i)}}
        \geq    \underline{\sigma} \norm{\theta-\theta_0} - \frac{G_0\overline{M}^{(2)}}{2} \norm{\theta-\theta_0}^2. \]
By consistency, with probability approaching one:
\[  \Norm{\expect{m(\plugin,\group_{0,i},W_i)}}
        \geq    \underline{\sigma} \norm{\plugin-\theta_0} + o_P(1). \]
It follows that:
\begin{align*}
    \norm{\plugin-\theta_0}
        \leq    &\frac{4}{\underline{\sigma}}\sup_{\theta\in\Theta}\Norm{\frac{1}{N}\sum_{i=1}^{N}m(\theta,\widehat{\group}_i,W_i)-\expect{m(\theta,\group_{0,i},W_i)}}&\\
        =       &O_P(\lambda_N^{1/2}+\rho_N^{1/2}+N^{-1/2}),&
\end{align*}
where the inequality holds with probability approaching one. \qed

\subsection{Proof of Proposition \ref{sec04 - proposition03}}\label{sec04 - proposition03proof}

Consider the following functional equation:
\begin{equation}\label{sec04 - proposition03proof - eq00}
    \dt{ \expectF{F_0}{ m(\theta, \smth{A}_i(\discfunc_\tau), W_i) } } = \dt{  \expectF{F_\tau}{ \phi(\theta, \discfunc_0, \vecmu_0, W_i) }}.
\end{equation}
We note that a function $\phi_N(\cdot)$, solving \eqref{sec04 - proposition03proof - eq00}, induces the score $\psi_N = m + \phi_N$ obeying the conditions \eqref{sec04 - neyman} \citep{dml_chernozhukov_et_al2022,dml_ahrens_et_al2026}.

We first calculate $\dt{ \expectF{F_0}{ m(\theta, \smth{A}_i(\discfunc_\tau), W_i) } }$. By construction, $m(\theta, \smth{A}_i(\discfunc_\tau), W_i)$ is written as:
\[ m(\theta, \smth{A}_i(\discfunc_\tau), W_i) = M_{G_0}(\theta,W_i) + \sum_{g=1}^{G_0-1}\left\{ M_g(\theta,W_i) - M_{G_0}(\theta,W_i) \right\} \Ker{ \discfunc_{\tau,g}(Z_i) }, \]
and taking the derivative with respect to $\tau\in [0,\overline{\tau})$ gives:
\begin{align*}
\dt{ \expectF{F_0}{ m(\theta, \smth{A}_i(\discfunc_\tau), W_i) } }
    =   &\sum_{g=1}^{G_0-1}\dt{\ExpectF{F_0}{\left\{ M_g(\theta,W_i) - M_{G_0}(\theta,W_i) \right\}\frac{1}{b_N}\nabla\KER{ \discfunc_{0,g}(Z_i)}^\top \discfunc_{\tau,g}(Z)} }&\\
    =   &\sum_{g=1}^{G_0-1}\dt{\ExpectF{F_0}{ \underset{ \equiv \beta_{N,g}(\theta,Z_i)}{  \underline{ \condexpectF{F_0}{M_g(\theta,W_i) - M_{G_0}(\theta,W_i)}{Z_i}\frac{1}{b_{N}}\nabla\KER{ \discfunc_{0,g}(Z_i)}^\top}} \discfunc_{\tau,g}(Z)} }.&
\end{align*}
By Assumption \ref{sec03 - assumption02}(d), Assumption \ref{sec04 - assumption04}(d) and the Cauchy-Schwarz inequality, $\beta_{N,g}(\cdot)$ is $\mathcal{L}^2(P)$-integrable with a fixed $b_N>0$. Therefore, we have:
\begin{equation}\label{sec04 - proposition03proof - eq01}
\dt{ \expectF{F_0}{ m(\theta, \smth{A}_i(\discfunc_\tau), W_i) } }
    =   \sum_{g=1}^{G_0-1}\dt{\expectF{F_0}{ \beta_{N,g}(\theta,Z_i) \discfunc_{\tau,g}(Z)} }.    
\end{equation}

For $\tau\in [0,1]$, the nuisance $\discfunc_{\tau}(Z)$ is a solution to:
\begin{equation}\label{sec04 - def04 - local first-stage}
    \expectF{F_\tau}{ \zeta(Z_i)( \pmb{T}_{i,g} - \discfunc_{\tau,g}(Z_i)) } = 0 \quad\textnormal{for any measurable }\zeta(\cdot)\in\cald.   
\end{equation}
So, if we choose $\zeta(\cdot) = \{\beta_{N,g}(\cdot) : g\in[G_{0} - 1]\}$ as the instrument function, then the first-stage \eqref{sec04 - def04 - local first-stage} gives:
\[     \sum_{g=1}^{G_0-1}\expectF{F_\tau}{ \beta_{N,g}(\theta,Z_i)(\vect_{i,g} - \discfunc_{\tau,g}(Z_i)} 
        =   0\quad\textnormal{ for every }\tau \in[0,\overline{\tau}), \overline{\tau}\in[0,1]. \]
It follows, by construction:
\begin{align*}
    \frac{d}{dt}\big( 0 \big)
        =   &\dt{\expectF{F_\tau}{ \beta_{N,g}(\theta,Z_i)(\vect_{i,g} - \discfunc_{\tau,g}(Z_i)) }} &\\
        =   &\dt{\expectF{F_\tau}{ \beta_{N,g}(\theta,Z_i)(\vect_{i,g} - \discfunc_{0,g}(Z_i)) }} - \dt{\expectF{F_0}{ \beta_{N,g}(\theta,Z_i)\discfunc_{\tau,g}(Z_i) }}, &
\end{align*}
meaning
\begin{equation}\label{sec04 - proposition03proof - eq02}
    \sum_{g=1}^{G_0-1}\dt{\expectF{F_0}{ \beta_{N,g}(\theta,Z_i)\discfunc_{\tau,g}(Z_i) }} = \sum_{g=1}^{G_0-1}\dt{\expectF{F_\tau}{ \beta_{N,g}(\theta,Z_i)(\vect_{i,g} - \discfunc_{0,g}(Z_i)) }}.
\end{equation}
By combining \eqref{sec04 - proposition03proof - eq01} and \eqref{sec04 - proposition03proof - eq02}, we obtain:
\begin{equation}
    \dt{ \expectF{F_0}{ m(\theta, \smth{A}_i(\discfunc_\tau), W_i) } } = \frac{d}{d\tau} \mathbb{E}_{F_\tau}\left[ \sum_{g=1}^{G_0-1}\beta_{N,g}(\theta,Z_i)(\vect_{i,g} - \discfunc_{0,g}(Z_i)) \right] \Bigg|_{\tau = 0}.
\end{equation}
In words, $\phi_{N}(\theta, \discfunc_{0}, \vecmu_{0}, W_i) = \sum_{g=1}^{G_0-1}\beta_{N,g}(\theta,Z_i)(\vect_{i,g} - \discfunc_{0,g}(Z_i))$ satisfies the equation \eqref{sec04 - proposition03proof - eq00}. \qed

\subsection{Proof of Proposition \ref{sec04 - proposition04}}
\label{sec04 - proposition04 - proof}

Fix $\theta\in\Theta_0$. To simplify notation, for every \(g\in[G_0-1]\), we write $\pmb D_g\equiv\discfunc_{0,g}(Z_i)$, $\iota_g\equiv \ind{}{\pmb D_g\geq\pmb 0},$ where the vector inequality is understood componentwise. Also, define
\[ R_g^{(0)}(\theta,\pmb{t})\equiv\expect{\{M_g(\theta,W_i)-M_{G_0}(\theta,W_i)\}\ind{}{\pmb D_g\geq\pmb t}}, \]
\[ R_g^{(1)}(\theta,\pmb{t})\equiv\expect{\{\nabla_\theta M_g(\theta,W_i)-\nabla_\theta M_{G_0}(\theta,W_i)\}\ind{}{\pmb D_g\geq\pmb t}}. \]
Then
\[ R_g^{(0)}(\theta,\pmb{t}) =  \int_{t_1}^{\infty}\cdots \int_{t_{G_0-1}}^{\infty} r_g^{(0)}(\theta,\pmb{s}) \,ds_1\cdots ds_{G_0-1}.  \]
\[ R_g^{(1)}(\theta,\pmb{t}) =  \int_{t_1}^{\infty}\cdots \int_{t_{G_0-1}}^{\infty} r_g^{(1)}(\theta,\pmb{s}) \,ds_1\cdots ds_{G_0-1}.  \]

We fix $j\in\{0,1\}$. By Assumption \ref{sec04 - assumption05}, $r_g^{(j)}(\theta, \cdot)$ is continuously differentiable on the open neighborhood $\mathcal{U}_{g}$. Because $\pmb{D}_g$ has compact support, $r_g^{(j)}(\theta,\cdot)$ and its first-order partial derivatives are bounded and integrable.

For every \(l\in[G_0 - 1]\), differentiation under the integral gives
\[ \frac{\partial}{\partial t_l}R_g^{(j)}(\theta, \pmb{t}) = - \int_{\pmb{s}_{-l}\geq\pmb{t}_{-l}} r_g^{(j)}(\theta, t_l,\pmb{s}_{-l}) \,d\pmb{s}_{-l}. \]

Differentiating once more,
\[ \frac{\partial^2}{\partial t_l^2}R_g^{(j)}(\theta, \pmb{t}) = - \int_{\pmb{s}_{-l}\geq\pmb{t}_{-l}} \frac{\partial}{\partial t_l} r_g^{(j)}(\theta, t_l,\pmb{s}_{-l})\,d\pmb{s}_{-l}, \]
\[ \frac{\partial^2}{\partial t_l\partial t_{l'}} R_g^{(j)}(\theta, \pmb{t}) = \int_{\pmb{s}_{-(l,l')} \geq \pmb{t}_{-(l,l')}} r_g^{(j)}(\theta, t_l, t_{l'}, \pmb{s}_{-(l,l')}) \,d\pmb{s}_{-(l,l')} \quad\textnormal{ for } l \ne l'. \]
Because $r_g^{(j)}(\theta,\cdot)$ and its first-order partial derivatives are continuous, bounded, and integrable, the dominated convergence theorem implies that $R_g^{(j)}(\theta,\cdot)$ is twice continuously differentiable on a neighborhood of zero. Since \(G_0\) is fixed, there exist positive constants $C^{(j)} < \infty$ such that:
\[ \sup_{\theta\in\Theta_0}\sup_{g\in[G_0-1]} \sup_{\pmb{t}\in \mathcal{U}_{g}} \norm{\nabla_{\pmb{t}}^2 R_g^{(j)}(\theta, \pmb{t})} \leq C^{(j)}.\]

By Assumption \ref{sec04 - assumption04}, there exist random vectors $\pmb{V}_g = (V_{g1}, ...,V_{g,G_{0}-1})^\top$ for $g\in[G_0 - 1]$, independent of \((W_i,Z_i,T_i)\), whose components are mutually independent and have common CDF $k(\cdot)$. Then $\pmb{V}_g$ has joint CDF $\pmb{K}(\pmb{t}) = \prod_{g=1}^{G_0 - 1}k(t_{g})$, and $\pmb{V}_g\in[-S,S]^{G_0 - 1}$ almost surely by Assumption \ref{sec04 - assumption04}(c). Moreover, the symmetry of $k'(t)$ by Assumption \ref{sec04 - assumption04}(c) gives $\expect{\pmb{V}_g}=0$ and bounded support gives $\expect{\norm{\pmb{V}_g}^2} \leq (G_0 - 1)S^2<\infty$.

Conditional on $\pmb{D}_g$:
\[ \KER{ \pmb{D}_g } = \condProb{ \pmb{V}_g\leq\frac{\pmb{D}_g}{b_N} }{ \pmb{D}_g } = \condProb{ b_N \pmb{V}_g \leq \pmb{D}_g }{ \pmb{D}_g }.\]
By the law of iterated expectations and Fubini's theorem,
\begin{align*}
    \expect{ m(\theta,\classifier{A}_i(\discfunc_0),W_i) }
        =   & \sum_{g=1}^{G_0-1} \expect{ \{ M_g(\theta,W_i)-M_{G_0}(\theta,W_i) \}\cdot\{\Ker{ \pmb{D}_g } -\iota_g\} }&\\
        =   & \sum_{g=1}^{G_0-1} \int_{[-S,S]^{G_0-1}} \left\{ R_g^{(0)}(\theta, b_N\pmb{v})-R_g^{(0)}(\theta, 0) \right\} \pmb{K}(d\pmb{v}).
\end{align*}

Because \(b_N\to0\), for $\varepsilon>0$, there exists \(N_0<\infty\) such that $N\geq N_0$ implies $b_N S <\varepsilon$, meaning,
\[ b_N\pmb{v} \in \mathcal{U}_{g}\quad\textnormal{for every }\pmb{v}\in[-S,S]^{G_0 - 1}, \]
and the entire line segment joining 0 and $b_N\pmb{v}$ is contained in $\mathcal{U}_{g}$. Therefore, Taylor's theorem gives, for each $\pmb{v}\in[-S,S]^{G_0-1}$,
\[ R_g^{(j)}(\theta, b_N \pmb{v}) - R_g^{(j)}(\theta, 0) = b_N\nabla_{\pmb{t}}R_g(0 )^\top\pmb{v} + \frac{b_N^2}{2} \pmb{v}^\top \nabla_{\pmb{t}}^2R_g(\pmb{t}_{g,N}^{\dagger}) \pmb{v},\]
where $\pmb{t}_{g,N}^{\dagger} \in \mathcal{U}_{g}$. Integrating with respect to the distribution of \(\pmb V_g\),
\begin{align*}
\int_{[-S,S]^{G_0 - 1}} \left\{ R_g^{(j)}(\theta, b_N\pmb{v})-R_g^{(j)}(\theta,0) \right\} \pmb{K}(d\pmb{v})\
    =   & \int_{[-S,S]^{G_0-1}} \Big\{ b_N\nabla_{\pmb{t}}R_g^{(j)}(\theta,0)^\top \pmb{v} + \frac{b_N^2}{2} \pmb{v}^\top \nabla_{\pmb{t}}^2R_g^{(j)}(\theta, \pmb{t}_{g,N}^{\dagger}) \pmb{v}  \Big\}\pmb{K}(d \pmb{v})&\\
\end{align*}
Because $\expect{\pmb{V}_g}=0$, the first-order term vanishes. Thus,
\begin{align*}
\Norm{ \int_{[-S,S]^{G_0 -1}} \left\{ R_g^{(j)}(\theta, b_N\pmb{v})-R_g^{(j)}(\theta,0) \right\} \pmb{K}(d\pmb{v})  }
    \leq    & \frac{b_N^2}{2} \sup_{\pmb{t}\in\mathcal{U}_g} \norm{\nabla_{\pmb{t}}^2R_g^{(j)}(\theta, \pmb{t})} \int_{[-S,S]^{G_0 - 1}} \norm{\pmb{v}}^2\pmb{K}(d\pmb{v})&\\
    \lesssim    & \frac{b_N^2}{2} \expect{\norm{\pmb{V}_g}^2} = O(b_N^2),&
\end{align*}
uniformly over \(g\in[G_0-1]\). Since \(G_0\) is fixed, we obtained result (i).

We now verify results (ii) and (iii). By the affine structure of the moment function,
\begin{align*}
    &\expect{\nabla_\theta m(\theta,\smth{A}_i(\discfunc_0),W_i)}-\expect{\nabla_\theta m(\theta,\classifier{A}_i(\discfunc_0),W_i)}\\
    &\qquad=\sum_{g=1}^{G_0-1}\expect{\left\{\nabla_\theta M_g(\theta,W_i)-\nabla_\theta M_{G_0}(\theta,W_i)\right\}\left\{\Ker{\pmb D_g}-\iota_g\right\}}\\
    &\qquad=\sum_{g=1}^{G_0-1}\int_{[-S,S]^{G_0-1}}\left\{R_g^{(1)}(\theta,b_N\pmb v)-R_g^{(1)}(\theta,\pmb 0)\right\}\pmb K(d\pmb v),
\end{align*}
where the second equality follows from the law of iterated expectations and Fubini's theorem. Applying the preceding Taylor expansion with $j=1$ gives
\begin{align*}
    &\sup_{\theta\in\Theta_0}\Norm{\expect{\nabla_\theta m(\theta,\smth{A}_i(\discfunc_0),W_i)}-\expect{\nabla_\theta m(\theta,\classifier{A}_i(\discfunc_0),W_i)}}_{\mathrm{op}}\\
    &\qquad\leq\frac{b_N^2}{2}\sum_{g=1}^{G_0-1}\sup_{\theta\in\Theta_0}\sup_{\pmb t\in\mathcal U_g}\norm{\nabla_{\pmb t}^2R_g^{(1)}(\theta,\pmb t)}_{\mathrm{op}}\expect{\norm{\pmb V_g}^2}=O(b_N^2),
\end{align*}
where $\norm{\pmb{A}}_{\mathrm{op}}$ denotes the Euclidean operator norm defined by $\norm{\pmb{A}}_{\mathrm{op}}\equiv \underset{ \|\pmb{v}\| \leq 1}{\textnormal{sup}}\, \norm{\pmb{A}\pmb{v}}$, and where
\[ \norm{\nabla_{\pmb t}^2R_g^{(1)}(\theta,\pmb t)}_{\mathrm{op}}\equiv\sup_{ \|\pmb{a}\|\leq1,\|\pmb{c}\| \leq1}\Norm{\sum_{l=1}^{G_0-1}\sum_{l'=1}^{G_0-1}a_lc_{l'}\frac{\partial^2R_g^{(1)}(\theta,\pmb t)}{\partial t_l\partial t_{l'}}}_{\mathrm{op}}. \]

Moreover, by construction,
\[ \expect{\phi_N(\theta,\discfunc_0,\vecmu_0,W_i)}=0 \quad\textnormal{for every }\theta\in\Theta_0, \]
because $\expect{\vect_{i,g}-\discfunc_{0,g}(Z_i)\mid Z_i}=0$. Differentiating this identity with respect to $\theta$ and interchanging differentiation and expectation under Assumption \ref{sec03 - assumption02}(c), we obtain
\[ \expect{\nabla_\theta\phi_N(\theta,\discfunc_0,\vecmu_0,W_i)}=0. \]
Consequently,
\begin{align*}
    \expect{\nabla_\theta\psi_N(\theta_0,\discfunc_0,\vecmu_0,W_i)}
    &=\expect{\nabla_\theta m(\theta_0,\smth{A}_i(\discfunc_0),W_i)}\\
    &=\expect{\nabla_\theta m(\theta_0,\classifier{A}_i(\discfunc_0),W_i)}+O(b_N^2),
\end{align*}
which proves result (ii).

Finally, using the inequality $\minsingular{ \pmb{A} + \pmb{B} } \geq \minsingular{\pmb{A}} - \norm{\pmb{B}}$ for two generic conformable matrices $\pmb{A}, \pmb{B}$, for every $\theta\in\Theta_0$, Assumption \ref{sec03 - assumption02}(c) implies the following:
\begin{align*}
    \minsingular{\expect{\nabla_\theta\psi_N(\theta_0,\discfunc_0,\vecmu_0,W_i)}}
    &\geq\minsingular{\expect{\nabla_\theta m(\theta_0,\classifier{A}_i(\discfunc_0),W_i)}}\\
    &\quad-\Norm{\expect{\nabla_\theta m(\theta_0,\smth{A}_i(\discfunc_0),W_i)}-\expect{\nabla_\theta m(\theta_0,\classifier{A}_i(\discfunc_0),W_i)}}_{\mathrm{op}}\\
    &\geq\underline{\sigma}-O(b_N^2).
\end{align*}
This proves result (iii) and completes the proof. \qed

\subsection{Proof of Theorem \ref{sec04 - theorem01}}\label{sec04 - theorem01 - proof}

\newcommand{\foldavg }[1]{ \frac{1}{L}\sum_{\ell=1}^{L}{#1} }
\newcommand{\avginfold}[1]{ \frac{1}{ | I_{\ell} | }\sum_{i\in I_{\ell} }{#1} }
\newcommand{ \foldnorm }[1]{ \| {#1} \|_{-\ell, 2} }
\newcommand{ \foldNorm }[1]{ \left\| {#1} \right\|_{-\ell, 2} }
For simplicity, let $\psi_{0,i} \equiv \psi(\theta_0, \discfunc_0, \vecmu_0, W_i)$ and $\widehat{\psi}_{\ell,i} \equiv \psi(\theta_0, \widehat{\discfunc}_\ell, \widehat{\vecmu}_\ell, W_i)$ and $\mathcal{W}_{-\ell}$ denote a $\sigma$-field induced by $\{W_i : i\in I_{-\ell}\}$ for each $\ell\in [L]$. Throughout this proof, $\norm{\cdot}_{-\ell,p}$ denotes the conditional $\mathcal{L}^{p}$-norm given $\mathcal{W}_{-\ell}$, defined by $\norm{X}_{-\ell,p} \equiv \big\{\condexpect{\norm{X}^{p}}{\mathcal{W}_{-\ell}} \big\}^{1/p}$.

By adding and subtracting $\psi_{0,i}$, we have:
\begin{align}
    \sqrt{Nb_{N}}\foldavg{ \avginfold{ \widehat{\psi}_{\ell,i} } } 
        =   \label{sec04 - theorem01 - proof - step0 - eq01-1}
            & \sqrt{Nb_{N}}\foldavg{ \avginfold{ \psi_{0,i} }}&\\
            \label{sec04 - theorem01 - proof - step0 - eq01-2}
            &+ \sqrt{Nb_{N}}\foldavg{  \condexpect{\widehat{\psi}_{\ell,i} - \psi_{0,i}}{\mathcal{W}_{-\ell}}  }&\\
            \label{sec04 - theorem01 - proof - step0 - eq01-3}
            &+ \sqrt{Nb_{N}}\foldavg{ \avginfold{ \big\{ \widehat{\psi}_{\ell,i} - \psi_{0,i} - \condexpect{\widehat{\psi}_{\ell,i} - \psi_{0,i}}{\mathcal{W}_{-\ell}} \big\} } }.&
\end{align}

In Step 1, we show $\norm{\psi_{0,i}}_2 = O( b_{N}^{-1/2} )$, meaning the appropriate scaling is $\sqrt{Nb_N}$.

In Step 2, we apply the Lindeberg-Feller central limit theorem and establish:
\[ \sqrt{Nb_{N}}\times \foldavg{ \avginfold{ \psi_{0,i} }} = \frac{\sqrt{b_{N}}}{\sqrt{N}}\sum_{i=1}^{N}\psi_{0,i} \overset{d}{\to} \mathcal{N}\big(0,\Sigma_0\big), \]
and then link the asymptotic distribution of $\sqrt{Nb_N}(\widehat{\theta}_N - \theta_0)$ to that of $\sqrt{Nb_{N}}\frac{1}{N}\sum_{i}\psi_{0,i}$.

Then we show that the remainder terms \eqref{sec04 - theorem01 - proof - step0 - eq01-2} (in Step 3) and \eqref{sec04 - theorem01 - proof - step0 - eq01-3} (in Step 4) vanish with probability approaching one.

\newcommand{\uniker}[1]{ k\left(  \frac{#1}{b_N}\right) }
\newcommand{\duniker}[1]{ k'\left(  \frac{#1}{b_N}\right) }
\newcommand{ \deldel }{ {\Delta\discfunc_{\ell}} }
\newcommand{ \delmu }{ {\Delta\vecmu_{\ell}} }

Here, we introduce useful lemmas induced by the restrictions on the kernel. For notational simplicity, let $\deldel \equiv \norm{\widehat{\discfunc}_{\ell}(\cdot) - \discfunc_{0}(\cdot)}_{\infty}$ and $\delmu \equiv \norm{\widehat{\vecmu}_{\ell}(\cdot) - \vecmu_{0}(\cdot)}_{\infty}$. For $g\ne g'\in [G_0]$ and $\ell\in[L]$, we define two indicators $\iota_{0,gg'}$ and $\iota_{\ell,gg'}$ as:
\begin{equation}\label{sec04 - theorem01 - proof - step1 - iota}
    \begin{aligned}
        \iota_{0,gg'}
            \equiv  &\mathbf{1}\left\{
            {|\delta_{0,gg'}(Z)| \leq     b_N S \textnormal{ and }}\,
            {\delta_{0,gh}(Z)    \geq    -b_N S \textnormal{ for }h\ne g,g' }  \right\} &\\
        \iota_{\ell,gg'}
            \equiv  &\mathbf{1}\left\{
            {|\delta_{0,gg'}(Z)| \leq     b_N S + \deldel\,\textnormal{ and }}\,
            {\delta_{0,gh}(Z)    \geq    -b_N S - \deldel \textnormal{ for }h\ne g,g' }  \right\}.&
    \end{aligned}
\end{equation}

\begin{lemma}\label{sec04 - theorem01 - proof - lemma01}
Under Assumptions \ref{sec03 - assumption03} and \ref{sec04 - assumption04}, for each $g\ne g'\in[G_0 - 1]$ and $\ell\in [L]$:
\begin{itemize}
    \item[(i)] $\iota_{0,gg'} \leq \iota_{\ell,gg'}$.
    \item[(ii)] $\Ker{\discfunc_{0,g}} \ne 0$ if and only if $\delta_{0,gh}(Z)>-b_NS$ for every $h\ne g$. Likewise, $\pmb{K}(\widehat{\discfunc}_{\ell,g}/b_N) \ne 0$ if and only if $\widehat{\delta}_{\ell,gh}(Z)>-b_NS$ for every $h\ne g$.
    \item[(iii)] $\{\nabla\Ker{\discfunc_{0,g}}\}_{g'} \ne 0$ implies $\iota_{0,gg'} = 1$. Likewise, $\{\nabla \pmb{K}(\widehat{\discfunc}_{\ell,g}/b_N) \}_{g'} \ne 0$ implies $\iota_{\ell,gg'} = 1$.
    \item[(iv)] $\foldnorm{ \iota_{0,gg'} } \leq \sqrt{b_N c_0 S}$ and $\foldnorm{ \iota_{\ell,gg'} } \leq \sqrt{c_0 \cdot ( b_N S + \deldel) }$    
\end{itemize}
\end{lemma}

\textbf{Proof of Lemma \ref{sec04 - theorem01 - proof - lemma01} :} See Appendix \ref{sec04 - theorem01 - proof - lemma01 - proof}.

\begin{lemma}\label{sec04 - theorem01 - proof - lemma02} 
Under Assumptions \ref{sec03 - assumption03}, \ref{sec04 - assumption04}, and \ref{sec04 - assumption06}, for each $g\in [G_0-1]$ and each $\ell\in[L]$, there exist positive constants $\overline{k}_{1},L_{K,1}, L_{K,2} <\infty$ , which are independent of $N$, fold index $\ell\in[L]$, data $\{W_i\}_{i\in[N]}$, such that
\begin{align}
    \Norm{\KER{\widehat{\discfunc}_{\ell,g}}-\KER{\discfunc_{0,g}}}
        \leq    \label{sec04 - theorem01 - proof - lemma02 - eq02-1}      
            &\frac{L_{K,1}}{b_N}\deldel\ind{}{\sum_{g'\ne g}\iota_{\ell,gg'}>0},&\\
    \Norm{\nabla\KER{\widehat{\discfunc}_{\ell,g}}-\nabla\KER{\discfunc_{0,g}}}
        \leq    \label{sec04 - theorem01 - proof - lemma02 - eq02-2}      
            &\frac{L_{K,2}}{b_N}\deldel\ind{}{\sum_{g'\ne g}\iota_{\ell,gg'}>0},&\\
    \Norm{\nabla\KER{\widehat{\discfunc}_{\ell,g}}}^2
        \leq    \label{sec04 - theorem01 - proof - lemma02 - eq02-3}      
            &\overline{k}_{1}^2\sum_{g'\ne g}\iota_{\ell,gg'},&\\
    \Norm{\nabla\KER{\discfunc_{0,g}}}^2
        \leq    \label{sec04 - theorem01 - proof - lemma02 - eq02-4}  
            &\overline{k}_{1}^2\sum_{g'\ne g}\iota_{0,gg'}.&
\end{align}
\end{lemma}

\textbf{Proof of Lemma \ref{sec04 - theorem01 - proof - lemma02} : }See Appendix \ref{sec04 - theorem01 - proof - lemma02 - proof}\\

Because $\iota_{\ell,gg'}$ is measurable with respect to $\discfunc_{0,g}(Z_i)$ conditional on $\mathcal{W}_{-\ell}$, for every fixed $\theta\in\Theta_0$, the tower property and Lemma \ref{sec04 - theorem01 - proof - lemma01}(iv) give:
\begin{align}
    \condexpect{\norm{M_g(\theta,W_i)-M_{G_0}(\theta,W_i)}^2\iota_{\ell,gg'}}{\mathcal{W}_{-\ell}}
        =           \nonumber
                &\condexpect{\condexpect{\norm{M_g(\theta,W_i)-M_{G_0}(\theta,W_i)}^2}{\discfunc_{0,g}(Z_i)}\iota_{\ell,gg'}}{\mathcal{W}_{-\ell}}&\\
        \leq        \nonumber
                &C_M\condexpect{\iota_{\ell,gg'}}{\mathcal{W}_{-\ell}}&\\
        \leq    \label{sec04 - theorem01 - proof - eq00 - M}
                &C_Mc_0(b_NS+\deldel).&
\end{align}
Since $\mu_{0,g}(\theta,Z_i)=\condexpect{M_g(\theta,W_i)-M_{G_0}(\theta,W_i)}{Z_i}$, conditional Jensen's inequality and the tower property yield:
\begin{align}
    \condexpect{\norm{\mu_{0,g}(\theta,Z_i)}^2\iota_{\ell,gg'}}{\mathcal{W}_{-\ell}}
        \leq        \nonumber
                &\condexpect{\condexpect{\norm{M_g(\theta,W_i)-M_{G_0}(\theta,W_i)}^2}{Z_i}\iota_{\ell,gg'}}{\mathcal{W}_{-\ell}}&\nonumber\\
        =           \nonumber
                &\condexpect{\norm{M_g(\theta,W_i)-M_{G_0}(\theta,W_i)}^2\iota_{\ell,gg'}}{\mathcal{W}_{-\ell}}&\\
                    \label{sec04 - theorem01 - proof - eq00 - mu2}   
        \leq    &C_Mc_0(b_NS+\deldel).&\\
    \condexpect{\norm{\mu_{0,g}(\theta,Z_i)}\iota_{\ell,gg'}}{\mathcal{W}_{-\ell}}
        \leq        \nonumber
                &\condexpect{\norm{\mu_{0,g}(\theta,Z_i)}^2\iota_{\ell,gg'}}{\mathcal{W}_{-\ell}}^{1/2}\condexpect{\iota_{\ell,gg'}}{\mathcal{W}_{-\ell}}^{1/2}&\\
        \leq        \label{sec04 - theorem01 - proof - eq00 - mu1}
                &\sqrt{C_M}c_0(b_NS+\deldel).&
\end{align}
The first-stage equation and Assumption \ref{sec03 - assumption03}(b) imply:
\begin{equation}\label{sec04 - theorem01 - proof - eq00 - t}
    \condexpect{\norm{\vect_{i,g}-\discfunc_{0,g}(Z_i)}^2}{Z_i}=\condexpect{\norm{\vect_{i,g}}^2}{Z_i}-\norm{\discfunc_{0,g}(Z_i)}^2\leq C_T.
\end{equation}
Lastly, we have:
\begin{align}
    \condexpect{\norm{\mu_{0,g}(\theta,Z_i)}^2\norm{\vect_{i,g}-\discfunc_{0,g}(Z_i)}^2\iota_{\ell,gg'}}{\mathcal{W}_{-\ell}}
        =           \nonumber
                &\condexpect{\norm{\mu_{0,g}(\theta,Z_i)}^2\iota_{\ell,gg'}\condexpect{\norm{\vect_{i,g}-\discfunc_{0,g}(Z_i)}^2}{Z_i}}{\mathcal{W}_{-\ell}}&\\
        \leq        \nonumber
                &C_T\condexpect{\norm{\mu_{0,g}(\theta,Z_i)}^2\iota_{\ell,gg'}}{\mathcal{W}_{-\ell}}&\\
        \leq        \label{sec04 - theorem01 - proof - eq00 - MuT}
                &C_MC_Tc_0(b_NS+\deldel).&
\end{align}
These bounds also hold at $\theta=\widetilde{\theta}_{\ell}$ conditional on $\mathcal{W}_{-\ell}$. The same calculations with $\iota_{0,gg'}$ give the bounds with $b_NS$ in place of $b_NS+\deldel$.

\paragraph{Step 1.}By the definition of $\psi_N(\cdot)$ and the triangle inequality, we have:
\begin{align*}
    \foldnorm{\psi_{0,i}}
        \leq    &(2G_0-1)\norm{\overline{M}}_2+\sum_{g=1}^{G_0-1}\foldNorm{\frac{\mu_{0,g}(\widetilde{\theta}_{\ell},Z_i)}{b_N}\nabla\KER{\discfunc_{0,g}}^\top(\vect_{i,g}-\discfunc_{0,g})}.&
\end{align*}
By Lemma \ref{sec04 - theorem01 - proof - lemma02} and \eqref{sec04 - theorem01 - proof - eq00 - MuT} with $\iota_{0,gg'}$,
\begin{align}
    \foldNorm{\frac{\mu_{0,g}(\widetilde{\theta}_{\ell},Z_i)}{b_N}\nabla\KER{\discfunc_{0,g}}^\top(\vect_{i,g}-\discfunc_{0,g})}^2
        \leq    &\frac{\overline{k}_{1}^2}{b_N^2}\sum_{g'\ne g}\condexpect{\norm{\mu_{0,g}(\widetilde{\theta}_{\ell},Z_i)}^2\norm{\vect_{i,g}-\discfunc_{0,g}}^2\iota_{0,gg'}}{\mathcal{W}_{-\ell}}&\nonumber\\
        \leq    &\frac{\overline{k}_{1}^2(G_0-1)C_MC_Tc_0S}{b_N}.&\label{sec04 - theorem01 - proof - step1 - eq01}
\end{align}
Since $G_0$ is fixed, $\foldnorm{\psi_{0,i}}=O(b_N^{-1/2})$ on the stated event. Because the upper bound \eqref{sec04 - theorem01 - proof - step1 - eq01} is independent of $\mathcal{W}_{\ell}$ for each $\ell\in[L]$, the same bound holds for $\norm{\psi_{0,i}}_2$ by the law of iterated expectations.

\paragraph{Step 2.} Recall that $\psi_{0,i}\equiv\psi_N(\theta_0,\discfunc_0,\vecmu_0,W_i)$. We first decompose $\sqrt{Nb_N}\avg{\psi_{0,i}}$ as the following:
\[ \sqrt{Nb_N}\avg{\psi_{0,i}}=\sqrt{Nb_N}\avg{\left\{\psi_{0,i}-\expect{\psi_{0,i}}\right\}}+\sqrt{Nb_N}\expect{\psi_{0,i}}. \]
Since $\expect{\phi_N(\theta_0,\discfunc_0,\vecmu_0,W_i)}=0$, Proposition \ref{sec04 - proposition04}(i) and Assumption \ref{sec03 - assumption01}(a) imply
\begin{align*}
    \expect{\psi_{0,i}}
        =   &\expect{m(\theta_0,\smth{A}_i(\discfunc_0),W_i)}&\\
        =   &\expect{m(\theta_0,\classifier{A}_i(\discfunc_0),W_i)} + O(b_N^2)&\\
        =   &\expect{m(\theta_0,\group_{0,i},W_i)} + O(\sqrt{\lambda_{N}}) + O(b_N^2) = O(\sqrt{\lambda_N} + b_N^2),&
\end{align*}
where the last equality comes from \eqref{sec04 - proposition02 - proof - eq02}. By Assumptions \ref{sec04 - assumption06}(b) and (c), the remainder term is:
\[ \sqrt{Nb_N}\expect{\psi_{0,i}} = O\left( \sqrt{Nb_{N}\lambda_{N}} \right) + O\left(\sqrt{Nb_{N}^5} \right) = o(1). \]

Define a function class $\mathcal{F}_{N,\ell}\equiv\left\{\psi_N\big(\theta,\widehat{\discfunc}_{\ell},\widehat{\vecmu}_{\ell}(\widetilde{\theta}_{\ell},\cdot),\cdot\big):\theta\in\Theta\right\}$.
We aim to show that, 
\[ \sup_Q\log N\left(\epsilon\norm{\overline{\psi}_{N,\ell}}_{Q,2},\mathcal{F}_{N,\ell},\norm{\cdot}_{Q,2}\right)\leq G_0\cdot A_0 \log\left(\frac{ G_0^2 A_1}{\epsilon}\right)\quad\textnormal{for all }0<\epsilon\leq1. \]
By construction,
\[ \psi_N\big(\theta,\widehat{\discfunc}_{\ell},\widehat{\vecmu}_{\ell}(\widetilde{\theta}_{\ell},\cdot),\cdot\big)=\sum_{g=1}^{G_0}\left(\eucvec{g}^\top\smth{A}_{i}(\widehat{\discfunc}_\ell)\right)M_g(\theta,\cdot)+\phi_N(\widetilde{\theta}_\ell,\widehat{\discfunc}_\ell,\widehat{\vecmu}_\ell(\widetilde{\theta}_\ell,\cdot),\cdot), \]
where the last term does not depend on $\theta$, and $\widehat{\discfunc}_{\ell}$ and $\widehat{\vecmu}_{\ell}(\widetilde{\theta}_{\ell},\cdot)$ are fixed conditional on $\mathcal{W}_{-\ell}$.

Because covering numbers are invariant to a common translation, $\phi_N(\cdot)$ does not affect the covering number of $\mathcal{F}_{N,\ell}$. Hence, it suffices to derive the entropy bound for the class induced by $\sum_{g=1}^{G_0}(\eucvec{g}^\top\smth{A}_{i}(\widehat{\discfunc}_\ell))M_g(\theta,\cdot)$. Since $\sup_{g\in[G_0]}\norm{\eucvec{g}^\top\smth{A}_{i}(\widehat{\discfunc}_\ell)}_{\infty}\leq G_0$, Assumption \ref{sec03 - assumption02}(d) implies that an envelope for the $\theta$-indexed component is $G_0^2\overline{M}$. Thus, an envelope for $\mathcal{F}_{N,\ell}$ is:
\[ \overline{\psi}_{N,\ell} \equiv G_0^2 \overline{M} + \norm{\phi_N(\widetilde{\theta}_\ell,\widehat{\discfunc}_\ell,\widehat{\vecmu}_\ell(\widetilde{\theta}_\ell,\cdot),\cdot)}, \]
which is $\mathcal{L}^2(P)$-integrable by Assumption\ref{sec03 - assumption02}(d). The bounded-multiplier and finite-sum covering inequalities imply:
\begin{align*}
    \sup_Q\log N\left(\epsilon\norm{\overline{\psi}_{N,\ell}}_{Q,2},\mathcal{F}_{N,\ell},\norm{\cdot}_{Q,2}\right)
        \leq   &G_0\sup_Q\log N\left(\frac{\epsilon}{G_0^2}\norm{\overline{M}}_{Q,2},\mathscr{M},\norm{\cdot}_{Q,2}\right)&\\
        \leq   &G_0\cdot A_0 \log\left(\frac{G_0^2A_{1}}{\epsilon}\right).
\end{align*}
By the analogous argument, a class $\{\psi_{N}(\theta, \discfunc_0, \vecmu_0(\widetilde{\theta}_\ell,\cdot), \cdot) : \theta\in \Theta\}$ also obey the same log-entropy bound. Equipped with the uniform log-entropy condition, applying the Lindeberg-Feller central limit theorem yields:
\[ \sqrt{Nb_{N}}\frac{1}{N}\sum_{i=1}^{N}\psi_{0,i} \overset{d}{\to} \caln\big(0, \Sigma_0 \big). \]

\newcommand{\lin}[1]{ {\Psi}_N( #1 ) }
Lastly, we define an estimator $\theta^*_N$ as a global minimizer of the following problem:
\[ \theta^*_{N} \equiv \argmin{\theta\in\Theta}{\norm{\lin{\theta}}}\quad\textnormal{where}\quad \lin{\theta} \equiv \Upsilon_{0}(\theta-\theta_0)+\avg{\psi_{0,i}}, \]
the linear term is multiplied by $\Upsilon_0$ because $\expect{ \nabla_\theta \psi_{N}( \theta, \discfunc_0, \vecmu_0, W_i ) } = \Upsilon_{0} + O(b_N^2)$ by Proposition \ref{sec04 - proposition04}(b). The resulting first-order condition gives
\[ \sqrt{Nb_N}(\theta^*_{N}-\theta_0)=-(\Upsilon_{0}^\top \Upsilon_{0})^{-1}\Upsilon_{0}^\top\sqrt{Nb_N}\avg{\psi_{0,i}}\overset{d}{\to}\caln(0,\Omega_0). \]
Therefore, it remains to show that $\norm{\widehat{\theta}_{N}-\theta^*_{N}}=o_P(1/\sqrt{Nb_N})$.

Together with the moment and first-stage rate bounds established in Steps 1, 2 and the earlier part of Step 3, the log-entropy bound yields the required stochastic equicontinuity conditional on each training sample. Consequently, over a shrinking neighborhood $\Theta_N$ of $\theta_0$ containing $\widehat{\theta}_N$ and $\theta_N^*$ with probability approaching one,
\[ \sup_{\theta\in\Theta_N}\Norm{ \foldavg{ \avginfold{ \psi_N(\theta, \widehat{\discfunc}_\ell, \widehat{\vecmu}_\ell, W_i) } } -  \lin{\theta} }=o_P\left(\frac{1}{\sqrt{Nb_N}}\right). \]
The standard minimum-distance comparison argument, as in the proof of Theorem 3.3 of \citet{pakespollard1989} and that of Theorem 3.3 of \citet{dml_chernozhukov_et_al2018}, then gives $\norm{\widehat{\theta}_N-\theta_N^*}=o_P( 1/\sqrt{Nb_N} )$.

\paragraph{Step 3.} Fix $\ell\in [L]$. By Proposition \ref{sec04 - proposition02} and since $L$ is fixed,
\[ \max_{\ell\in[L]}\norm{\widetilde{\theta}_{\ell}-\theta_0}=O_P(\lambda_N^{1/2}+\rho_N^{1/2}+N^{-1/2})=o_P(1). \]
Conditional on $\mathcal{W}_{-\ell}$, the primitive estimator and both correction coefficients remain fixed at $\widetilde{\theta}_{\ell}$. $\condexpect{\widehat{\psi}_{\ell,i}-\psi_{0,i}}{\mathcal{W}_{-\ell}}$ is decomposed as:
\begin{align*}
    \condexpect{\widehat{\psi}_{\ell,i}-\psi_{0,i}}{\mathcal{W}_{-\ell}}
        =   &\sum_{g=1}^{G_0-1}\condExpect{ \mu_{0,g}(\widetilde{\theta}_{\ell},Z_i) \left\{\KER{\widehat{\discfunc}_{\ell,g}} - \KER{\discfunc_{0,g}}-\frac{1}{b_N}\nabla\KER{\widehat{\discfunc}_{\ell,g}}^\top(\widehat{\discfunc}_{\ell,g}-\discfunc_{0,g})\right\}}{\mathcal{W}_{-\ell}}&\\
            &+\sum_{g=1}^{G_0-1}\condExpect{\big\{M_g(\theta_0,W_i) - M_{G_0}(\theta_0,W_i)\big\}\left\{\KER{\widehat{\discfunc}_{\ell,g}}-\KER{\discfunc_{0,g}}\right\}}{\mathcal{W}_{-\ell}}&\\
            &-\sum_{g=1}^{G_0-1}\condExpect{\big\{ M_g(\widetilde{\theta}_{\ell},W_i) - M_{G_0}(\widetilde{\theta}_{\ell},W_i)\big\}\left\{\KER{\widehat{\discfunc}_{\ell,g}}-\KER{\discfunc_{0,g}}\right\}}{\mathcal{W}_{-\ell}}&\\
            &-\sum_{g=1}^{G_0-1}\condExpect{\frac{\widehat{\mu}_{\ell,g}(\widetilde{\theta}_{\ell},Z_i)-\mu_{0,g}(\widetilde{\theta}_{\ell},Z_i)}{b_N}\nabla\KER{\widehat{\discfunc}_{\ell,g}}^\top(\widehat{\discfunc}_{\ell,g}-\discfunc_{0,g})}{\mathcal{W}_{-\ell}}.&
\end{align*}
Taylor's theorem and the Lipschitz condition in Lemma \ref{sec04 - theorem01 - proof - lemma02} gives:
\begin{align*}
    &\Norm{\KER{\widehat{\discfunc}_{\ell,g}}-\KER{\discfunc_{0,g}}-\frac{1}{b_N}\nabla\KER{\widehat{\discfunc}_{\ell,g}}^\top(\widehat{\discfunc}_{\ell,g}-\discfunc_{0,g})}&\\
        &\hspace{+1cm}\leq  
                \frac{1}{b_N}\int_{0}^{1}\Norm{  \left\{\nabla\KER{\discfunc_{0,g}+\tau(\widehat{\discfunc}_{\ell,g}-\discfunc_{0,g})} - \nabla\KER{\widehat{\discfunc}_{\ell,g}}\right\}^\top(\widehat{\discfunc}_{\ell,g}-\discfunc_{0,g})}d\tau&\\
        &\hspace{+1cm}\leq  
                \frac{L_{K,2}}{b_N^2}\deldel^2\int_0^1(1-\tau)d\tau\ind{}{\sum_{g'\ne g}\iota_{\ell,gg'}>0}&
\end{align*}
therefore, we have:
\begin{equation*}
    \Norm{\KER{\widehat{\discfunc}_{\ell,g}}-\KER{\discfunc_{0,g}}-\frac{1}{b_N}\nabla\KER{\widehat{\discfunc}_{\ell,g}}^\top(\widehat{\discfunc}_{\ell,g}-\discfunc_{0,g})}
    \leq  
    \frac{L_{K,2}}{2b_N^2}\deldel^2\sum_{g'\ne g}\iota_{\ell,gg'}.
\end{equation*}

By Lemma \ref{sec04 - theorem01 - proof - lemma02}, we have:
\[ \Norm{ \nabla \KER{ \widehat{\discfunc}_{\ell,g}}^\top(\widehat{\discfunc}_{\ell,g} - \discfunc_{0,g})} \leq \overline{k}_{1}\deldel\sum_{g'\ne g}\iota_{\ell,gg'}. \]

Since $\theta_0$ lies in the interior of $\Theta_0$, the segment joining $\theta_0$ and $\widetilde{\theta}_{\ell}$ lies in $\Theta_0$ with probability approaching one. By Taylor's theorem and Assumption \ref{sec03 - assumption02}(c), integration along this segment gives:
\begin{align*}
    &\foldNorm{M_g(\theta_0,W_i)-M_{G_0}(\theta_0,W_i)-M_g(\widetilde{\theta}_{\ell},W_i)+M_{G_0}(\widetilde{\theta}_{\ell},W_i)}&\\
        &\hspace{+1cm}\leq\norm{\widetilde{\theta}_{\ell}-\theta_0}\int_0^1\foldNorm{\nabla_\theta M_g(\theta_0+\tau(\widetilde{\theta}_{\ell}-\theta_0),W_i)-\nabla_\theta M_{G_0}(\theta_0+\tau(\widetilde{\theta}_{\ell}-\theta_0),W_i)}d\tau&\\
        &\hspace{+1cm}\leq2\overline{M}^{(1)}\norm{\widetilde{\theta}_{\ell}-\theta_0}.&
\end{align*}
By Lemmas \ref{sec04 - theorem01 - proof - lemma01}(iv) and \ref{sec04 - theorem01 - proof - lemma02}, the Cauchy--Schwarz inequality therefore gives:
\begin{align}
    &\Norm{\condExpect{\big\{M_g(\theta_0,W_i)-M_{G_0}(\theta_0,W_i)-M_g(\widetilde{\theta}_{\ell},W_i)+M_{G_0}(\widetilde{\theta}_{\ell},W_i)\big\}\big\{\KER{\widehat{\discfunc}_{\ell,g}}-\KER{\discfunc_{0,g}}\big\}}{\mathcal{W}_{-\ell}}}&\nonumber\\
        &\hspace{+1cm}\leq\frac{2\overline{M}^{(1)}L_{K,1}\deldel}{b_N}\norm{\widetilde{\theta}_{\ell}-\theta_0}\sqrt{(G_0-1)c_0(b_NS+\deldel)}.&\label{sec04 - theorem01 - proof - step3 - eq01}
\end{align}

Combining \eqref{sec04 - theorem01 - proof - eq00 - mu1}, and \eqref{sec04 - theorem01 - proof - step3 - eq01} with the two kernel bounds above yields, on the event $\deldel\leq b_N$ (by Assumption \ref{sec04 - assumption06}),
\begin{align*}
    \norm{\condexpect{\widehat{\psi}_{\ell,i}-\psi_{0,i}}{\mathcal{W}_{-\ell}}}
        \leq    & \frac{(G_0-1)^2c_0\sqrt{C_M}L_{K,2}(S+1)}{2b_N}\deldel^2&\\
                & + (G_0-1)^2c_0\overline{k}_{1}(S+1)\deldel\delmu&\\
                & + 2(G_0-1)\overline{M}^{(1)}L_{K,1}\sqrt{(G_0-1)c_0(S+1)}\frac{\deldel}{\sqrt{b_N}}\norm{\widetilde{\theta}_{\ell}-\theta_0}.&
\end{align*}
Here, $\delmu$ controls $\norm{\widehat{\vecmu}_{\ell}(\widetilde{\theta}_{\ell},\cdot)-\vecmu_0(\widetilde{\theta}_{\ell},\cdot)}_\infty$, as follows when the nuisance rate is uniform over $\Theta_0$. Using the rate of the plug-in estimator in Proposition \ref{sec04 - proposition02}, we obtain
\begin{align*}
    \sqrt{Nb_N}\norm{\condexpect{\widehat{\psi}_{\ell,i}-\psi_{0,i}}{\mathcal{W}_{-\ell}}}
        =       &o_P\left(\sqrt{\frac{N}{b_N}}\rho_N^2\right)+o_P\left(\sqrt{Nb_N}\rho_N\nu_N\right)&\\
                &+o_P\left(\sqrt{N}\rho_N\lambda_N^{1/2}\right)+o_P\left(\sqrt{N}\rho_N^{3/2}\right)+o_P(\rho_N).&
\end{align*}

Since $N\rho_N^3=O(1)$ and $\rho_N/b_N = O(1)$ by Assumption \ref{sec04 - assumption06}(c), 
\[ \sqrt{\frac{N}{b_N}}\rho_N^2=\sqrt{N}\rho_N^{3/2}\sqrt{\frac{\rho_N}{b_N}} = O(1). \]

By Assumption \ref{sec04 - assumption06}(d), $\sqrt{Nb_N}\rho_N\nu_N=O(1)$.

Since $\sqrt{N/b_N}\rho_N^2=O(1)$ and since $\sqrt{Nb_N}\lambda_N = o(1)$ by Assumption \ref{sec04 - assumption06}(e), it follows that
\[ \sqrt{N\lambda_N}\rho_N=\left\{\left(\sqrt{\frac{N}{b_N}}\rho_N^2\right)\left(\sqrt{Nb_N}\lambda_N\right)\right\}^{1/2}=o(1). \]
Thus, the conditional bias remainder is $o_P(1)$. Since $L$ is fixed,
\[ \sqrt{Nb_N}\foldavg{\condexpect{\widehat{\psi}_{\ell,i}-\psi_{0,i}}{\mathcal{W}_{-\ell}}}=o_P(1). \]

\paragraph{Step 4.} Fix $\ell\in [L]$. By construction of the orthogonal score, $\widehat{\psi}_{\ell,i} - \psi_{0,i}$ can be decomposed as:
\[ \widehat{\psi}_{\ell,i} - \psi_{0,i} = \sum_{g\in[G_0-1]}\big( r_{\ell,g,i}^{1} + r_{\ell,g,i}^{2} + r_{\ell,g,i}^{3} \big), \]
where
\begin{align*}
    r_{\ell,g,i}^{1}
        \equiv  &   \big\{ \widehat{\mu}_{\ell,g}(\widetilde{\theta}_{\ell}, Z_i) - \mu_{0,g}(\widetilde{\theta}_{\ell}, Z_i) \big\}
                    \frac{1}{b_N}\nabla \KER{ \discfunc_{0,g}}^\top\cdot (\vect_{i,g} - \discfunc_{0, g}),&\\
    r_{\ell,g,i}^{2}
        =       &   \{M_g(\theta_0,W_i)-M_{G_0}(\theta_0,W_i)\} \left\{\KER{\widehat{\discfunc}_{\ell,g}}-\KER{\discfunc_{0,g}} \right\}&\\
                &   +   \frac{\mu_{0,g}(\widetilde{\theta}_{\ell},Z_i)}{b_N} \left\{ \nabla\KER{\widehat{\discfunc}_{\ell,g}}-\nabla\KER{\discfunc_{0,g}}\right\}^\top (\vect_{i,g}-\discfunc_{0,g})
                    -   \frac{\mu_{0,g}(\widetilde{\theta}_{\ell},Z_i)}{b_N}\nabla\KER{\widehat{\discfunc}_{\ell,g}}^\top(\widehat{\discfunc}_{\ell,g}-\discfunc_{0,g}).&\\
    r_{\ell,g,i}^{3}
        \equiv  &   \big\{ \widehat{\mu}_{\ell,g}(\widetilde{\theta}_{\ell}, Z_i) - \mu_{0,g}(\widetilde{\theta}_{\ell}, Z_i) \big\}
                    \frac{1}{b_N} \left\{ \nabla \KER{ \widehat{\discfunc}_{\ell,g} }^\top
                    \cdot (\vect_{i,g} - \widehat{\discfunc}_{\ell, g}) - \nabla \KER{ \discfunc_{0,g}}^\top\cdot (\vect_{i,g} - \discfunc_{0, g})\right\}.&
\end{align*}
Then the triangle inequality gives:
\[ b_N\foldnorm{\widehat{\psi}_{\ell,i}-\psi_{0,i}}^2\leq 3(G_0-1)b_N\sum_{g=1}^{G_0-1}\left\{\foldnorm{r_{\ell,g,i}^{1}}^2+\foldnorm{r_{\ell,g,i}^{2}}^2+\foldnorm{r_{\ell,g,i}^{3}}^2\right\}. \]

For the first remainder, \eqref{sec04 - theorem01 - proof - eq00 - t}, Lemma \ref{sec04 - theorem01 - proof - lemma02}, and Lemma \ref{sec04 - theorem01 - proof - lemma01}(iv) give
\begin{align}
    \foldnorm{r_{\ell,g,i}^{1}}^2
        \leq        \nonumber
                &\frac{\norm{\widehat{\mu}_{\ell,g}(\widetilde{\theta}_{\ell},\cdot)-\mu_{0,g}(\widetilde{\theta}_{\ell},\cdot)}_\infty^2}{b_N^2}\condExpect{\Norm{\nabla\KER{\discfunc_{0,g}}}^2\norm{\vect_{i,g}-\discfunc_{0,g}}^2}{\mathcal{W}_{-\ell}}&\\
        \leq        \nonumber
                &\frac{C_T\overline{k}_{1}^2}{b_N^2}\delmu^2\sum_{g'\ne g}\condexpect{\iota_{0,gg'}}{\mathcal{W}_{-\ell}}&\\
        \leq        \label{sec04 - theorem01 - proof - step4 - eq01}
                &\frac{(G_0-1)c_0SC_T\overline{k}_{1}^2}{b_N}\delmu^2.&
\end{align}
The second remainder term is bounded by:
\begin{equation}\label{sec04 - theorem01 - proof - step4 - eq02}
    \foldnorm{r_{\ell,g,i}^{2}}^2
        \leq    3(G_0-1)c_0C_M(b_NS+\deldel)\deldel^2 \left\{\frac{L_{K,1}^2+\overline{k}_{1}^2}{b_N^2}+\frac{C_TL_{K,2}^2}{b_N^4} \right\}.
\end{equation}
This is because applying Lemma \ref{sec04 - theorem01 - proof - lemma02} and \eqref{sec04 - theorem01 - proof - eq00 - M}, \eqref{sec04 - theorem01 - proof - eq00 - MuT}, and \eqref{sec04 - theorem01 - proof - eq00 - mu2}, respectively, gives:
\begin{align*}
    &\foldNorm{\{M_g(\theta_0,W_i)-M_{G_0}(\theta_0,W_i)\} \left\{\KER{\widehat{\discfunc}_{\ell,g}}-\KER{\discfunc_{0,g}} \right\}}^2&\\
        &\hspace{+1cm}\leq\frac{L_{K,1}^2\deldel^2}{b_N^2}\sum_{g'\ne g}\condexpect{\norm{M_g(\theta_0,W_i)-M_{G_0}(\theta_0,W_i)}^2\iota_{\ell,gg'}}{\mathcal{W}_{-\ell}}&\\
        &\hspace{+1cm}\leq (G_0-1)c_0C_ML_{K,1}^2\frac{b_NS+\deldel}{b_N^2}\deldel^2,&\\
    &\foldNorm{\frac{\mu_{0,g}(\widetilde{\theta}_{\ell},Z_i)}{b_N}\{\nabla\KER{\widehat{\discfunc}_{\ell,g}}-\nabla\KER{\discfunc_{0,g}}\}^\top(\vect_{i,g}-\discfunc_{0,g})}^2&\\
        &\hspace{+1cm}\leq\frac{L_{K,2}^2\deldel^2}{b_N^4}\sum_{g'\ne g}\condexpect{\norm{\mu_{0,g}(\widetilde{\theta}_{\ell},Z_i)}^2\norm{\vect_{i,g}-\discfunc_{0,g}}^2\iota_{\ell,gg'}}{\mathcal{W}_{-\ell}}&\\
        &\hspace{+1cm}\leq (G_0-1)c_0C_MC_TL_{K,2}^2\frac{b_NS+\deldel}{b_N^4}\deldel^2,&\\
    &\foldNorm{\frac{\mu_{0,g}(\widetilde{\theta}_{\ell},Z_i)}{b_N}\nabla\KER{\widehat{\discfunc}_{\ell,g}}^\top(\widehat{\discfunc}_{\ell,g}-\discfunc_{0,g})}^2&\\
        &\hspace{+1cm}\leq\frac{\overline{k}_{1}^2\deldel^2}{b_N^2}\sum_{g'\ne g}\condexpect{\norm{\mu_{0,g}(\widetilde{\theta}_{\ell},Z_i)}^2\iota_{\ell,gg'}}{\mathcal{W}_{-\ell}}&\\
        &\hspace{+1cm}\leq (G_0-1)c_0C_M\overline{k}_{1}^2\frac{b_NS+\deldel}{b_N^2}\deldel^2.&
\end{align*}

For the third remainder, using \eqref{sec04 - theorem01 - proof - eq00 - t} and Lemma \ref{sec04 - theorem01 - proof - lemma01}(iv) yields
\begin{align}
    \foldnorm{r_{\ell,g,i}^{3}}^2
        \leq        \nonumber
                &2 \delmu^2 \times \left\{\frac{L_{K,2}^2\deldel^2}{b_N^4}\sum_{g'\ne g}\condexpect{\norm{\vect_{i,g}-\discfunc_{0,g}}^2\iota_{\ell,gg'}}{\mathcal{W}_{-\ell}} + \frac{\overline{k}_{1}^2\deldel^2}{b_N^2}\sum_{g'\ne g}\condexpect{\iota_{\ell,gg'}}{\mathcal{W}_{-\ell}}\right\}&\\
        \leq        \label{sec04 - theorem01 - proof - step4 - eq03}
                &2(G_0-1)c_0(b_NS+\deldel)\deldel^2\delmu^2\left\{\frac{C_TL_{K,2}^2}{b_N^4}+\frac{\overline{k}_{1}^2}{b_N^2}\right\}.&
\end{align}

On the event $\deldel\leq b_N$ (by Assumption \ref{sec04 - assumption06} (c)), \eqref{sec04 - theorem01 - proof - step4 - eq01}, \eqref{sec04 - theorem01 - proof - step4 - eq02}, and \eqref{sec04 - theorem01 - proof - step4 - eq03} imply:
\begin{align*}
    b_N\foldnorm{\widehat{\psi}_{\ell,i}-\psi_{0,i}}^2
        =   &O_{P}\big( \delmu^2 + (1 +b_{N}^{-2})\deldel^2\left\{ 1+ \delmu^2\right\} \big)&\\
        =   &o_{P}\big( \nu_{N}^2 \big) + o_{P}\left( \rho_{N}^2 \right) + o_{P}\left( \rho_{N}^2 \nu_{N}^2 \right) + o_{P}\left( \frac{\rho_{N}^2}{b_N^2} \right) + o_{P}\left( \frac{\nu_{N}^2\rho_{N}^2}{b_{N}^2} \right)&
\end{align*}
and therefore this is $o_{P}(1)$ by Assumptions \ref{sec04 - assumption06}(a) and (c).

Conditional on $\mathcal{W}_{-\ell}$, we have:
\begin{align*}
    \condExpect{\Norm{\sqrt{Nb_N}\avginfold{ \big\{\widehat{\psi}_{\ell,i}-\psi_{0,i}-\condexpect{\widehat{\psi}_{\ell,i}-\psi_{0,i}}{\mathcal{W}_{-\ell}} \big\}}}^2}{\mathcal{W}_{-\ell}}
        \leq    &\frac{Nb_N}{|I_\ell|}\condExpect{\Norm{\widehat{\psi}_{\ell,i}-\psi_{0,i}-\condexpect{\widehat{\psi}_{\ell,i}-\psi_{0,i}}{\mathcal{W}_{-\ell}}}^2}{\mathcal{W}_{-\ell}}&\\
        \leq    &\frac{Nb_N}{|I_\ell|}\foldnorm{\widehat{\psi}_{\ell,i}-\psi_{0,i}}^2.&
\end{align*}
For every $\epsilon>0$, the conditional Markov inequality and the law of iterated expectations give:
\begin{align*}
    \Prob{\Norm{\sqrt{Nb_N}\avginfold{ \left\{\widehat{\psi}_{\ell,i}-\psi_{0,i}-\condexpect{\widehat{\psi}_{\ell,i}-\psi_{0,i}}{\mathcal{W}_{-\ell}} \right\} }}>\epsilon}
        \leq    \,\Expect{\min\left\{1,\frac{Nb_N}{\epsilon^2|I_\ell|}\foldnorm{\widehat{\psi}_{\ell,i}-\psi_{0,i}}^2\right\}}.&
\end{align*}
Because $N/|I_\ell|\to L$ is restricted and the random variable inside the expectation converges to zero in probability, therefore,
\[ \sqrt{Nb_N}\foldavg{\avginfold{ \big\{ \widehat{\psi}_{\ell,i}-\psi_{0,i}-\condexpect{\widehat{\psi}_{\ell,i}-\psi_{0,i}}{\mathcal{W}_{-\ell}} \big\} } }=o_P(1), \]
which completes the proof. \qed

\subsubsection{Proof of Lemma \ref{sec04 - theorem01 - proof - lemma01}}\label{sec04 - theorem01 - proof - lemma01 - proof}

The statement (i) follows directly from $\deldel\geq0$ and the definitions of $\iota_{0,gg'}$ and $\iota_{\ell,gg'}$.

For the statement (ii), Assumption \ref{sec04 - assumption04} implies that $k(t)>0$ if and only if $t>-S$. Since
\[ \Ker{\discfunc_{0,g}}=\prod_{h\ne g}k\left(\frac{\delta_{0,gh}(Z)}{b_N}\right)\quad\textnormal{and}\quad\pmb{K}(\widehat{\discfunc}_{\ell,g}/b_N)=\prod_{h\ne g}k\left(\frac{\widehat{\delta}_{\ell,gh}(Z)}{b_N}\right), \]
each product is nonzero if and only if every factor is positive, establishing the statement (ii).

For the statement (iii), the product rule gives
\[ \{\nabla\Ker{\discfunc_{0,g}}\}_{g'}=k'\left(\frac{\delta_{0,gg'}(Z)}{b_N}\right)\prod_{h\ne g,g'}k\left(\frac{\delta_{0,gh}(Z)}{b_N}\right). \]
Since $k'(t)=0$ for $|t|\geq S$, a nonzero component implies
\[ |\delta_{0,gg'}(Z)|<b_NS\quad\textnormal{and}\quad\delta_{0,gh}(Z)>-b_NS\quad\textnormal{for }h\ne g,g', \]
and therefore $\iota_{0,gg'}=1$. Analogously, a nonzero estimated gradient component implies
\begin{equation}\label{sec04 - theorem01 - proof - lemma01 - proof - eq01}
    |\widehat{\delta}_{\ell,gg'}(Z)|<b_NS\quad\textnormal{and}\quad\widehat{\delta}_{\ell,gh}(Z)>-b_NS\quad\textnormal{for }h\ne g,g'.
\end{equation}
On this event, the definition of $\deldel$ gives
\begin{equation}\label{sec04 - theorem01 - proof - step1 - eq00 - 3}
    \begin{aligned}
        |\delta_{0,gg'}(Z)|
            \,\,\leq\,\,    &|\widehat{\delta}_{\ell,gg'}(Z)|+|\widehat{\delta}_{\ell,gg'}(Z)-\delta_{0,gg'}(Z)|&\\
            \,\,\leq\,\,    &b_NS+\deldel,&\\
        \delta_{0,gh}(Z)
            \,\,=\,\,       &\widehat{\delta}_{\ell,gh}(Z)+(\delta_{0,gh}(Z)-\widehat{\delta}_{\ell,gh}(Z))&\\
            \,\,\geq\,\,    &-b_NS-\deldel.&
    \end{aligned}
\end{equation}
Thus, $\iota_{\ell,gg'}=1$, establishing the statement (iii).

Lastly, conditional on $\mathcal{W}_{-\ell}$, $\deldel$ is fixed and the evaluation observation has its original distribution. Since each indicator takes values in $\{0,1\}$, the margin condition of Assumption \ref{sec03 - assumption03}(e) gives
\begin{align*}
    \foldnorm{\iota_{0,gg'}}^2
        =       &\condexpect{\iota_{0,gg'}}{\mathcal{W}_{-\ell}}&\\
        \leq    &\condprob{|\delta_{0,gg'}(Z)|\leq b_NS}{\mathcal{W}_{-\ell}}
        \leq c_0b_NS,&\\
    \foldnorm{\iota_{\ell,gg'}}^2
        =       &\condexpect{\iota_{\ell,gg'}}{\mathcal{W}_{-\ell}}&\\
        \leq    &\condprob{|\delta_{0,gg'}(Z)|\leq b_NS+\deldel}{\mathcal{W}_{-\ell}}
        \leq c_0(b_NS+\deldel).&
\end{align*}
Taking square roots establishes the statement (iv). \qed

\subsubsection{Proof of Lemma \ref{sec04 - theorem01 - proof - lemma02}}\label{sec04 - theorem01 - proof - lemma02 - proof}
By Assumption \ref{sec04 - assumption04}, $k(t)$ is constant outside $[-S,S]$ and differentiable at $t=\pm S$, so that $k'(-S)=k'(S)=0$. There exist positive constants $\overline{k}_{1},\overline{k}_{2}<\infty$ such that, for every $u,v\in\mathbb{R}$,
\[ |k(u)-k(v)|\leq\overline{k}_{1}|u-v|\quad\textnormal{and}\quad|k'(u)-k'(v)|\leq\overline{k}_{2}|u-v|. \]
The bounds follow from the mean value theorem on $(-S,S)$ and the constant extension outside $[-S,S]$.

Since $k(\cdot)\in[0,1]$, the product inequality gives, for every $\pmb{u},\pmb{v}\in\mathbb{R}^{G_0-1}$,
\begin{equation}
    \norm{\pmb{K}(\pmb{u})-\pmb{K}(\pmb{v})}
        \leq    \sum_{g=1}^{G_0-1}|k(u_g)-k(v_g)|
        \leq    \overline{k}_{1}\sum_{g=1}^{G_0-1}|u_g-v_g|
        \leq \sqrt{G_0-1}\overline{k}_{1}\norm{\pmb{u}-\pmb{v}}.
        \label{sec04 - theorem01 - proof - lemma02 - proof - eq01-1}
\end{equation}
Hence, we take $L_{K,1}=\sqrt{G_0-1}\cdot \overline{k}_{1}$.

For every $g\in[G_0-1]$,
\begin{align*}
    \left|k'(u_g)\prod_{h\ne g}k(u_h)-k'(v_g)\prod_{h\ne g}k(v_h)\right|
        \leq    &|k'(u_g)-k'(v_g)|\prod_{h\ne g}|k(u_h)|+|k'(v_g)|\left|\prod_{h\ne g}k(u_h)-\prod_{h\ne g}k(v_h)\right|&\\
        \leq    &\overline{k}_{2}|u_g-v_g|+\overline{k}_{1}^2\sum_{h\ne g}|u_h-v_h|.&
\end{align*}
The matrix with zero diagonal and unit off-diagonal entries has Euclidean operator norm $G_0-2$. Therefore,
\begin{equation}\label{sec04 - theorem01 - proof - lemma02 - proof - eq01-2}
    \norm{\nabla\pmb{K}(\pmb{u})-\nabla\pmb{K}(\pmb{v})}\leq\left\{\overline{k}_{2}+(G_0-2)\overline{k}_{1}^2\right\}\norm{\pmb{u}-\pmb{v}},
\end{equation}
and we take $L_{K,2}=\overline{k}_{2}+(G_0-2)\overline{k}_{1}^2$. These constants depend only on $G_0$ and the kernel bounds.

By Lemma \ref{sec04 - theorem01 - proof - lemma01}, if $\sum_{g'\ne g}\iota_{\ell,gg'}=0$, the gradient vanishes along the entire segment joining $\discfunc_{0,g}/b_N$ and $\widehat{\discfunc}_{\ell,g}/b_N$. Integration along this segment gives $\Ker{\widehat{\discfunc}_{\ell,g}}=\Ker{\discfunc_{0,g}}$, and both endpoint gradients are zero. Combining this observation with \eqref{sec04 - theorem01 - proof - lemma02 - proof - eq01-1} and \eqref{sec04 - theorem01 - proof - lemma02 - proof - eq01-2} yields
\begin{align*}
    \Norm{\KER{\widehat{\discfunc}_{\ell,g}}-\KER{\discfunc_{0,g}}}
        \leq    &\frac{L_{K,1}}{b_N}\deldel\ind{}{\sum_{g'\ne g}\iota_{\ell,gg'}>0},&\\
    \Norm{\nabla\KER{\widehat{\discfunc}_{\ell,g}}-\nabla\KER{\discfunc_{0,g}}}
        \leq    &\frac{L_{K,2}}{b_N}\deldel\ind{}{\sum_{g'\ne g}\iota_{\ell,gg'}>0}.&
\end{align*}

Finally, each gradient component is bounded in absolute value by $\overline{k}_{1}$. The support implication above at $\tau=1$ gives
\[  \Norm{\nabla\KER{\widehat{\discfunc}_{\ell,g}}}^2
        \leq    \overline{k}_{1}^2\sum_{g'\ne g}\iota_{\ell,gg'}\quad\text{and}\quad
    \Norm{\nabla\KER{\discfunc_{0,g}}}^2
        \leq    \overline{k}_{1}^2\sum_{g'\ne g}\iota_{0,gg'}. \]
\qed

\subsection{Proof of Proposition \ref{sec04 - proposition05}}\label{sec04 - proposition05 - proof}

Let $\widehat{\psi}_{\ell,i}\equiv\psi_{N,\ell}(\widehat{\theta}_N,W_i)$ and $\psi_{0,i}\equiv\psi_N(\theta_0,\discfunc_0,\vecmu_0,W_i)$. Then, by Proposition \ref{sec04 - proposition04}(b), $\widehat{\Upsilon} - \Upsilon_{0} = \widehat{\Upsilon} - \expect{ \nabla \psi_{0,i} } - O(b_N^2)$ is:
\begin{equation*}
    \widehat{\Upsilon} - \Upsilon_{0}
        =   \foldavg{\left(\avginfold{\nabla_\theta \widehat{\psi}_{\ell,i}} -\condexpect{\nabla_\theta\widehat{\psi}_{\ell,i} }{ \mathcal{W}_{-\ell}}\right)}
            +\left( \foldavg{ \condexpect{\nabla_\theta\widehat{\psi}_{\ell,i}}{\mathcal{W}_{-\ell}} - \expect{\nabla_\theta\psi_{0,i}}} \right) - O(b_N^2).
\end{equation*}
The first term is $o_P(1)$ by the conditional uniform law of large numbers and the entropy condition established in Step 4 for proof of Theorem \ref{sec04 - theorem01}. The second term is $o_P(1)$ by $\widehat\theta_N \overset{P}{\to}\theta_0$, the first-stage rate conditions, and the Lipschitz properties in Lemma \ref{sec04 - theorem01 - proof - lemma02}, therefore meaning  $\widehat{\Upsilon} = \Upsilon_{0} + o_P(1)$.

By Assumption \ref{sec04 - assumption07}, $\condexpect{ b_N \psi_{0,i}\psi_{0,i}^\top }{\mathcal{W}_{-\ell}} \to \Sigma_0$. We decompose $\widehat{\Sigma} - \Sigma_0$ by:
\[  \widehat{\Sigma}-\Sigma_0
        =   \foldavg{b_N\left(\avginfold{\widehat\psi_{\ell,i}\widehat\psi_{\ell,i}^{\top}}-\condexpect{\widehat\psi_{\ell,i}\widehat\psi_{\ell,i}^{\top}}{\mathcal{W}_{-\ell}}\right)}  + \foldavg{b_N\condexpect{\widehat\psi_{\ell,i}\widehat\psi_{\ell,i}^{\top}-\psi_{0,i}\psi_{0,i}^{\top}}{\mathcal{W}_{-\ell}}},  \]
for each $b_N$. The first term is $o_P(1)$ by the conditional law of large numbers for the triangular array and $Nb_N\to\infty$. For the second term, using an inequality of $\norm{\pmb{u}\pmb{u}^\top-\pmb{v}\pmb{v}^\top} \leq \norm{\pmb{u}-\pmb{v} } \cdot \big(\norm{\pmb{u} }+\norm{\pmb{v} }\big)$, the Cauchy--Schwarz inequality gives:
\begin{align*}
    b_N\Norm{\condexpect{\widehat\psi_{\ell,i}\widehat\psi_{\ell,i}^{\top}-\psi_{0,i}\psi_{0,i}^{\top}}{\mathcal{W}_{-\ell}}}
        \leq    &\left(\sqrt{b_N}\norm{\widehat\psi_{\ell,i}-\psi_{0,i}}_{2}\right)\left(\sqrt{b_N}\norm{\widehat\psi_{\ell,i}}_{2}+\sqrt{b_N}\norm{\psi_{0,i}}_{2}\right)=o_P(1),
\end{align*}
where the last equality follows from the first-stage rate conditions and
\[ \sqrt{b_N}\norm{\widehat\psi_{\ell,i}-\psi_{0,i}}_{2}=o_P(1),\qquad \sqrt{b_N}\norm{\widehat\psi_{\ell,i}}_{2}+\sqrt{b_N}\norm{\psi_{0,i}}_{2}=O_P(1). \]
Thus, $\widehat{\Sigma} = {\Sigma_0} + o_P(1)$. 

Finally, Proposition \ref{sec04 - proposition04}(iii) implies that $\Upsilon_{0}$ has full column rank for all sufficiently large $N$. Hence, the continuous mapping theorem establishes the consistency of the variance estimator. \qed

\end{appendix}
\end{document}